\documentclass[utf8,sort&compress,table,hidelinks]{frontiersFPHY} 

\setcitestyle{square} 
\usepackage{url,hyperref,lineno,microtype,subcaption}
\usepackage[labelfont={bf,sf},%
labelsep=period,%
justification=centering,
labelformat=parens,labelsep=quad,skip=3pt,font=scriptsize]{caption}
\usepackage{graphicx}
\usepackage[onehalfspacing]{setspace}
\usepackage{multicol}
\usepackage[encoding,filenameencoding=utf8]{grffile}
\usepackage{xspace}
\usepackage{multirow}
\usepackage{booktabs}

\newcommand{\thesi}{SI\xspace}
\newcommand{\tbd}{\textcolor{red}{TBD}\xspace}

\def\keyFont{\fontsize{8}{11}\helveticabold }
\def\firstAuthorLast{Coupette {et~al.}} 
\def\Authors{Corinna Coupette\,$^{1,\dagger}$, Janis Beckedorf\,$^{2,\dagger}$, Dirk Hartung\,$^{3,4,*}$,\\ Michael Bommarito\,$^{5}$, and Daniel Martin Katz\,$^{3,4,5}$}
\def\Address{%
$^{\dagger}$These authors have contributed equally to this work and share first authorship.\\
$^{1}$Max Planck Institute for Informatics, Saarbr\"ucken, Germany\\ 
$^{2}$Ruprecht-Karls-Universit\"at Heidelberg, Heidelberg, Germany\\
\mbox{$^{3}$Center for Legal Technology and Data Science, Bucerius Law School, Hamburg, Germany}\\
$^{4}$CodeX -- The Stanford Center for Legal Informatics, Stanford Law School, CA,  USA\\
$^{5}$Illinois Tech -- Chicago Kent College of Law, Chicago, IL, USA
}
\def\corrAuthor{Dirk Hartung}

\def\corrEmail{dirk.hartung@law-school.de}

\graphicspath{ {./figures/} }

\begin{document}
\onecolumn
\firstpage{1}

\title[Measuring Law Over Time]{Measuring Law Over Time\newline\vspace*{-6pt}{\Large A network analytical framework with an application\newline to statutes and regulations in the United States and Germany}}

\author[\firstAuthorLast]{\Authors} 
\address{} 
\correspondance{} 

\extraAuth{}

\maketitle

\begin{abstract}
\section{}
How do complex social systems evolve in the modern world? 
This question lies at the heart of social physics, 
and network analysis has proven critical in providing answers to it.
In recent years, network analysis has also been used to gain a quantitative understanding of law as a complex adaptive system, 
but most research has focused on legal documents of a single type, 
and there exists no unified framework for quantitative legal document analysis using network analytical tools.
Against this background, we present a comprehensive framework for analyzing legal documents as multi-dimensional, dynamic document networks. 
We demonstrate the utility of this framework by applying it to an original dataset of statutes and regulations from two different countries, the United States and Germany, spanning more than twenty years ($1998$--$2019$).
Our framework provides tools for assessing the size and connectivity of the legal system as viewed through the lens of specific document collections as well as for tracking the evolution of individual legal documents over time. 
Implementing the framework for our dataset, we find that at the federal level, the United States legal system is increasingly dominated by regulations, whereas the German legal system remains governed by statutes.
This holds regardless of whether we measure the systems at the macro, the meso, or the micro level.

\tiny
 \keyFont{ \section{Keywords:} legal complexity, evolution of law, quantitative legal studies, empirical legal studies, legal data science, network analysis, natural language processing, complex systems} 
\end{abstract}

\section{Introduction}
\label{sec:introduction}

Originating from mathematics and physics, complexity science has been successfully applied in the study of social phenomena \cite{mitchell2009,miller2009}. 
More recently, it was introduced as an approach to gain a quantitative understanding of the structure and evolution of law \cite{ruhl2017}. 
While legal scholars have long used concepts and terminology from complexity science in legal theory \cite{murray2018,ruhl1995,scott1993},
research has also called for the development of computational models, methods, and metrics to describe how law evolves in practice \cite{ruhl2015}.

Network analysis, a critical tool for understanding many complex systems \cite{amaral2004,albert2002,watts1998}, has proven particularly useful for scientific work answering this call.
It has been used, inter alia, to analyze network data derived from decisions by national courts \cite{coupette2019,winkels2019,black2013,lupu2013,bommarito2011,cross2010,fowler2007},
international courts \cite{olsen2020,alschner2018,larsson2017,tarissan2016a,panagis2015,pelc2014,lupu2012},
and international tribunals \cite{charlotin2017, langford2017},
as well as from statutes \cite{katz2020,coupette2019a,boulet2018,koniaris2018,li2015,katz2014,bommarito2010},
constitutions \cite{lee2019,rutherford2018,rockmore2017},
and international treaties \cite{boulet2019,alschner2016,kim2013,kinne2013,saban2010}.
Relevant work in this context explored, for example, which characteristics of complex systems occur in statutory law \cite{katz2020,koniaris2018,li2015}, how references to judicial decisions are used to shape legal arguments \cite{larsson2017,black2013,lupu2013}, or where social dynamics exist between judges or international arbitrators \cite{katz2010,langford2017}.
The network analytical methods employed include centrality measures, clustering, and degree distributions \cite{katz2020,coupette2019,winkels2019,lee2019,alschner2016}.
However, while all studies examine network representations of legal document collections, the data models and methods employed vary widely, which makes it hard to assess the quality of individual results and compare findings across studies.
Furthermore, most of this research considers one legal document type only, 
and some important categories of legal documents, most prominently regulations (i.e., rules promulgated by the executive branch of government with authorization of the legislative branch of government), have---to the best of our knowledge---not received any network analytic attention.

This points to two gaps in the literature: 
First, on the methodological side, there exists no comprehensive framework for quantitative legal document analysis using network analytical tools.
Such a framework should be flexible in three ways:
It should (1) produce sensible results for different document types, countries, and time periods, 
(2) allow us to explore document collections of vastly different sizes, 
and (3) offer insights on the global (\emph{macro}), intermediate (\emph{meso}), and local (\emph{micro}) level of analysis. 
Second, on the empirical side, there is a lack of studies that combine multiple legal document types or include regulations.

In this article, we take a step toward filling both gaps.
We offer a comprehensive framework for analyzing legal documents as multi-dimensional, dynamic document networks 
and demonstrate its utility by applying it to an original dataset of statutes and regulations from two different countries, the United States and Germany, that spans more than twenty years ($1998$--$2019$).
Our framework provides tools for assessing the size and connectivity of the legal system as viewed through the lens of specific document collections as well as for profiling individual legal documents over time. 
It goes beyond the existing literature, inter alia, by adapting network analytical methods to the peculiarities of legal documents, allowing the joint examination of multiple document types, and enabling temporal analysis.
Implementing the framework for our dataset, we find that the United States legal system is increasingly dominated by regulations, 
whereas the German legal system remains governed by statutes, 
regardless of whether we measure the systems at the macro, the meso, or the micro level.

The remainder of the paper is structured as follows.
In Section~\ref{sec:data}, we specify our network model of legal documents and detail how we instantiate it to analyze statutes and regulations in the United States and Germany.
Section~\ref{sec:methods} describes our methodological framework,
and the results of applying this framework to our original dataset are presented in Section~\ref{sec:results}.
We conclude by discussing the strengths and weaknesses of our approach in Section~\ref{sec:discussion}, where we also identify avenues for future research. 
Our exposition uses the basic terminology of graphs and networks; for textbook introductions, see \cite{easley2010,barabasi2016,newman2018}.

\section{Data}
\label{sec:data}

In this section, we introduce our network model of legal documents (\ref{subsec:data:model}) and instantiate it for our original dataset (\ref{subsec:data:instance}).

\vspace*{6pt}
\subsection{Data Model Specification}\label{subsec:data:model}

\begin{figure}
    \centering
    \includegraphics[width=\textwidth]{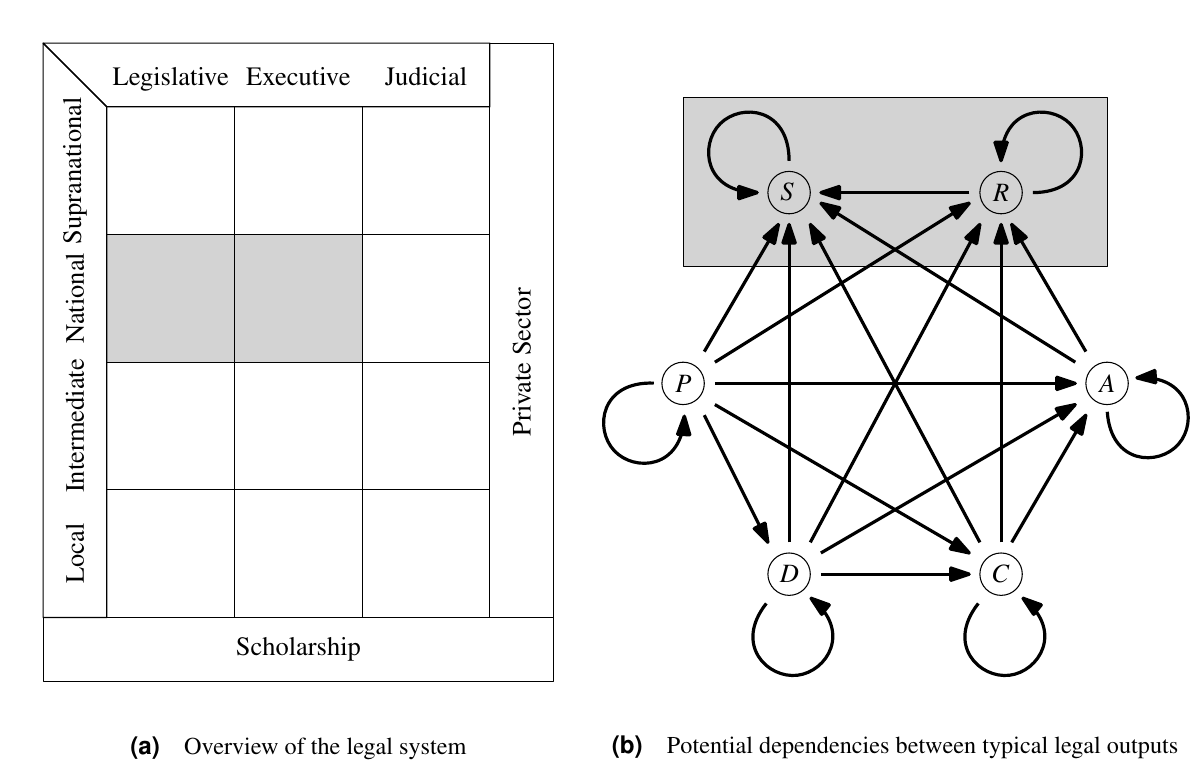}
    \caption{Two-dimensional overview of the legal system (left) and dependencies between its typical outputs (right), with the areas covered by the dataset introduced in this work highlighted in grey. 
    The legal outputs in Panel~(b) are (clockwise from the top left and in reverse topological order) \textbf{S}tatutes, \textbf{R}egulations, Administrative \textbf{A}cts, \textbf{C}ontracts, (Court) \textbf{D}ecisions, and (Scholarly) \textbf{P}ublications.
    The arrows illustrate typical dependencies between the document types, e.g., through explicit references (although in reality, most dependencies can be bidirectional, e.g., some courts also cite legal scholarship). 
    Note that the sources covered by our dataset lie at the end of all typical dependency chains (i.e., statutes and regulations are last in the topological order of the dependency graph).
    }\label{fig:legal-system}
\end{figure}

As visualized in Figure~\ref{fig:legal-system}~(a), the legal system consists of multiple levels: 
the local (e.g., municipal) level, the intermediate (e.g., state or province) level, the national (e.g., federal) level, and the supranational (including international) level.
Horizontally, it is usually subdivided into the legislative, executive, and judicial branches of government. 
These public parts are framed by the private sector, which operates on all levels, and the research community, which studies all parts of the legal system (including itself \cite{katz2011,newton2011,schwartz2011,george2006,ellickson2000}).
In all parts of the legal system, agents of varying sizes produce different types of outputs that create, modify, delete, apply, debate, or evaluate legal rules.
These agents and their typical outputs are summarized in Table~\ref{tab:legal-system}.

 \begin{table}
	\centering
	\renewcommand{\arraystretch}{1.5}
\small
    \begin{tabular}{rllllll}
		\toprule&\multicolumn{3}{c}{\bfseries Branch of Government}&\bfseries Private Sector&&\bfseries Scholarship\\
		&\bfseries Legislative & \bfseries Executive & \bfseries Judicial&&&\\\midrule
		\bfseries Typical Large Agent&Parliament&Agency&Court&Firm&&Institute\\
		\bfseries Typical Small Agent&Parliamentarian&Bureaucrat&Judge&Individual&&Scholar\\
		\bfseries Typical Output(s)&Statute&Regulation&Decision&Contract&&Publication\\[-6pt]
		&&Administrative Act&&\\
		\bottomrule
	\end{tabular}

	\caption{Overview of agents and outputs in the legal system.}\label{tab:legal-system}
\end{table}

As the agents interact, they consciously interconnect their outputs. 
For example, court decisions regularly contain references to statutes, regulations, contracts, and other court decisions.
Figure~\ref{fig:legal-system}~(b) gives an overview of the classic dependencies between the typical outputs of agents in the legal system.
It illustrates that the \emph{documented} part of the legal system constitutes a multilayered document network, 
which is changing over time as the agents continue producing or amending their outputs.
Since the connections between the legal documents are placed deliberately by the agents, they encode valuable information about the content and the context of these documents. 
A lot of this information cannot be inferred from the documents' language alone (reliably or at all).
Therefore, investigating the dynamic document network representation of a legal system using network analytical tools promises insights into its structure and evolution that would be hard or impossible to obtain via other methods. 

To perform network analysis of a dynamic network of legal documents, we need to represent it as a series of graphs. 
Here, we build on a generalizable network model of statutory materials \cite{katz2020} and exploit that the typical outputs listed in Table~\ref{tab:legal-system} have three common features (beyond the obvious characteristic that they all contain \emph{text}):
\begin{enumerate}
    \item They are hierarchically structured (\emph{hierarchy}).
    \item Their text is placed in containers that are sequentially ordered and can be sequentially labeled (\emph{sequence}).
    \item Their text may contain explicit citations or cross-references (henceforth: references) to the text in other legal documents or in other parts of the same document (\emph{reference}).
\end{enumerate}

\begin{figure}
	\centering
	\includegraphics[width=\textwidth]{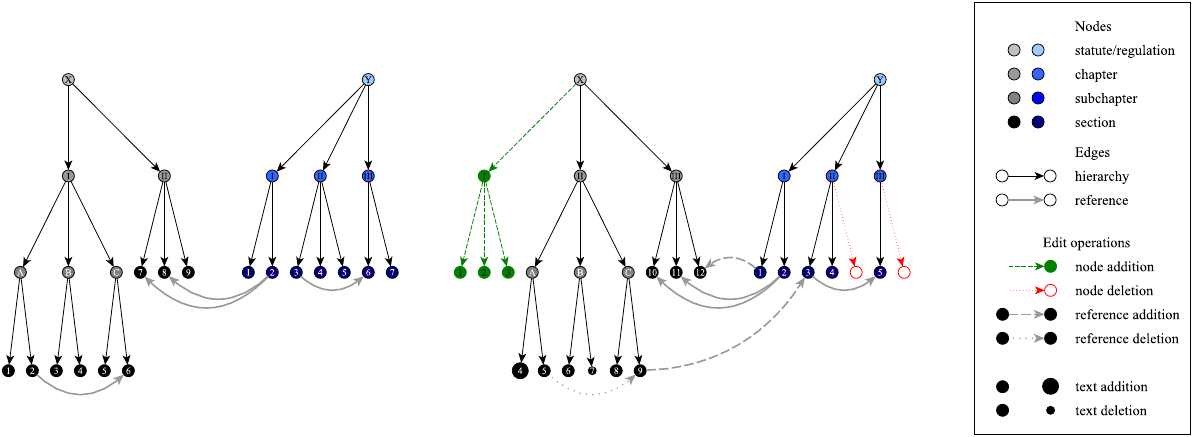}
	\caption{Our legal network data model (adapted from \cite{katz2020}), illustrated for a dummy graph containing one statute and one regulation: initial configuration (left) and potential edit operations (right).}\label{fig:model}
\end{figure}

Therefore, each document at a given point in time (henceforth: \emph{snapshot}) is represented as a (sub)graph, with its \emph{hierarchy} modeled as a tree using \emph{hierarchy edges}.
We capture a document's \emph{reference} using \emph{reference edges} at the level corresponding to the document's \emph{sequence}, which, inter alia, avoids that the graph induced by these references becomes too sparse (thereby eliminating some noise in the data and facilitating its analysis).
The result is a directed multigraph, as illustrated in Figure~\ref{fig:model} for documents that are statutes or regulations (whose sequence level is the \emph{section} level). 
Depending on the analytical focus, other edge types can be included (e.g., \emph{authority edges} pointing from regulations to statutes can indicate which statutes delegated the rule-making power used to create which regulations), 
and depending on the document types considered, different types of edit operations are possible (e.g., court decisions and scholarly articles are only seldom changed after their initial publication), 
but the general model applies to all outputs listed in Table~\ref{tab:legal-system}.

\vspace*{6pt}
\subsection{Data Model Instantiation}\label{subsec:data:instance}
To illustrate the power of the methodological framework laid out in Section~\ref{sec:methods} and produce the results presented in Section~\ref{sec:results}, we instantiate the data model described in Section~\ref{subsec:data:model} with the outputs of the legislative and executive branches of government at the national level in the United States and Germany, i.e., federal statutes and regulations, over the $22$~years from $1998$ to $2019$ (inclusive).
The rules contained in these sources are universally binding and directly enforceable through public authority (the combination of which distinguishes them from the other outputs listed in Table~\ref{tab:legal-system}).
In the United States and Germany, they are arranged into edited collections: the United States Code (USC) and the Code of Federal Regulations (CFR) in the United States, and the federal statutes and federal regulations (which have no special name) in Germany.
These collections are actively maintained to reflect the latest consolidated state of the law (though, in the United States, the consolidation may lag several years behind the actual law).
As such, they are a best-effort representation of all universally binding and directly enforceable rules at the federal level in their country at any point in time, commonly referred to as \emph{codified law}.

For the United States, we obtain the annual versions (reflecting the state of the codified law at the \emph{end} of the respective year) of the United States Code (USC) and the Code of Federal Regulations (CFR) from the United States Office of the Law Revision Council\footnote{\url{https://uscode.house.gov/download/annualhistoricalarchives/downloadxhtml.shtml}.} and the United States Government Publishing Office\footnote{\url{https://www.govinfo.gov/bulkdata/CFR}.}, respectively.
For Germany, we create a parallel set of annual snapshots for all federal statutes and regulations in effect at the \emph{end} of the year in question based on documents from Germany’s primary legal data provider, \emph{juris GmbH}.\footnote{
This differs from the approach taken in \cite{katz2020}, where the annual snapshots represented the law in effect at the \emph{beginning} of the year in question.
}
These data sources are the most complete presently available, 
but they may still be incomplete.
They also reflect choices made by and events affecting the agents in charge of their maintenance, e.g., varying rates and orders of updates, purposeful or unintentional omissions or modifications, and changes in the agents' composition (e.g., as a consequence of elections).

We perform several preprocessing steps on the raw input data, detailed in the Supplementary Information (\thesi), to extract the hierarchy, sequence, and reference structure contained in each collection.
The results are directed multigraphs, one per country and year, akin to those illustrated in Figure~\ref{fig:model}. 
These graphs contain all structural elements of the USC and the CFR (in the United States) or the federal statutes and regulations (in Germany) as nodes 
and all direct inclusion relationships (\emph{hierarchy}) and atomic references (\emph{reference}) as edges, where the references are resolved to the section level (\emph{sequence}).
Each graph represents the \emph{codified law} of a particular country in a particular year, containing documents of two \emph{document types} (statutes and regulations) at the federal level.

When modeling codified law as just described, we take a couple of design decisions that limit the scope of the results presented in Section~\ref{sec:results}. 
First, we focus on \emph{codified} law, i.e., \emph{law in books}, excluding other legal materials listed in Table~\ref{tab:legal-system}, especially those representing \emph{law in action} (in the sense of \cite{pound1910}),  
or even other representations of legislative materials such as the United States Statutes at Large or the German Federal Law Gazette (Bundesgesetzblatt).
These materials all merit investigation, and they need to be included in an all-encompassing assessment of the legal system. 
Our current work also serves as a preparatory step toward realizing this larger vision.

Second, we extract \emph{atomic} and \emph{explicit} references that follow a specified set of common patterns only, i.e., references including---in a typical format---a particular section (called ``Paragraph'' or ``Artikel'' in German law), a list of sections, or a range of sections.
With this procedure, we exclude \emph{container} references (e.g., references to an entire chapter of the USC), \emph{pinpoint} references (e.g., references to a codified Act of the United States Congress by its popular name), \emph{implicit} references (e.g., the use of a certain term implying its definition), and \emph{explicit} references following \emph{uncommon} patterns.
As sketched in Section~3.3 of the \thesi, there are plenty of such references, especially in the CFR, and including them would produce results different from those presented in Section~\ref{sec:results}.
However, the graph representation of such references is inherently ambiguous, 
and their extraction is inherently more challenging than the extraction of atomic citations.
Solving these problems falls outside the scope of this paper but presents an interesting opportunity for future work.

Third, we resolve the atomic references we extract to the level of sections, rather than the smallest referenced unit (which might be a subsection or even an item in an enumeration), thereby effectively discarding potentially valuable information.
Since for statutes and regulations, the section level corresponds to the documents' sequence level, this is consistent with our data model.
It also reflects a focus on the perspective of the user, who tends to navigate the law on the section level because it is the only level at which the individual German laws or their United States counterparts, the chapters of the USC and the CFR, are uniquely sequentially labeled. 
Finally, it ensures a certain degree of comparability because sections are the only structural elements in which text is (with very few exceptions) guaranteed to be contained (albeit the amount of text varies widely across sections).
Therefore, resolving references to the section level is reasonable for our purposes, but further research is needed on how the choice of the resolution level impacts the analysis of legal networks.

\section{Methods}
\label{sec:methods}

Since the legal system produces a diverse set of outputs, many of which are rich in internal structure, a methodological framework for its dynamic network analysis must allow for many different foci and units of analysis. 
Our methods are designed to match this need, enabling us to describe the legal system and its evolution in its entirety (\emph{macro level}), through selected sets of legal documents (\emph{meso level}), or using individual documents and their substructures \emph{(micro level}), 
all while integrating documents of potentially different types.
More precisely, we provide tools for measuring 
\begin{enumerate}
	\item the \emph{growth} of the legal system (\ref{subsec:methods:growth}),
	\item the macro-level, meso-level, and micro-level \emph{connectivity} of the legal system (\ref{subsec:methods:connectivity}), and 
	\item the \emph{evolution} of individual units of law (e.g., single statutes, single regulations, or their sections or chapters) in the legal system (\ref{subsec:methods:profiles}).
\end{enumerate}
While most network analytical tools we employ are conceptually simple, the challenge lies in selecting adequate units of analysis and metrics allowing for substantive interpretation.
In the following, we refer to the object of study as \emph{the legal system} for brevity, 
but one should keep in mind that this system is explored through the window provided by the document collection underlying the analysis (cf.~Figure~\ref{fig:legal-system}), 
and therefore can also refer to a \emph{national legal system}.

\vspace*{6pt}
\subsection{Growth}
\label{subsec:methods:growth}

As detailed in Section~\ref{subsec:data:model}, despite their diversity, the typical outputs of the legal system have \emph{hierarchy}, \emph{sequence}, and \emph{reference} as common structural features, and they also all contain \emph{text}. 
Therefore, to assess the growth of the legal system, we track the number of tokens (roughly corresponding to words), the number of structural elements (i.e., hierarchical structures), and the number of references between documents of the same type at the documents' sequence level (e.g., the section level for statutes and regulations) across all documents in the collection, separately for each document type and across all temporal snapshots (e.g., all years). 
For the token counts, we concatenate the text of all materials for one snapshot and document type and split on whitespace characters. 
For a given document type, the structural element counts reflect the number of nodes of that type, and the reference counts reflect the number of reference edges between nodes of that type in our graphs.
These measures give a first, high-level idea of how the legal system evolves, and the aggregation by document type allows us to uncover, e.g., differences in growth rates. 

\begin{figure} 
	\centering
	\includegraphics[width=0.5\textwidth]{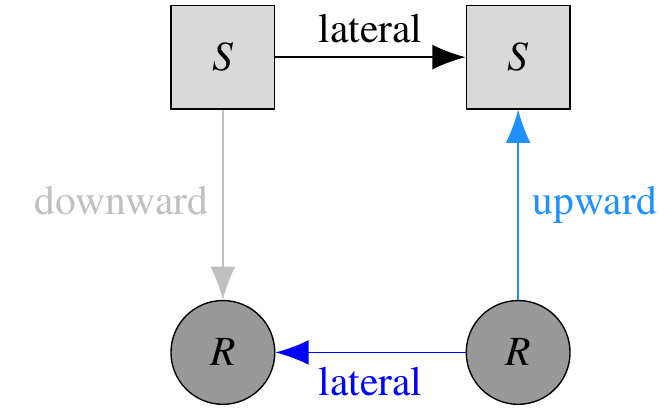}
	\caption{Differentiation of reference types for a document collection containing two types of documents: statutes and regulations. 
	Light grey squares marked $S$ represent sections of statutes, dark grey circles marked $R$ represent sections of regulations. 
	Given that statutes stand above regulations in the hierarchy of legal rules, 
	a reference from a statute to a regulation is \emph{downward} (silver), a reference from a regulation to a statute is \emph{upward} (light blue), 
	and references between documents of the same type are \emph{lateral} (black and blue).}\label{fig:crossref-differentiation}
\end{figure}

To explore the relevance of references between documents of different types, if the document collection contains $x$ types of documents, we distinguish between $x^2$ types of references. 
Figure~\ref{fig:crossref-differentiation} illustrates the idea for $x = 2$ with statutes and regulations as document types.
We count the number of references of each type for each temporal snapshot.

Raw reference counts do not show how the incoming and outgoing references are distributed across the individual sequence-level items. 
The user experience of a legal system, however, depends crucially on these distributions: 
If the typical item on the sequence level has very few outgoing references, the \emph{expected} cost of navigating the law is much smaller than if the outgoing references are uniformly distributed, 
and if the distribution of incoming references is very skewed, when reading the law, we are much more likely to pass a few prominent sequence-level items than a large number of less prominent ones
(of course, the \emph{actual} cost of the user also depends on the size of the items to be navigated, which can vary widely, e.g., amongst the sections in the USC).
Therefore, we inspect the evolution of the in-degree distribution and the out-degree distribution of the subgraph induced by the reference edges.
We compute these distributions separately for each combination of reference edge types (e.g., considering any combination of the reference types depicted in Figure~\ref{fig:crossref-differentiation} for a document collection containing statutes and regulations) and across all snapshots.
This allows us to evaluate whether growth in the number of references further amplifies differences in the prominence of different parts of the law, 
which would be reflected in a lengthening or thickening of the distributions' tails, and to assess how this affects the navigability of a legal system.

\vspace*{6pt}
\subsection{Connectivity}
\label{subsec:methods:connectivity}

When exploring the connectivity of the legal system over time, we distinguish between macro-level connectivity (\ref{subsubsec:methods:connectivity:macro}), meso-level connectivity (\ref{subsubsec:methods:connectivity:meso}), and micro-level connectivity (\ref{subsubsec:methods:connectivity:micro}). 

\vspace*{6pt}
\subsubsection{Macro-level connectivity}
\label{subsubsec:methods:connectivity:macro}

Investigating connectivity at the macro level helps us understand how information in the legal system is organized and processed. 
As the basis of all analyses, we consider the graph induced by the structural items on the documents' sequence level (referred to as \emph{seqitems} in \cite{katz2020}) as nodes and the references between them as edges. 
For each snapshot, we count the number of non-trivial connected components (i.e., components with more than one node). 
Furthermore, we compute the fraction of nodes in the largest (weakly) connected component, 
the fraction of nodes in satellites, i.e., non-trivial components that are not the largest connected component, 
and the fraction of isolated nodes. 
We do this for the graph containing nodes of all document types as well as for the graphs containing only nodes of a single document type. 
These statistics provide a high-level overview of the system's information infrastructure and how it changes over time, 
and they enable a differentiated assessment of the role of documents of different types. 

For a more detailed picture, we draw on concepts introduced in the study of the Web graph \cite{broder2000}, which have also proven useful in the analysis of complex and self-organizing systems, e.g., in biology \cite{friedlander2015,csete2004,ma2003}. 
More precisely, we analyze the largest connected component of each of the sequence-level reference graphs, 
tracking the fraction of nodes contained in its strongly connected component, its in-only component (i.e., the nodes which can reach \emph{to} but cannot be reached \emph{from} the strongly connected component), its out-only component (i.e., the nodes which can be reached \emph{from} but cannot reach \emph{to} the strongly connected component), and its tendrils and tubes (whatever remains), 
again across all snapshots.
We ask to which extent the legal system has a bowtie structure (i.e., a small strongly connected core joined by larger in-only and out-only components), which has been associated with ``effective trade-offs among efficiency, robustness and evolvability'' \cite{csete2004}, inter alia, in complex biological systems, and whether any empirical deviations from that structure are characteristic of legal information processing.

\vspace*{12pt}
\subsubsection{Meso-level connectivity}
\label{subsubsec:methods:connectivity:meso}

One fundamental question concerning a legal system's connectivity at the meso level is how it self-organizes into areas of law.
Existing taxonomies categorizing the law into distinct fields are largely based on tradition (e.g., the titles of the USC) or intuition (e.g., the thematic categories used by some legal database providers). 
Exploiting the connectivity provided by references between legal documents at the meso level, network analytical methods provide an alternative, data-driven approach to mapping the law.
To implement such an approach, 
we follow a multi-step procedure:
\begin{enumerate}
	\item We preprocess the graphs for each snapshot by taking the quotient graph at the granularity we are interested in (e.g., at the level of individual chapters for an analysis of the USC and the CFR).
	That is, we remove all nodes above and below that level and reroute all references outgoing from or incoming to a lower-level node to the node's unique ancestor that lies on the level of interest.\label{step:quotient} 
	\item We cluster each of the \emph{undirected} versions of the graphs from Step~\ref{step:quotient} separately using the \emph{Infomap} algorithm \cite{rosvall2008,rosvall2009} with a parametrization that mirrors domain knowledge, and passes sensitivity and robustness checks. 
	Leveraging the randomness inherent in this algorithm, we increase the robustness of the clustering for each graph by computing the \emph{consensus clustering} \cite{lancichinetti2012} of $1000$ \emph{Infomap} runs with different seeds, where two nodes are put into the same cluster if they are in the same cluster in $95~\%$ of all runs.
	We choose the \emph{Infomap} algorithm as our clustering algorithm because it is scalable, has a solid information-theoretic foundation, and mirrors the process in which users like lawyers navigate law (inter alia, by identifying a relevant section of a statute, reading that section, then potentially following a reference).\label{step:clustering} 
	\item We compute pairwise alignments between the clusterings of all temporally adjacent snapshots based on the nodes of the \emph{unpreprocessed} graphs that wrap text (for details on our alignment procedure, see Section~4.3.1 in the \thesi).
	This is most relevant for collections containing documents that can change over time (e.g., statutes and regulations), and it allows us to assess, inter alia, what amount of text from a cluster $A$ in year $y$ is contained in a cluster $B$ in year $y+1$.\label{step:alignment}   
	\item We use the clusterings from Step~\ref{step:clustering} and the alignments from Step~\ref{step:alignment} to define a \emph{cluster family graph} as introduced in \cite{katz2020}. 
	This graph contains all clusters from all snapshots as nodes, and two clusters $A$ and $B$ are connected by a (weighted) edge if $A$ lies in snapshot $y$, $B$ lies in snapshot $y+1$, at least $p~\%$ of the tokens from $A$ are contained in $B$, and at least $p~\%$ of the tokens from $B$ are contained in $A$, where $p$ is chosen based on the analytical resolution we are interested in.\label{step:familygraph}
	\item We define a \emph{cluster family} as a connected component in a cluster family graph from Step~\ref{step:familygraph} and compute, for each cluster family in each year, the number of tokens it contains from each document type.\label{step:family}
\end{enumerate}
This process is a variant of the family graph construction developed for statutes in \cite{katz2020}, 
with the modification that we now allow for input data containing documents of different types.
It results in a dynamic, data-driven map of the legal system that accounts for the information provided by the references between its documents.

\vspace*{6pt}
\subsubsection{Micro-level connectivity}
\label{subsubsec:methods:connectivity:micro}

At the micro level, the connectivity created by references allows us to assess the roles of individual units of law. 
Regardless of the level at which we aggregate the references between documents, 
the shapes of the nodes' neighborhoods at that level contain valuable information about their function in the legal system. 
This information is partly accounted for in the meso-level connectivity assessment, which leverages local \emph{density}.
While local density can help us find out which nodes interact strongly, local \emph{sparsity} lets us identify nodes that play a particularly prominent role for the information flow in the network: 
If a node's neighbors are themselves only very sparsely connected (i.e., their neighbors almost form an independent set), the node provides an important \emph{bridge} between them.
We call the ego graph of such a node a \emph{star}, 
with the node as its \emph{hub} and the node's neighbors as the \emph{spokes}.

\begin{figure}
	\centering
	\includegraphics{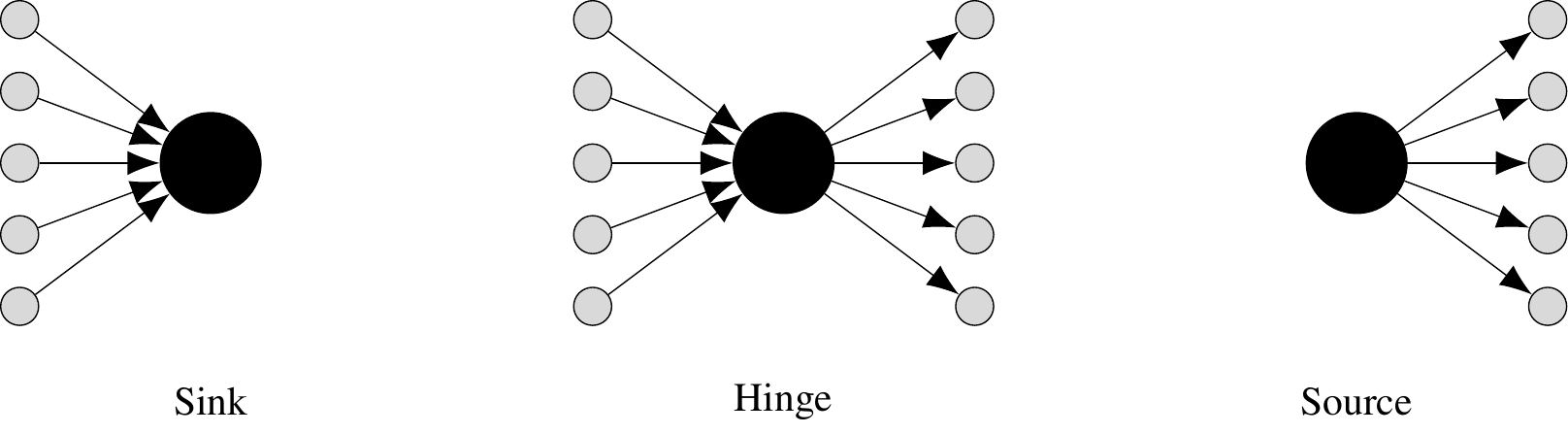}
	\caption{Star types. 
	If the hub's in-degree is at least ten times its out-degree, the star is a \emph{sink}. 
	If the hub's out-degree is at least ten times its in-degree, the star is a \emph{source}. 
	Otherwise, the star is a \emph{hinge}.}\label{fig:star-types}
\end{figure}

In directed graphs, we can classify stars according to the ratio between the hub's in-degree and the hub's out-degree as depicted in Figure~\ref{fig:star-types}.
More precisely, we define the type of a star $s$ with hub $v$ as follows:
\begin{align*}
	\text{type}(s) := \begin{cases}
		\frac{\delta^-(v)}{\delta^+(v)} \geq 10&\text{sink}\\
		\frac{1}{10} < \frac{\delta^-(v)}{\delta^+(v)} < 10&\text{hinge}\\
		\frac{\delta^+(v)}{\delta^-(v)} \geq 10&\text{source},
	\end{cases}
\end{align*}
where $\delta^+(v)$ is $v$'s out-degree and $\delta^-(v)$ is $v$'s in-degree.
In the legal system, the type of a star captures the hub's role in mediating the information flow in its neighborhood.

To identify and classify stars in the legal system at the documents' sequence level,
for each snapshot, we create the ego graph for each node $v$ in the graph induced by the reference edges (where we exclude parallel edges). 
We then iteratively remove the node $w$ that is connected to most of $v$'s neighbors while $w$ is connected to more than $5~\%$ of $v$'s neighborhood (excluding $w$), and keep the ego graph if it has a certain minimum size determined by the size of the collection (e.g., $10$ nodes for collections with several thousands of items on the sequence level).
The stars produced in this way contain no spoke that is connected to more than $5~\%$ of the other spokes, and we classify them to identify those sequence-level items that are vital to the information flow in their neighborhoods and to describe the type of mediation they perform.
To find stars at levels above the documents' sequence level, we can apply the methodology just described on graphs that aggregate references at those levels (e.g., on the quotient graphs described in Section~\ref{subsubsec:methods:connectivity:meso}).

\vspace*{12pt}
\subsection{Profiles}
\label{subsec:methods:profiles}

To assess the evolution of individual units of law (e.g., individual court decisions or chapters of a regulation) in the legal system, we create profiles of these units covering all temporal snapshots in the document collection under study.
More specifically, based on the quotient graphs that are created on the level of our unit of interest and contain only reference edges (like the preprocessed graphs described in Step~\ref{step:quotient}, Section~\ref{subsubsec:methods:connectivity:meso}), 
we track ten statistics in five groups (note that not all of these statistics can change over time for units of all legal document types):
\begin{enumerate}
	\item the number of tokens and the number of unique tokens,
	\item the number of items above, on, and below the sequence level (provided our unit of analysis lies above the sequence level),
	\item the number of self-loops,
	\item the weighted in- and out-degree (accounting for parallel edges), and
	\item the binary in- and out-degree (excluding parallel edges).
\end{enumerate} 

These statistics capture how the unit of law in focus evolves in size (number of tokens), 
topical breadth (number of unique tokens), 
structure (number of items above, on, and below the sequence level), 
self-referentiality (number of self-loops), 
scope of interdependence within the legal system (weighted in-degree and weighted out-degree), 
and diversity of interdependence within the legal system (binary in-degree and binary out-degree).
Finally, by constructing the ego graphs of the profiled unit for its out-neighborhood and its in-neighborhood and following the evolution of these ego graphs across snapshots,
we assess to which extent a profiled unit references which other units (\emph{reliance}) 
and to which extend it is referenced by which other units (\emph{responsibility}).

\section{Results}
\label{sec:results}

In the following, we apply the framework presented in Section~\ref{sec:methods} to the data introduced in Section~\ref{subsec:data:instance}, i.e., codified law comprising federal statutes and regulations in the United States and Germany over the $22$ years from $1998$ to $2019$ (inclusive).
We start by examining the growth of the United States and German national legal systems (henceforth: the national legal systems) as viewed through the lens of our data (Section~\ref{subsec:results:growth}).
Next, we investigate the macro-level, meso-level, and micro-level connectivity of these national legal systems (Section~\ref{subsec:results:connectivity}).
Finally, we explore the evolution of selected chapters of the USC and the CFR and selected German statutes and regulations within their national legal systems in a case study focusing on financial regulation (Section~\ref{subsec:results:profiles}).
The results we report are mostly descriptive, and as discussed in Section~\ref{sec:discussion}, identifying causal factors behind the dynamics we observe or interpreting our results using a qualitative approach is left to future research.

\vspace*{30pt}
\subsection{Growth}
\label{subsec:results:growth}

\begin{figure}
	\centering
	\includegraphics[width=\textwidth]{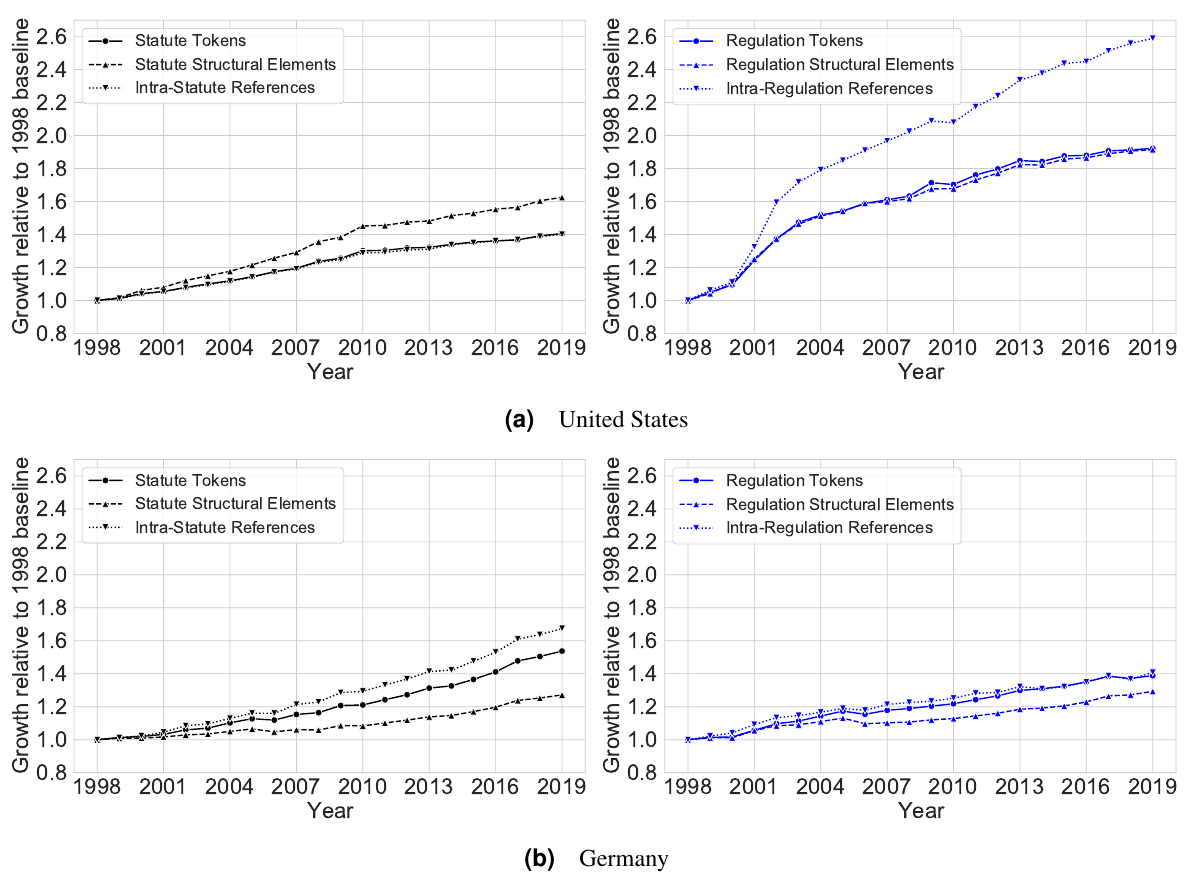}
	\caption{Growth relative to the $1998$ baseline for statutes (left) and regulations (right) in the United States (top) and Germany (bottom).}\label{fig:basic-statistics}
\end{figure}

Figure \ref{fig:basic-statistics} summarizes the growth of the United States legal system and the German legal system as measured by the tokens, structural elements, and lateral references contained in their codified law. 
Each row of the figure corresponds to a country, and each column corresponds to a document type. 
All counts are divided by their value in $1998$, i.e., Figure~\ref{fig:basic-statistics} depicts growth relative to the $1998$ baseline.
Supplementing the time series data, 
Table~\ref{tab:basic-statistics} provides the absolute counts for $1998$ and $2019$ and the total percentage change between these years ($\Delta$).

\begin{table}
	\centering
			\renewcommand{\arraystretch}{1.5}
				\begin{tabular}{lrrrrrr}
\toprule & \multicolumn{3}{c}{\textbf{Statutes}} & \multicolumn{3}{c}{\textbf{Regulations}} \\
            & 1998   & 2019   &   $\Delta$ & 1998   & 2019   &   $\Delta$ \\
\midrule
 \textbf{Tokens}     & 15.2~M               & 21.4~M               &                      41 & 43.9~M                  & 84.3~M                  &                         92 \\
 \textbf{Structures} & 516.2~K              & 838.8~K              &                      63 & 1.4~M                   & 2.7~M                   &                         91 \\
 \textbf{References} & 80.1~K               & 112.1~K              &                      40 & 134.6~K                 & 348.4~K                 &                        159 \\
\bottomrule
\end{tabular}
				
				{\vspace*{6pt}\small \textbf{\textsf{(a)}}\quad United States}\vspace*{12pt}
			
				\begin{tabular}{lrrrrrr}
\toprule & \multicolumn{3}{c}{\textbf{Statutes}} & \multicolumn{3}{c}{\textbf{Regulations}} \\
            & 1998   & 2019   &   $\Delta$ & 1998   & 2019   &   $\Delta$ \\
\midrule
 \textbf{Tokens}     & 5.0~M                & 7.7~M                &                      54 & 3.9~M                   & 5.4~M                   &                         39 \\
 \textbf{Structures} & 130.6~K              & 166.0~K              &                      27 & 87.9~K                  & 113.7~K                 &                         29 \\
 \textbf{References} & 86.4~K               & 144.6~K              &                      67 & 33.5~K                  & 47.1~K                  &                         41 \\
\bottomrule
\end{tabular}
				
				{\vspace*{6pt}\small \textbf{\textsf{(b)}}\quad Germany} 

	\caption{(Rounded) size of the national legal systems of the United States (top) and Germany (bottom) as measured by the tokens, structural elements, and references in their codified law in $1998$ and $2019$, including the total percentage change between these years ($\Delta$).}\label{tab:basic-statistics}
\end{table}

Figure~\ref{fig:basic-statistics} and Table~\ref{tab:basic-statistics} show that over the last two decades, the legal systems of both countries have grown substantially.  
In the United States, the USC (containing codified statutes) has over $60$ new structural elements (e.g., chapters, parts, or sections) in $2019$ for every $100$ such elements it had in $1998$.
Notably, as evident from the upper left panel of Figure~\ref{fig:basic-statistics}, the growth rate of the USC appears to have experienced two distinct periods when measured by its structural elements: 
one period with a monotonic growth rate of approximately $4~\%$ per year ($1998$--$2010$), followed by another period with a decelerated monotonic growth rate of approximately $2~\%$ per year ($2010$--$2019$). 
At a slightly lower level, this trend also occurs for both the number of tokens and the number of intra-USC references. 
For example, there are approximately $40$ new tokens or references in $2019$ for every $100$ tokens or references that existed in $1998$.
Shifting the focus for the United States to the CFR (containing codified regulations), as observable from the upper right panel of Figure~\ref{fig:basic-statistics}, the quantity of regulations has increased by an even greater factor.  
For every $100$ structural elements or tokens that were present in $1998$, approximately $90$ additional elements or tokens exist in $2019$. 
This increase is even more extreme for intra-CFR references, where there are almost $160$ \emph{new} references in $2019$ for every $100$ that existed in $1998$.  
Apart from brief intervals of stagnation or slight decrease ($2009$--$2010$, $2013$--$2014$), 
these increases have been monotonic.

Corresponding trends for German statutes and regulations are presented in the bottom row of Figure \ref{fig:basic-statistics}.  
Growth in the German legal system has been qualitatively similar to that in the United States legal system but quantitatively less pronounced and of different functional shape. 
For both German statutes and German regulations, there are approximately $30$ new structural elements in $2019$ for every $100$ that existed in $1998$. 
Unlike in the United States, however, this growth has been non-monotonic: 
When measured through structural elements, both statutes and regulations experienced some periods of shallow decline between $2005$ and $2010$.
These shrinking periods are generally not mirrored by the token and lateral reference counts, with one notable exception:
In the period from $2005$ to $2006$, \emph{all} German statistics decreased. 
This is likely due to statutes aiming to cleanse the law (\emph{Rechtsbereinigungsgesetze}), eight of which were introduced in $2006$ (recall that our $2006$ snapshot represents the law at the \emph{end} of $2006$).\footnote{%
These statutes are: 
(1) Erstes Gesetz über die Bereinigung von Bundesrecht im Zuständigkeitsbereich des Bundesministeriums des Innern vom 19. Februar 2006 (BGBl.~I~S.~334),
(2) Gesetz zur Bereinigung des Bundesrechts im Zuständigkeitsbereich des Bundesministeriums für Ernährung, Landwirtschaft und Verbraucherschutz vom 13. April 2006 (BGBl.~I~S.~855),
(3) Erstes Gesetz über die Bereinigung von Bundesrecht im Zuständigkeitsbereich des Bundesministeriums der Justiz vom 19. April 2006 (BGBl.~I~S.~866),
(4) Erstes Gesetz zur Bereinigung des Bundesrechts im Zuständigkeitsbereich des Bundesministeriums für Wirtschaft und Technologie und im Zuständigkeitsbereich des Bundesministeriums für Arbeit und Soziales vom 19. April 2006 (BGBl.~I~S.~894),
(5) Gesetz zur Änderung und Bereinigung des Lastenausgleichsrechts vom 21. Juni 2006 (BGBl.~I~S.~1323),
(6) Gesetz über die Bereinigung von Bundesrecht im Zuständigkeitsbereich des Bundesministeriums für Arbeit und Soziales und des Bundesministeriums für Gesundheit vom 14. August 2006 (BGBl.~I~S.~1869),
(7) Erstes Gesetz über die Bereinigung von Bundesrecht im Zuständigkeitsbereich des Bundesministeriums für Verkehr, Bau und Stadtentwicklung vom 19. September 2006 (BGBl.~I~S.~2146), and
(8) Zweites Gesetz über die Bereinigung von Bundesrecht im Zuständigkeitsbereich des Bundesministeriums des Innern vom 2. Dezember 2006 (BGBl.~I~S.~2674).
}
In total, there are approximately $55$ new statute tokens in Germany in $2019$ for every $100$ such tokens that existed in $1998$. 
Like \emph{regulations} in the United States, German \emph{statutes} experienced a greater increase in the quantity of lateral references than in other metrics: 
For every $100$ references in $1998$, there are approximately $70$ new references in $2019$. 
For German \emph{regulations}, as for \emph{statutes} in the United States, the rate of change has been more similar across metrics, with growth varying roughly between $30~\%$ and $40~\%$ (as noted in Table~\ref{tab:basic-statistics}).
At a high level, the growth of the German legal system thus seems to be driven by statutes, whereas the growth of the United States legal system appears to be driven by regulations.

\begin{figure}
	\centering
	\includegraphics[width=\textwidth]{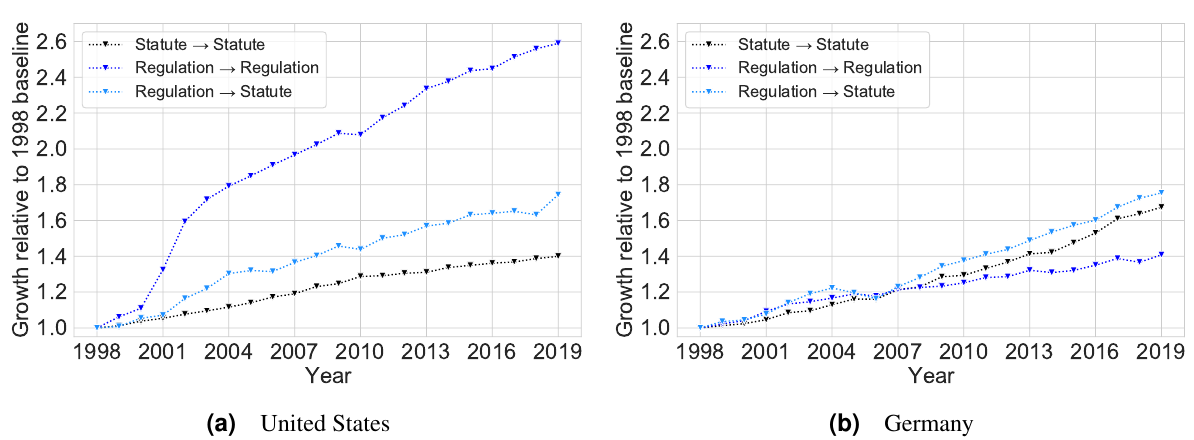}
	\caption{Growth relative to the $1998$ baseline for lateral and upward references in the United States (left) and Germany (right).}\label{fig:crossref-evolution}
\end{figure}

Figure~\ref{fig:basic-statistics} and Table~\ref{tab:basic-statistics} only account for \emph{lateral} references, excluding references between documents of different types. 
Therefore, Figure~\ref{fig:crossref-evolution} shows growth relative to the $1998$ baseline for lateral and \emph{upward} (i.e., regulation-to-statute) references. 
We exclude \emph{downward} references because they are very few in number (which means that even a small absolute increase results in a large relative increase) 
but note that, contrary to the legal theory intuition, they \emph{do} occur.\footnote{%
The total number of downward references in the United States increases from $24$ in $1998$ to $90$ in $2019$.
In Germany, it rises from $305$ to $833$.
}
As evident from Figure~\ref{fig:crossref-evolution}, upward references have grown at similar rates in both countries, with approximately $80$ new upward references existing in $2019$ for every $100$ upward references that existed in $1998$. 
This relative increase is larger than that of the lateral references in both countries, with the exception of lateral regulation references in the United States, whose growth rate dwarfs all others. 
Since the token and structural element growth rates of German regulations are lower than or similar to those of German statutes, this means that over the period under study, connectivity between statutes and regulations in Germany has grown faster than connectivity within statutes or within regulations. 

\begin{figure}
	\centering
	\includegraphics[width=\textwidth]{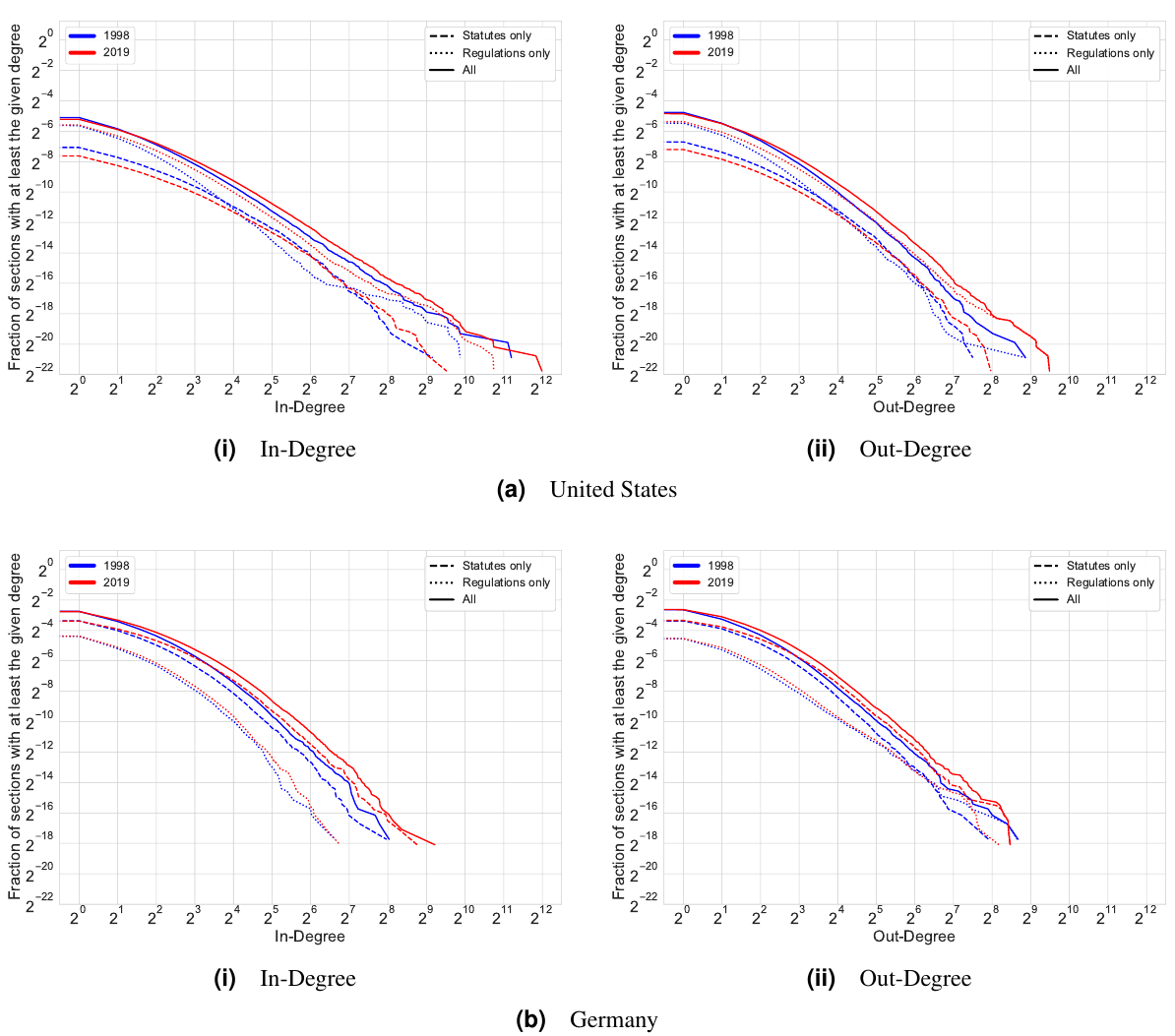}
	\caption{In-degree (left) and out-degree (right) distributions for the United States (top) and Germany (bottom) in $1998$ (blue) and $2019$ (red) when considering statutes only (dashed line), regulations only (dotted line), or statutes and regulations (solid line).}\label{fig:degree-distribution}
\end{figure}

To evaluate how the growth in the number of references affects the differences in the prominence of individual sections of codified law in the legal systems under study, in Figure~\ref{fig:degree-distribution}, we examine the in-degree distribution and the out-degree distribution of the graphs induced by the reference edges in $1998$ and $2019$ (an analogous figure normalizing section degrees by section size in tokens can be found in Section~4.1 of the \thesi).  
Since these distributions are highly skewed (as in many graphs arising from complex systems), we plot them on a log-log scale. 

All distributions plotted in Figure~\ref{fig:degree-distribution} demonstrate features common among graphs arising from bibliometric dynamics. 
For example, most sections of statutes and regulations in the United States and Germany are referenced very few times (if at all). 
The detailed characteristics of the distributions, however, differ between distribution types, document types, and countries:
For the United States, the out-degree distributions exhibit less right skew than their in-degree counterparts, 
while in Germany, we observe the opposite: 
All out-degree distributions are either within an order of magnitude of or have a longer and thicker right tail than their in-degree counterparts. 
Similarly, the sections contained in United States regulations exhibit a higher degree of skew in their in-degree distributions than the sections contained in United States statutes 
(e.g., a higher fraction of these sections has more than $500$ ingoing references), 
but in Germany, the opposite phenomenon occurs at a lower absolute level: 
There are many statute sections with more than $100$ ingoing references but hardly any regulation sections clearing that threshold. 
These national divergences might be partly due to the differing ratio between statutes and regulations, but they could also point to peculiarities in United States and German drafting style.

Comparing \emph{all} distributions across countries, we observe that the United States legal system exhibits more extreme statistics than the German legal system (which might, at least in part, be due to its larger size).
Finally, we see that from $1998$ to $2019$, most distributions shift to the right, i.e., the tails become both longer and thicker, 
which is in line with bibliometric preferential attachment dynamics \cite{merton1968,price1976}.
This indicates that reference growth has disparate impact, amplifying the differences in relevance between the individual sections contained in United States and German statutes and regulations.
As a consequence, the difficulty of navigating the law increases more slowly than the growth of the reference count may suggest. 

\vspace*{12pt}
\subsection{Connectivity}
\label{subsec:results:connectivity}

When exploring the connectivity of the national legal systems of the United States and Germany over time, we distinguish between the macro level, the meso level, and the micro level as suggested in Section~\ref{subsec:methods:connectivity}.

\vspace*{6pt}
\subsubsection{Macro-level connectivity}
\label{subsubsec:results:connectivity:macro}

\begin{figure}
	\centering
	\includegraphics[width=\textwidth]{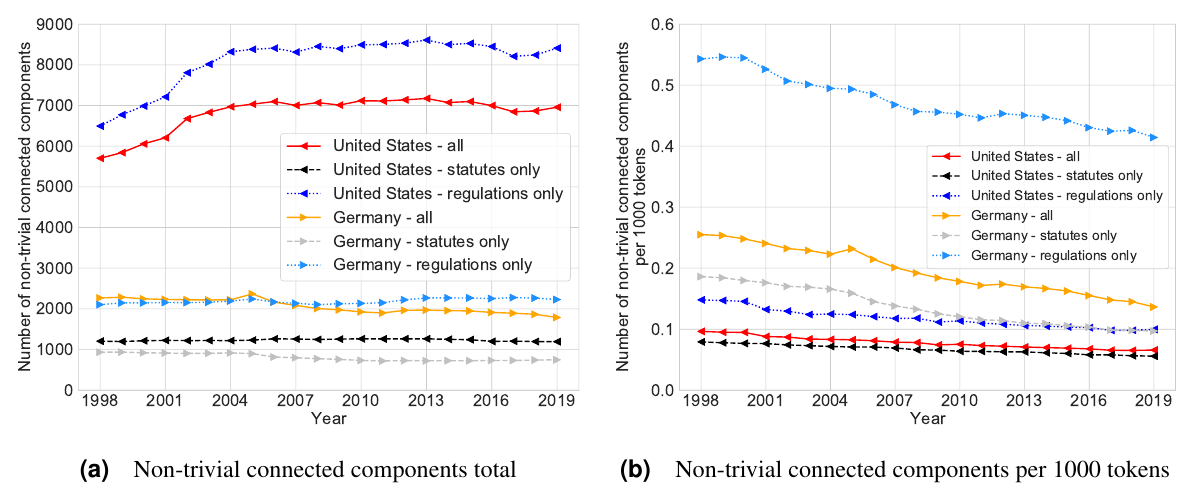}
	\caption{Development of reference connectivity as measured by the absolute number of non-trivial (weakly) connected components (left) and the number of non-trivial (weakly) connected components per $1000$ tokens (right) in the graphs induced by reference edges between all sections (solid lines), statute sections only (dashed lines), and regulation sections only (dotted lines) in the United States (left-pointing triangles) and Germany (right-pointing triangles).}\label{fig:connectivity-components}
\end{figure}

To understand how the United States legal system and its German counterpart organize and process information, we investigate the connectivity of the graphs containing code sections as nodes and references between them as edges. 
Figure~\ref{fig:connectivity-components} displays the number of non-trivial (weakly) connected components (i.e., components with more than one node) in these graphs over time, 
in absolute terms and per $1000$ tokens.
It shows that the connectivity in the graphs containing only statute sections is generally higher than that in the graphs containing only regulation sections or sections of both document types.
Furthermore, while the United States system seems more fragmented than the German system (Figure~\ref{fig:connectivity-components}~(a)) when considering absolute numbers, it turns out to be relatively less fragmented when normalizing for system size (Figure~\ref{fig:connectivity-components}~(b)).

\begin{figure}
	\centering
	\includegraphics[width=\textwidth]{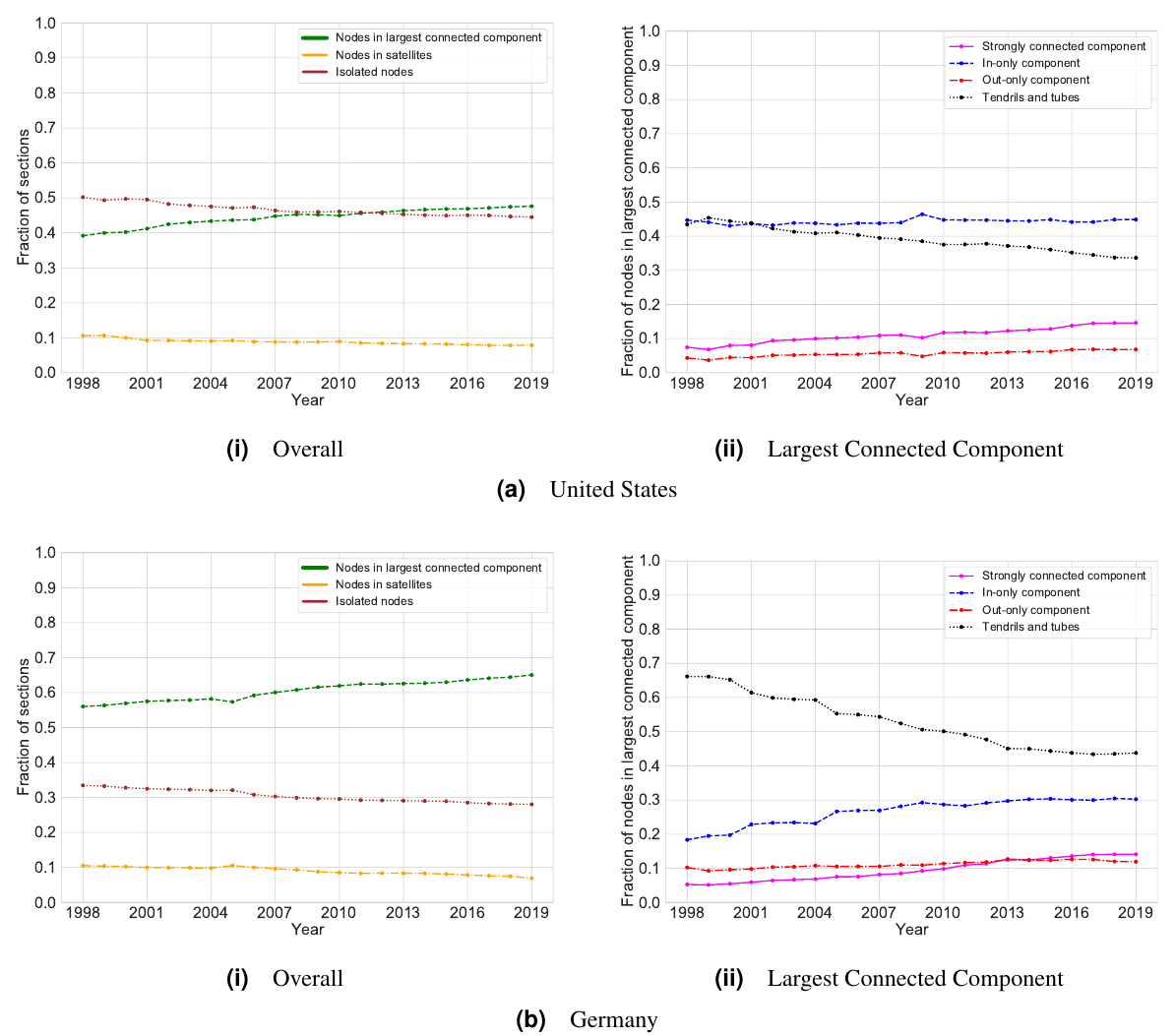}
	\caption{Development of reference connectivity in the United States (top) and Germany (bottom) as measured by the fraction of sections contained in the largest connected component (left), along with the internal structure of that component (right).}\label{fig:connectivity-all}
\end{figure}

For a more granular connectivity assessment over time, Figure~\ref{fig:connectivity-all} shows, for each year from $1998$ to $2019$, what fraction of statute sections and regulation sections in each country is contained in the largest connected component, satellite components, or isolates, 
and how the largest connected component is composed internally. 
The underlying graphs do not distinguish between sections of different document types; 
analogous figures considering statute sections only and regulation sections only can be found in Section~4.2 of the \thesi.
In both the United States and Germany, the largest connected component is growing as the fraction of sections contained in both satellites and singletons decreases, 
and the difference between the largest connected component fraction in $1998$ and that in $2019$ is around $10~\%$. 
However, the relative size of these largest connected components varies substantially between the two countries: 
In the United States, the largest connected component has grown from about $40~\%$ to nearly $50~\%$, 
while in Germany, its size has increased from circa $55~\%$ to roughly $65~\%$.
Furthermore, the fraction of isolates (sections that neither reference another section nor get referenced by another section) is larger in the United States (around $45~\%$ in $2019$) than in Germany (below $30~\%$ in $2019$).

When focusing on the largest connected component and taking edge directions into account, the differences between the two countries become even more pronounced.  
In the United States, the fraction of the largest connected component contained in the in-only component is almost equal to that contained in its tendrils and tubes in $1998$, and both lie around $45~\%$.
Over time, these fractions diverge as the strongly connected component and the out-only component grow and the in-only component stagnates. 
In Germany, however, tendrils and tubes dominate in $1998$, accounting for more than $65~\%$ of nodes, 
but by $2019$, their fraction has declined to less than $45~\%$, while the strongly connected component and the out-only component have grown at low levels and the in-only component has gained more than $50~\%$ in fractional size (growing from less than $20~\%$ to over $30~\%$).

\begin{figure}[h]
	\centering
	\includegraphics{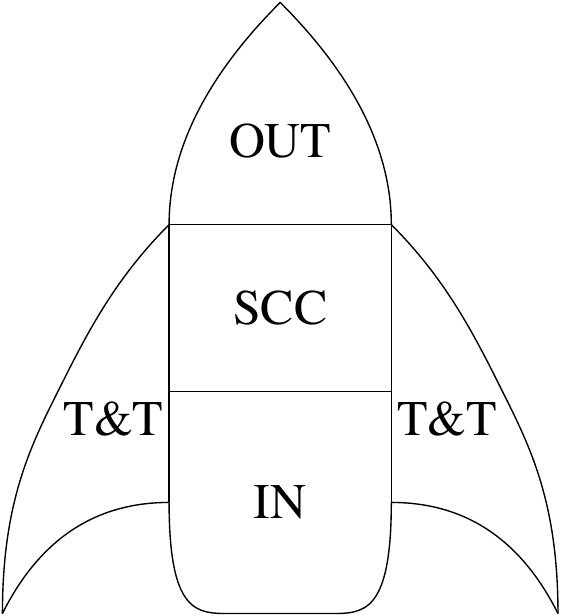}
	\caption{Rocket structure of a legal system when viewed through the lens of macro-level connectivity, with the in-component (IN) as the rocket's base, the strongly connected component (SCC) as the body, the out-component (OUT) as the nose cone, and tendrils and tubes (T\&T) as the fins.}\label{fig:rocket}
\end{figure}

Notably, in both legal systems, the out-only component accounts for the smallest fraction of sections in $2019$ (around $7~\%$ in the United States and around $12~\%$ in Germany), followed by the strongly connected component, with none of them containing more than $15~\%$ of all sections, while the in-component is twice as large in Germany and thrice as large in the United States.
Hence, at least when considering code sections as nodes and references between them as edges, both national legal systems do not exhibit the bowtie structure observed in biological systems (small strongly connected component with larger in-only and out-only components \cite{friedlander2015}) or that found in early measurements of the World Wide Web (all components, including tendrils and tubes, of roughly the same size, with a slightly larger strongly connected component \cite{broder2000}).
Rather, the legal systems we study are shaped more like rockets, with the in-only component as their base, tendrils and tubes as their fins, the strongly connected component as their body, and the out-component as their nose cone (see Figure~\ref{fig:rocket} for an illustration).
The rocket structure mirrors both the hierarchical structure of legal systems (large in-only component, small out-only component) and the fact that some areas of law function relatively independently (many tendrils and tubes; also evident from the nontrivial fraction of nodes outside the largest connected component). 
This suggests that it might be characteristic of legal systems in general, but more research is needed to corroborate this hypothesis.
Similarly, it would be interesting to investigate how our observations change if we include, e.g., non-atomic references (which, by definition, interconnect multiple sections).

\vspace*{12pt}
\subsubsection{Meso-level connectivity}
\label{subsubsec:results:connectivity:meso}

When analyzing the connectivity of the United States and German legal systems at the meso level, our goal is to create a dynamic, data-driven map of their codified law.
To this end, for both the United States and Germany, we compute cluster families as described in Section~\ref{subsubsec:methods:connectivity:meso}, 
using quotient graphs on the chapter level in the United States and on the statute or regulation level (or the book level, if available) in Germany. 
Here, we choose $100$ as the preferred number of \emph{Infomap} clusters and $15~\%$ of tokens as the edge threshold for constructing the cluster family graph (for details on how we handle text that does not lie in a chapter as well as a sensitivity analysis of the parameter choice, see Sections~4.3.1 and~4.3.4 of the \thesi). 
We calculate how many tokens from statutes and regulations these families contain in each year from $1998$ to $2019$.
By construction, our cluster families unite sets of related rules that can be thought of as different areas of law, where---unlike in, e.g., the title structure of the USC or the German finding aids' subject classification (Fundstellennachweise, FNA)---the categorization is based solely on the empirically observed reference relationships between the legal documents in our data.

\begin{figure}
	\centering
	\vspace*{-9pt}\includegraphics[width=0.75\textwidth]{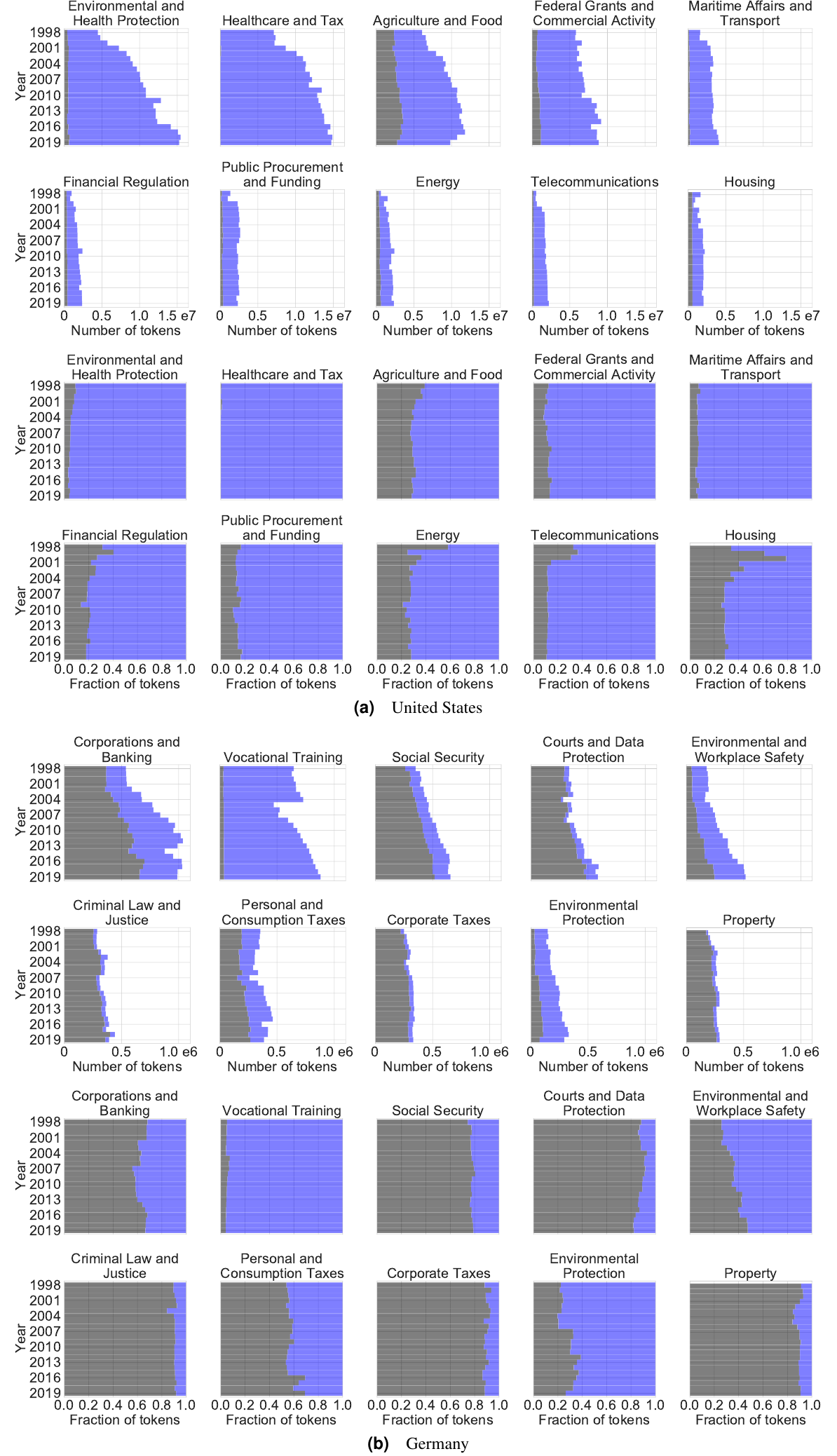}
	\caption{Development of the top ten cluster families from $1998$ to $2019$ as measured by their absolute size in tokens (rows $\{1,2,5,6\}$) and their document type composition (rows $\{3,4,7,8\}$) in the United States (top) and Germany (bottom).
	Black areas represent tokens from statutes and blue areas represent tokens from regulations.}\label{fig:families}
\end{figure}

Figure~\ref{fig:families} shows the evolution ($1998$--$2019$) of the ten cluster families with the largest number of tokens in $2019$ (henceforth: top ten cluster families) for each country, which we label leveraging our subject matter expertise (details on the labeling procedure and complementary linguistic statistics can be found in Section~4.3.2 of the \thesi).
Most families either represent a traditional field of law (e.g., property law or financial regulation) or concern a real-life domain (e.g., energy or vocational training).
A few families hold clusters from more diverse backgrounds and are therefore hard to interpret at first sight (e.g., a family containing military, public finance, and research regulation in the United States or a family containing court procedure, data security, and telecommunications in Germany).
However, a more detailed examination of the individual clusters constituting these families uncovers nuanced underlying topics (e.g., grants and commercial activity by the federal government in the example from the United States, and data protection in public [including court] proceedings in the example from Germany).
Hence, in summary, the method sketched in Section~\ref{subsubsec:methods:connectivity:meso} produces an informative map of the codified law for both countries we investigate (at the resolution level determined by our parametrization).

Inspecting the panels in Figure~\ref{fig:families} in more detail, we observe that the families' ratios of statute tokens to regulation tokens span the whole possible range: 
Some families are \emph{statute-heavy} (i.e., contain mostly statute tokens), 
others are \emph{regulation-heavy} (i.e., contain mostly regulation tokens),
and yet others are \emph{mixed} (i.e., lie between the aforementioned extremes).
For a robust categorization of the ten largest families, the data suggests a threshold of an average $80~\%$ (i.e., an average ratio of $4:1$) over the entire investigation period to classify a family as \emph{x-heavy} for $x\in\{\text{statute}, \text{regulation}\}$.
In the United States, this leads to four mixed families (Agriculture and Food; Financial Regulation; Energy; Housing) and six regulation-heavy families.
In Germany, we find four statute-heavy families (Courts and Data Protection; Criminal Law and Justice; Corporate Taxes; Property), one regulation-heavy family (Vocational Training) and five mixed families.
The overall situation remains similar even if we adopt a simple majority for the classification (eliminating the \emph{mixed} category): 
With the exception of three singular years in two families (Energy in $1998$, and Housing in $1999$ and $2000$), all top families in the United States contain a majority of regulation tokens.
The German data then presents three regulation-heavy families (Vocational Training; Environmental and Workplace Safety; Environmental Protection) and seven statute-heavy families.
A particularly striking example of a statute-heavy family is the Property family in Germany, in which there are nine times as many statute tokens as there are regulation tokens in all years except between $2002$ and $2006$.
The United States cluster family concerning Healthcare and Tax (two topics connected, inter alia, via the tax-based funding of Medicare and Medicaid) represents the opposite extreme, containing almost no statute tokens over the entire period under study.
Interestingly, no family in either country is constantly balanced between statutes and regulations, with the family concerning Personal and Consumption Taxes in Germany coming closest in the period from $1998$ to $2015$.

As Figure~\ref{fig:families} traces the development of the top ten cluster families in each country over time, we can also observe changes in the families' composition.
Extending the terminology adopted above, we can classify the families' growth based on the fraction of growth that is attributable to each of our document types. 
We say that a family is \emph{x-driven} for $x\in \{\text{statute}, \text{regulation}\}$ if tokens from $x$ account for at least $80~\%$ of the family's net growth when comparing $1998$ and $2019$, otherwise, we say that its growth is \emph{mixed}. 
Using these categories, we can classify all of the United States top ten families as regulation-driven
and half of the German top ten families as statute-driven (Social Security; Personal and Consumption Taxes; Criminal Law and Justice; Corporate Taxes; Property), while only one German family is classified as regulation-driven (Vocational Training). 
The full categorization of all top ten families for both countries, both in terms of their average composition and in terms of their growth, can be found in Section~4.3.3 of the~\thesi, 
where we further show that the general tendencies described above also hold for the entire population (although the trends are neither monotone nor universal and their extent differs from family to family).

Overall, the dynamics of the largest cluster families reflect the growth patterns documented in Figure~\ref{fig:basic-statistics}.
In absolute terms, regulations outgrow statutes by large margins in all of the top ten United States families, and statutes moderately outgrow regulations in most of the top ten German families. 
In relative terms, regulations still dominate in the United States, 
and statutes still dominate in Germany (although they are less prominent than they appear when considering absolute numbers).
In summary, based on the top ten families depicted in Figure~\ref{fig:families}, the United States seems to favor rule making via regulations, while Germany seems to favor rule making via statutes, and both countries' preferences appear to get stronger over time.

Finally, to evaluate how federal regulations impact our data-driven map of the United States and German legal systems, we compare the cluster families depicted in Figure~\ref{fig:families} to those derived in prior work that considers only federal statutes \cite{katz2020}.
For the United States, the top ten cluster families based on statutes only have topics similar to those derived from statutes and regulations combined, 
including Environmental and Health Protection, Public Health and Social Welfare, Taxes, Agriculture and Food, Financial Regulation, Public Procurement, Telecommunications, and Federal Grants and Commercial Activity including Small Business Aid.
The topic of Education makes the top ten in the statutes-only data but not in the data containing regulations, while Maritime Affairs and Transport as well as Energy only rise to prominence in the combined data.
In Germany, topics such as Financial Regulation, Taxes, Social Security, Environmental Protection, Criminal Law and Justice, and Property represent sizeable cluster families based on both datasets. 
The topic of Public Servants, Judges, and Soldiers features prominently only in the results excluding regulations, while the families of Vocational Training and of Environmental and Workplace Safety make the top ten only in the combined data.

First and foremost, however, comparing our results to those from \cite{katz2020} demonstrates that adding federal regulations to the data results in a more accurate map of law.
For example, the German data from \cite{katz2020} features a family on Market and Network Regulation that includes a leading cluster on (renewable) energy law, while no comparable family exists in the United States.
Having added federal regulations, we now see such a family in the top ten also in the United States, whose prominent position is explained by its mixed composition (including more than $70~\%$ regulation tokens on average).
At the same time, a cluster concerning (renewable) energy law is now part of the Environmental and Workplace Safety family in Germany because its regulations connect it more closely to rules concerning the protection of the environment than its statutes alone.
This suggests that adding yet further document types, e.g., federal court decisions, to our data will continue to improve the legal maps produced using our methodology, making this a promising avenue for further research.

\vspace*{6pt}
\subsubsection{Micro-level connectivity}
\label{subsubsec:results:connectivity:micro}

We analyze the connectivity of the United States and German legal systems on the micro level in order to identify those code sections that play a particularly important role in mediating the information flow between the sections which they reference and the sections by which they are referenced.
More precisely, we apply the method sketched in Section~\ref{subsec:methods:connectivity} to the graphs induced by the reference edges, where we keep a star if it has at least ten nodes in total. 
We classify these stars (and their hubs) into sinks, hinges, and sources depending on the ratio between their hubs' in-degree and their hubs' out-degree, and hypothesize that hubs of the same type have a similar function within the legal system. 
We explore the merits of this hypothesis by identifying and classifying the stars of each type in $1998$ and $2019$ and analyzing the content of the top five stars (i.e., those with the largest number of nodes) of each type in $2019$ as shown in Table~\ref{tab:structures}.

\begin{table}
	\centering
\renewcommand{\arraystretch}{1.5}
\renewcommand{\footnotesize}{\fontsize{6.5pt}{8pt}\selectfont}
\footnotesize
	\begin{tabular}{rrrrlp{0.35\textwidth}p{0.35\textwidth}}
\toprule
   $n$ &   $m_S$ &   $\delta^+$ &   $\delta^-$ & \textbf{Type}   & \textbf{Hub}                                                                      & \textbf{Description}   \\
\midrule
  2721 &     933 &            2 &         2719 & Sink            & 5 USC 552 Public information;\newline agency rules, opinions, orders, records, and proceedings                            & Authority to delegate agency rules, records, etc. to regulations                       \\
  \rowcolor{lightgray!30}1702 &      26 &            1 &         1700 & Sink            & 40 CFR 721.125 Recordkeeping requirements                                   & Authority to require particular records to be kept                        \\
  \rowcolor{lightgray!30}1684 &      28 &            3 &         1680 & Sink            & 40 CFR 721.185\newline Limitation or revocation of certain notification requirements& Criteria and procedure for limitation or revocation of notifications by an agency                        \\
  1173 &     298 &            3 &         1171 & Sink            & 5 USC 552a Records maintained on individuals                          & General definitions and procedure for keeping records on individuals                        \\
  \rowcolor{lightgray!30}1023 &      13 &            0 &         1022 & Sink            & 40 CFR 721.80 Industrial, commercial, and consumer activities                & Definition of a new use of a regulated substance                        \\
   283 &      74 &           31 &          254 & Hinge           & 8 USC 1101 Definitions                                          & Definitions for subchapter on immigration and nationality                         \\
   \rowcolor{lightgray!30}218 &     150 &          114 &          213 & Hinge           & 49 CFR 171.7 Reference Material                                              & Collection of materials to be incorporated by reference in other subchapters                       \\
   \rowcolor{lightgray!30}141 &      34 &           34 &          127 & Hinge           & 49 CFR 172.101 Purpose and use of hazardous materials table                  & Collection of substances deemed hazardous materials                       \\
   138 &      18 &           20 &          117 & Hinge           & 10 USC 101 Definitions                                                   & Definitions including bundling of statutes                       \\
   127 &       8 &           24 &          103 & Hinge           & 15 USC 637 Additional Powers                                              & Authority to carry out actions required throughout the chapter                      \\
   \rowcolor{lightgray!30}215 &      23 &          213 &            1 & Source          & 7 CFR 2.22\newline Under Secretary for Marketing and Regulatory Programs            & Enumeration of stand-in duties contained in other statutes                       \\
   \rowcolor{lightgray!30}177 &      13 &          174 &            2 & Source          & 7 CFR 2.21\newline Under Secretary for Research, Education, and Economics            &  Enumeration of stand-in duties contained in other statutes                      \\
   \rowcolor{lightgray!30}150 &      36 &          149 &            0 & Source          & 19 CFR 178.2 Listing of OMB control numbers                                 & Mapping of documents in other parts to control numbers from the Office of Management and Budget                      \\
   \rowcolor{lightgray!30}133 &      16 &          128 &            4 & Source          & 7 CFR 2.16 Under Secretary for Farm Production and Conservation              & Enumeration of stand-in duties contained in other statutes                     \\
   \rowcolor{lightgray!30}129 &       8 &          128 &            0 & Source          & 7 CFR 2.79 Administrator, Agricultural Marketing Service                     & Enumeration of stand-in duties contained in other statutes                          \\
\bottomrule
\end{tabular}
			
{\vspace*{3pt}\small \textbf{\textsf{(a)}}\quad United States}\vspace*{6pt}
	
\begin{tabular}{rrrrlp{0.35\textwidth}p{0.35\textwidth}}
\toprule
   $n$ &   $m_S$ &   $\delta^+$ &   $\delta^-$ & \textbf{Type}   & \textbf{Hub}                                                                            & \textbf{Description}   \\
\midrule
   256 &      19 &            0 &          255 & Sink            & § 36 Gesetz über Ordnungswidrigkeiten                                                   & Determination of the competent authority to prosecute misdemeanor   \\
   224 &       6 &            0 &          223 & Sink            & § 4 Berufsbildungsgesetz                                                                & Authority to delegate vocational training regulations (professions)                      \\
   194 &       0 &            1 &          192 & Sink            & § 25 Gesetz zur Ordnung des Handwerks                                                   & Authority to delegate vocational training regulations (crafts)                        \\
   191 &       3 &            0 &          190 & Sink            & § 1 Berufsbildungsgesetz                                                                & Goal and definitions for vocational training                      \\
   180 &      24 &           16 &          168 & Sink            & § 1 Gesetz über das Kreditwesen                                                         & Definitions for financial and banking regulation                       \\
   131 &      27 &           88 &           46 & Hinge           & § 3 Einkommensteuergesetz                                                               & Enumeration of tax-free income types                       \\
    88 &       0 &           74 &           13 & Hinge           & Art 229 Weitere Überleitungsvorschriften\newline Einführungsgesetz zum Bürgerlichen Gesetzbuche & Transitional provisions of the civil code                       \\
    86 &       1 &           18 &           68 & Hinge           & § 60 Lebensmittel-, Bedarfsgegenstände-\newline und Futtermittelgesetzbuch                      & Misdemeanors in food and feed safety                       \\
    84 &       0 &           73 &           10 & Hinge           & Art 97 Übergangsvorschriften\newline Einführungsgesetz zur Abgabenordnung                       & Transitional provisions of the fiscal code                       \\
    82 &       0 &           61 &           20 & Hinge           & § 100a Strafprozeßordnung                                                               & Definition of particularly serious crimes\newline allowing for telecommunication surveillance                       \\
    \rowcolor{lightgray!30}76 &       0 &           75 &            0 & Source          & § 69a Straßenverkehrs-Zulassungs-Ordnung                                                & Misdemeanors in traffic and road safety                       \\
    73 &       1 &           71 &            2 & Source          & § 340 Kapitalanlagegesetzbuch                                                           & Misdemeanors in the capital investment code                        \\
    59 &       0 &           56 &            2 & Source          & § 194 Gesetz zum Schutz vor der schädlichen Wirkung\newline ionisierender Strahlung             & Misdemeanors in the radiation protection statute                      \\
    \rowcolor{lightgray!30}48 &       0 &           47 &            0 & Source          & § 184 Verordnung zum Schutz vor der schädlichen Wirkung\newline ionisierender Strahlung         & Misdemeanors in the radiation protection regulation                       \\
    48 &       1 &           45 &            3 & Source          & § 120 Gesetz über den Wertpapierhandel                                                  & Misdemeanors and authority to delegate in the securities trading act                       \\
\bottomrule
\end{tabular}

{\vspace*{3pt}\small \textbf{\textsf{(b)}}\quad Germany}

	\caption{%
		Top five reference stars of each type in $2019$ for the United States (top) and Germany (bottom), 
		with stars whose hubs are contained in regulations marked grey.
		Edge and degree counts exclude multi-edges; 
		$m_S$ is the number of edges between spokes, 
		$\delta^+$ is the hub's out-degree, 
		and $\delta^-$ is the hub's in-degree.
	}\label{tab:structures}
	\vspace*{-6pt}
\end{table}

In the United States, we find that hubs of the same type indeed play similar roles in the legal system: 
\emph{Sinks} contain delegation of authority and general procedures, e.g., for record keeping, that are relevant for and therefore referenced by many other sections.
\emph{Hinges} connect entire collections, enumerations, and definitions to one another. 
49~CFR~§~171.7 is an example, as its only function is the incorporation of material collections by external parties (such as the American National Standards Institute) into other sections of the CFR like 49~CFR~§~173.306, which itself serves as a hub for other sections. 
\emph{Sources} enumerate duties contained in other statutes (four of the top five stem from the CFR title on Agriculture, in which this drafting technique seems to be popular) or provide a document map for their respective chapter. 

In Germany, the results paint a similar picture:
\emph{Sinks} contain provisions for delegation of legislative authority (as expected by legal theory), competencies, statements of goals, or definitions.
\emph{Hinges} contain transitional provisions, 
which are designed to bridge between old and new rules, as well as definitions.
The definition classified as a hinge (§~100a Strafprozeßordnung) establishes a well-known connection between the Criminal Code and its definition of crimes and investigative methods described both inside and outside the Code of Criminal Procedure.
All \emph{sources} (and one hinge) are collections of misdemeanors to sanction violations of rules contained in their respective statute or regulation, 
and they encompass activities as diverse as road traffic, securities trading, and handling radioactive materials.
Hence, our classification correctly identifies examples of this popular drafting technique.

As suggested in Section~\ref{subsubsec:methods:connectivity:micro} and confirmed for the largest stars, the type of a star contains information about a section's function within the legal system.
Examining the hundred largest stars, whose types are shown in Table~\ref{tab:star-statistics}, exposes different trends in both countries.
In the United States, sinks dominate both across document types and over time, accounting for three out of four stars in $1998$ and six out of seven stars in $2019$, which points to a pronounced drafting preference.
At the local level, sections of the United States codified law are mostly connected (only) by referencing the same section, which often contains a definition or the description of a procedure.
In Germany, the composition is more balanced to begin with, but sinks still make up the largest share in $1998$. 
Over time, though, the number of sinks and sources amongst the hundred largest stars decreases in favor of hinges. 
Hence, individual sections are no longer only connected by a reference to the same section, but the frequently referenced sections themselves increasingly reference other sections.
As a consequence, the number of sections that are reachable from any individual section in two hops (i.e., following two references) increases.
This makes the information flow via references more efficient, which could explain the reduced need for structural elements to guide information flow via hierarchy in Germany when compared to the United States.
But it also increases the prevalence of reference chains, possibly making the German legal system progressively harder to navigate.

\begin{table}
	\centering
\renewcommand{\arraystretch}{1.5}
\begin{multicols}{2}
	\begin{tabular}{rrrrr}
\toprule&\multicolumn{2}{c}{\textbf{1998}}&\multicolumn{2}{c}{\textbf{2019}}\\
        &   \textbf{S-Hub} &   \textbf{R-Hub} &   \textbf{S-Hub} &   \textbf{R-Hub} \\
\midrule
   \textbf{Sink} &           60 &           16 &           56 &           30 \\
  \textbf{Hinge} &            8 &            1 &            6 &            3 \\
 \textbf{Source} &            0 &           15 &            0 &            5 \\
\bottomrule
\end{tabular}

	{\vspace*{6pt}\small \textbf{\textsf{(a)}}\quad United States}\vspace*{6pt}
	\newpage
	\begin{tabular}{rrrrr}
\toprule&\multicolumn{2}{c}{\textbf{1998}}&\multicolumn{2}{c}{\textbf{2019}}\\
        &   \textbf{S-Hub} &   \textbf{R-Hub} &   \textbf{S-Hub} &   \textbf{R-Hub} \\
\midrule
   \textbf{Sink} &           42 &            0 &           33 &            0 \\
  \textbf{Hinge} &           30 &            0 &           52 &            1 \\
 \textbf{Source} &           17 &           11 &            8 &            6 \\
\bottomrule
\end{tabular}
	
	{\vspace*{6pt}\small \textbf{\textsf{(b)}}\quad Germany}
\end{multicols}

	\caption{%
		Types of the top hundred reference stars (i.e., those with the largest number of nodes) in $1998$ and $2019$ for the United States (left) and Germany (right). S-Hubs are hubs contained in statutes and R-Hubs are hubs contained in regulations.
	}\label{tab:star-statistics}
\end{table}

Mirroring the larger trends described in Section~\ref{subsubsec:results:connectivity:macro} and Section~\ref{subsubsec:results:connectivity:meso}, regulations play a more important role in the United States than in Germany at the micro level of connectivity as well. 
While the total share of regulation hubs in Germany is small, 
they make up almost half of the sources amongst the top hundred reference stars in $2019$, which again follows the larger pattern of regulations referencing rather than being referenced.
In the United States, regulation hubs account for just under $40~\%$ of the top hundred reference stars, 
but almost all of the largest stars are sinks, regardless of the document type of their hub.
This suggests that the United States drafting dynamics resulting in sinks affect both the executive and the legislative branches of government. 

\vspace*{6pt}
\subsection{Profiles}
\label{subsec:results:profiles}

\begin{figure}
	\centering
	\vspace*{-8pt}\includegraphics[width=\textwidth]{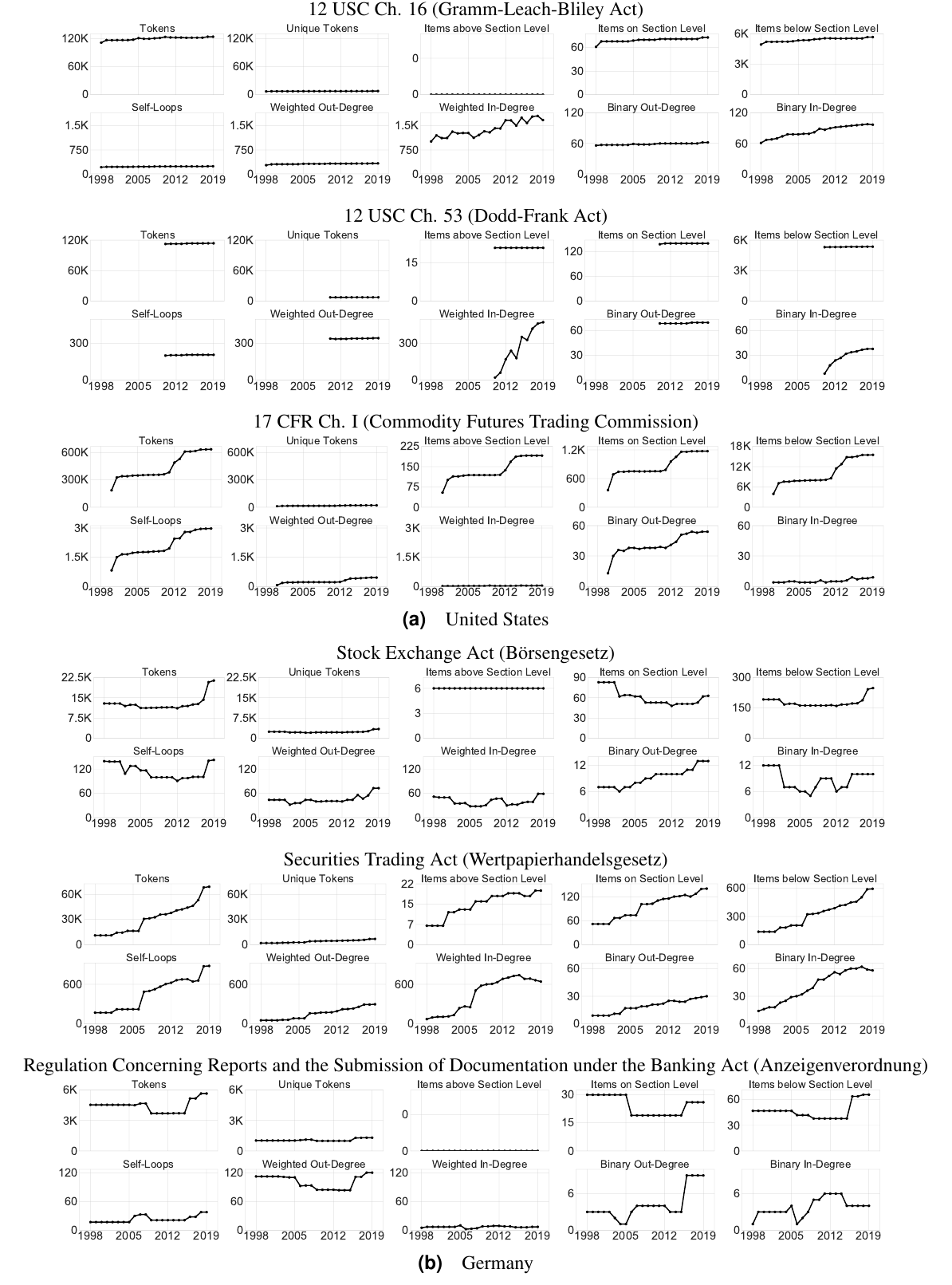}\vspace{-6pt}
	\caption{%
		Profiles tracking the evolution of selected laws related to financial regulation for the United States (top) and Germany (bottom) from $1998$ to $2019$.
	}\label{fig:profiles}
\end{figure}

In a step toward developing a dashboard for measuring and monitoring the law, 
we demonstrate the utility of the profiling procedure described in Section~\ref{subsec:methods:profiles} by applying it to selected statutes and regulations from the United States and Germany in a case study focusing on financial regulation. 
We profile a total of four statutes (two from each country) that constitute landmark legislation in this domain and trace their statistics over time in Figure~\ref{fig:profiles}, 
along with those of two additional regulations from the same area.

12~USC~Ch.~16, popularly known as the Gramm-Leach-Bliley Act or Financial Services Modernization Act of 1999 (GLBA), liberalized the United States financial market by allowing the combination of investment banks, commercial banks, and insurance companies in one institution. 
It has been in effect for nearly our entire investigation period ($1999$--$2019$) and, as indicated by nearly flat lines in all but the panels related to in-degree, has not materially changed.
However, the interaction of the GLBA with other parts of the legal system has been anything but static, with its initial weighted in-degree of $1000$ increasing by $60~\%$ between $1999$ and $2019$ due to incoming references from other statutes and regulations.
Unlike the growth trend in the weighted in-degree, the growth trend in the binary in-degree is nearly monotonic. 
This indicates that most of the fluctuations in the GLBA's regulatory environment occur within individual chapters of the USC or the CFR.
In summary, the GLBA can therefore be rightfully regarded as a landmark statute, which has required little engineering but has remained an important reference throughout the period under study.

The profile of 12~USC~Ch.~53, popularly known as the Dodd-Frank Act or the Wall Street Reform and Consumer Protection Act (DFA), shows similarities with the GLBA in most statistics we track.
Introduced in response to the Great Recession in $2010$, 
it is approximately half as old as the GLBA, 
and like the GLBA, it has barely changed in size, breadth, or structure.
But although the DFA is comparable to the GLBA in size, 
its interaction with the environment seems much more dynamic, 
with its weighted in-degree increasing by a factor of almost ten over its lifetime.
In absolute terms, however, the references increase by less than $500$, i.e., in the same order of magnitude as for the GLBA.
This is in line with the fact that both statutes are part of the same legal domain, and it highlights how much the evolution of individual pieces of legislation is influenced by their initial conditions, e.g., whether they are strongly connected with their environment already at birth.
For the DFA, the growth of the weighted in-degree again is not monotonic, 
but it is visibly steeper before $2017$ than afterwards, leveling off in the last years of the period under study.
The binary in-degree, whose gradient is almost monotonically decreasing from the start, anticipates this deceleration by several years.
This suggests the existence of an onboarding period, in which the DFA is integrated into the regulatory environment before finding its place in the United States legal system (see the related discussion in \cite{mclaughlin2021}).

The profiles of the two German statutes we examine are starkly different from those of the United States statutes.
Statutes under the names of both the Stock Exchange Act (SEA) and the Securities Trading Act (STA) have been in effect for the entire observation period.
As indicated by their unique token count, they are both constantly narrow in thematic scope (with the STA an order of magnitude more narrow than the SEA to start with), 
and their token count increases over time.
While the SEA and the STA start at comparable sizes in $1998$, the STA grows by a factor of seven, while the SEA merely doubles.
Possibly as a result, the SEA largely maintains its original number of structures,
while the number of structural elements in the STA triples.
This is accompanied by an expected, nearly parallel increase in the number of STA sections, 
and even a decrease of about $25~\%$ in the number of SEA sections.
Beyond the general growth trends present in almost all STA statistics,
the period from $2006$ to $2007$ stands out, as most of its statistics experience a relatively steep increase between these years.
The doctrinal investigation prompted by this observation reveals that the source of the increases is a legislative project translating extensive transparency requirements mandated by the European Union into German law (Transparenzrichtlinie-Umsetzungsgesetz), which came into effect in January $2007$.
This finding also shows how our methods can complement doctrinal legal scholarship, as has been demonstrated, e.g., for the development of the STA over its entire lifetime \cite{coupette2019a}.

Our statistics produce interesting insights not only for statutes but also for regulations.
For example, there is a noticeable increase in the self-referentiality of the CFR chapter about the Commodity Futures Trading Commission (CFTCR),
and the German Reports and Documents Regulation (RDR) shows structural changes between $2005$ and $2006$ as well as between $2015$ and $2016$.
As our framework enables the joint modeling of data from different document types, its application can surface characteristic differences between these types, too.
Examining our exemplary regulations and statutes in Figure~\ref{fig:profiles},
we find that the CFTCR experiences noticeable growth (about $200~\%$),
while the size of the featured statutes barely changes.
At the same time, the regulation's weighted in-degree is several orders of magnitude smaller than that of the DFA or the GLBA, supporting the intuition that statutes should be referenced more frequently than regulations for this particular case. 
In Germany, the RDR is smaller than the featured German statutes,
and it covers less diverse content (as would be expected for a regulation from a legal theory perspective).
Its structural organization is minimal, as is its self-referentiality.
This confirms that smaller units of law require less internal organization by both structure and references.
The RDR references, and is referenced by, a small number of different documents, indicating homogeneity in its regulatory environment.
Its weighted out-degree falls between the SEA and STA, i.e., it has a non-negligible number of references to a limited number of targets.
In summary, the RDR has most characteristics expected from a German regulation, and its overall profile can clearly be distinguished from that of the featured German statutes.

Our framework enables comparisons not only across document types but also across countries.
When examining the DFA and the STA, we see that the STA starts at a size of roughly a third of the DFA but grows to two thirds of the DFA over time.
Both statutes have similar degrees of structure at the section level and above,
but the DFA contains ten to fifteen times the amount of items below section level, indicating a vastly more granular hierarchical organization.
Conversely, the STA contains three to four times more self-loops than the DFA, with its weighted in-degree about $1.5$ times and its binary in-degree between $1.5$ and $2$ times higher than those of the DFA after the first couple of years.
This mirrors the more general finding that rule-making agents in the United States and Germany favor different mechanisms to handle the token growth of their statutory corpora: 
Americans like adding hierarchical structure, while Germans prefer adding references \cite{katz2020}.

\begin{figure} 
	\centering
	\vspace*{-8pt}\includegraphics[width=\textwidth]{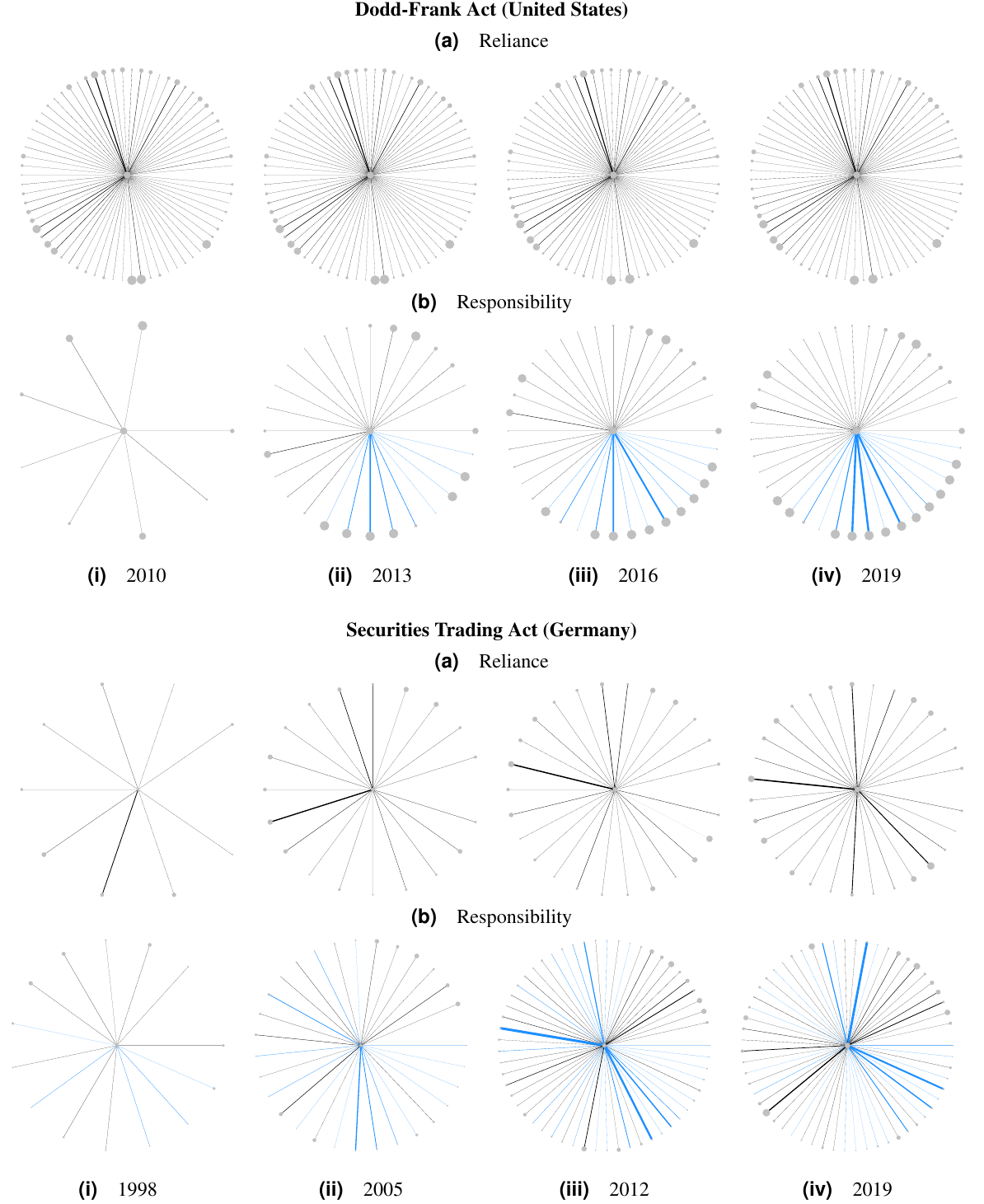}
	\caption{Reliance and responsibility of the Dodd-Frank Act (DFA, top) and the Securities Trading Act (STA, Wertpapierhandelsgesetz, bottom) from $1998$ to $2019$.
	Leaf nodes are chapters of the USC or the CFR that are cited by (reliance) or cite (responsibility) the DFA (top), 
	and statutes or regulations (or their books, if applicable) that are cited by (reliance) or cite (responsibility) the STA (bottom).
	Edge types indicate reference types as used in Figure~\ref{fig:crossref-differentiation} (black for lateral statute references, light blue for upward references, and silver for downward references).
	Node size is proportional to the node's number of tokens;
	edge width is proportional to the number of references represented by the edge.
	}\label{fig:profile-graphs}
\end{figure}

Figure~\ref{fig:profile-graphs} combines profile statistics concerning size and interdependence to enable direct visual comparisons. 
Here, we compare the ego graphs of the DFA and the STA for every quarter of their existence during our investigation period.
Note that the distance between the snapshots is different for both statutes, 
as the DFA was adopted only in the middle of our study period, but both series end in $2019$.
Complementing the statistics presented in Figure~\ref{fig:profiles}, the visualization attributes the references to their actual sources and targets, indicating their number by the weight of the edges.
For the DFA, its \emph{reliance} (i.e., how much and how diversely it references other statutes and regulations) barely changes,
while its \emph{responsibility} (i.e., how much and how diversely it is referenced by other statutes and regulations) discernibly increases, 
as the DFA becomes more and more integrated with its environment.
In contrast, both the reliance and the responsibility of the STA increase over time, with its responsibility starting nearly twice as high and increasing at a much faster rate than its reliance.
As shown by the edge colors, the diverse responsibility of the STA concerns both statutes and regulations,
and the DFA is most intensively responsible for regulations.
In line with the expectation derived from legal theory,
both the DFA and the STA rely mostly on statutes.

\section{Discussion}
\label{sec:discussion}

We have introduced an analytical framework for the dynamic network analysis of legal documents and demonstrated its utility by applying it to a dataset comprising federal statutes and regulations in the United States and Germany over a period of more than $20$ years.
The limitations of this work concern two separate areas: 
the methods introduced in Section~\ref{sec:methods} and the results presented in Section~\ref{sec:results}.

To gravitate toward its ideal formulation, our framework requires further refinement based on experiences from applications to diverse datasets.
Our model is deliberately document type and country agnostic, such that it can be easily instantiated for new data. 
Similar studies using legal documents from a variety of jurisdictions would be of immense value for improving our framework, 
and they could provide further context for the results reported in Section~\ref{sec:results}.
Furthermore, our network analytical framework could be complemented by a framework for natural legal language processing, 
as the combination of relational information and linguistic information will likely lead to insights that would not be possible using either of these sources alone.

When preparing this article, we found that combining documents of different types in one graph representation raises many conceptual questions.
Some of these questions relate to the presentation of our results, e.g., 
whether to depict dynamics in absolute or relative terms (thereby either impairing comparisons across document types of different sizes or visually overstating dynamics for small baselines).
Others concern design decisions when defining our methods, 
e.g., whether tokens from documents of different types should have the same weight when determining cluster families even if there is a striking imbalance between the total number of tokens in documents of these types (as is the case in our United States data).
Here, one alternative would be to rescale the token counts before constructing the cluster family graph, 
such that the total influence of tokens from one specific document type is equal across all types.
While this would change the results to a certain extent, 
it is difficult to assess whether the modified method would be superior because comparable investigations of multimodal legal document networks currently do not exist.

The results stated in Section~\ref{sec:results} are limited in geographic scope (United States and Germany), temporal scope ($1998$--$2019$), and institutional scope (legislative and executive branch on the federal level).
Most importantly, our findings cover only codified law.
As the United States and Germany are typically assumed to follow distinct legal traditions (common law and civil law), 
which are often thought to differ, inter alia, in their reliance on court precedent, including court decisions might have disparate impact on our results for both countries.
However, it could also provide empirical evidence against the traditional classification.
Irrespective of legal traditions, unlocking and integrating judicial data is an important direction for future work.

Regarding both growth and connectivity, the next steps consist in eliminating the uncertainties and limitations affecting our data.
For example, as highlighted in Section~\ref{subsec:data:instance}, one important stride toward a more comprehensive picture of the connectivity between legal documents is the extraction and resolution of non-atomic references.
At the macro level, connectivity could also be evaluated at other resolutions (e.g., the chapter level) or when including hierarchy edges, 
and our analysis could be expanded using further statistics, such as motif counts and their evolution over time.
Furthermore, applying our methods to other document types or other countries would help us assess whether the rocket structure we found in our data is characteristic of legal systems in general.
When assessing connectivity at the meso level, the dynamic map of law provided by our cluster families could be further refined, especially at other resolution levels.
At the micro level of connectivity, a more fine-grained star taxonomy might be in order because in both countries, there exists some functional overlap between hinge stars and sink stars.
For the profiles, a sensible step forward would be to apply the tracing methodology at other levels of resolution (e.g., at the level of individual sections), and the statistics we track could be complemented by similarity measures allowing us to compare between the different units of law we analyze.

Beyond the specific opportunities for further research outlined above, 
our work raises three larger questions to be explored in the future:
\begin{enumerate}
	\item \emph{When quantitatively analyzing legal documents, how should we choose the unit of analysis?}\\
	On the one hand, no clear consensus exists as to what constitutes a \emph{unit of law} or a \emph{legal rule}. 
	But on the other hand, the choice has far-reaching consequences for all analyses.
	Furthermore, even analyzing all documents at the same structural level presents problems: 
	Legal rules come in various sizes, and at times, a single paragraph might be longer than the average document due to drafting decisions by the agents in the legal system.
	This complicates comparisons and creates countless opportunities for erroneous interpretations.
	Detailing the full rationale behind all choices we made when presenting our results in Section~\ref{sec:results} is beyond the scope of this article. 
	However, an extensive exposition of the possible choices and the tradeoffs surrounding them would benefit the research community at large and, therefore, constitutes a fecund field for future findings.
	\item \emph{How can we measure the \emph{regulatory energy} of statutes?}\\
	The analysis of individual statutes such as the Gramm-Leach-Bliley Act and the Dodd-Frank Act suggests that legislative outputs impact their environments at potentially different rates (e.g., by prompting further rule making), 
	i.e., that they have a certain \emph{regulatory energy} that they emit over time. 
	This hypothesis could be validated, inter alia, using external data on regulatory relevance, e.g., the filings concerning regulatory risk that are required for annual and transition reports pursuant to sections 13 or 15(d) of the Securities Exchange Act of $1934$ under 17~CFR~249.310 -- Form 10-K \cite{bommarito2017}. 
	However, other approaches are equally possible and merit further investigation.
	\item \emph{How can we connect our empirical findings to established theories in law and political science?}\\
	Although beyond the scope of this work, some of our findings can be combined with analyses using established theories on the composition and evolution of codified law in both legal scholarship and political science.
	The most prominent example here is the question of delegation: 
	How does it happen and what are its limits, in theory and in practice?
	This touches the heart of democratic legitimacy, and it presents a promising opportunity for empirical legal studies to contribute to mainstream legal and political science discourse that we are planning to seize in the future.
\end{enumerate}

\section*{Conflict of Interest Statement}

DMK and MB are affiliated with the start-up LexPredict, which is now part of Elevate Services.
CC, JB, and DH declare that the research was conducted in the absence of any commercial or financial relationships that could be construed as a potential conflict of interest.

\section*{Author Contributions}

All authors conceived of the research project. CC, JB, and MB performed the computational analysis in consultation with DMK and DH.
CC, DH, and MB drafted the manuscript, which was revised and reviewed by all authors. All authors gave final approval for publication and agree to be held accountable for the work performed therein.




\section*{Supplemental Data}

The supplemental material for this paper is archived under the following DOI:\\ 
\tbd

\section*{Data Availability Statement}

For the United States, the raw input data used in this study is publicly available from the Annual Historical Archives published by the Office of the Law Revision Counsel of the U.S. House of Representatives and the United States Government Publishing Office, and is also available from the authors upon reasonable request. 
For Germany, the raw input data used in this study was obtained from \emph{juris GmbH} but restrictions apply to the availability of this data, which was used under license for the current study, and so is not publicly available. 
For details, see Section~1 of the \thesi. 
The preprocessed data used in this study (for both the United States and Germany) is archived under the following DOI:\\ \url{https://doi.org/10.5281/zenodo.4660133}

\section*{Code Availability Statement}

The code used in this study is available on GitHub in the following repositories: 
\begin{itemize}
	\item Paper: \url{https://github.com/QuantLaw/Measuring-Law-Over-Time}\\
	DOI of publication release: \url{https://doi.org/10.5281/zenodo.4660191}
	\item Data preprocessing: \url{https://github.com/QuantLaw/legal-data-preprocessing}\\
	DOI of publication release: \url{https://doi.org/10.5281/zenodo.4660168}
	\item Clustering: \url{https://github.com/QuantLaw/legal-data-clustering}\\
	DOI of publication release: \url{https://doi.org/10.5281/zenodo.4660184}
\end{itemize}


\bibliographystyle{frontiersinHLTH&FPHY} 
\bibliography{bibliography}

\end{document}


\onecolumn
\firstpage{1}

\title[Supplementary Information]{{\helveticaitalic{Supplementary Information}}}

\maketitle

\vspace*{6pt}
\section{Data Sources}
\label{sec:data}

\vspace*{6pt}
\subsection{United States Statutes}

For statutes in the United States, 
we use the United States Code (USC) as our data source. 
The USC is a compilation of the general and permanent laws of the United States on the federal level, excluding state legislation.
The Office of the Law Revision Counsel of the U.S. House of Representatives (henceforth: the Office) updates the Code continuously and publishes annual versions (although the codification process can lag behind the state of the law by several years). 
When Congress passes new legislation, 
this legislation is initially published as a \emph{Slip Law} in the \emph{United States Statutes at Large}. 
If the new legislation is considered general and permanent law, the Office integrates the law into the USC.
Depending on the Titles of the Code that are modified, 
the work of the Office is approved by Congress, 
whereby the \emph{Slip Law} is repealed and replaced by the USC as the new primary source of the law. 
Even if the work of the Office is not approved by Congress, the USC is still presumed to be the correct consolidation of the law.

We base our work on the Annual Historical Archives published by the Office, 
which are available on its website.\footnote{\url{https://uscode.house.gov/download/annualhistoricalarchives/annualhistoricalarchives.htm}.}
The USC is provided in a documented (X)HTML.\footnote{\url{https://uscode.house.gov/download/resources/USLM-User-Guide.pdf}.} 
This format is flexible and offers a wide variety of styles to closely represent the printed code. 
The Code is split into single files per year and Title.
The Annual Historical Archives date back until $1994$, 
and they are published with delay. 
As of January $2021$, therefore, $2019$ is the latest available edition.

While surveying and validating the raw data, we observe and correct the following obvious errors:
\begin{itemize}
	\item double-closing \texttt{<div>}-Tags in Title~40 in $2008$, and in Title~42 in $1994$ and $1995$,
	\item inconsistent metadata in the appendix of Title~28 in $2017$,
	\item a duplicate of Title~12 in $1998$ that is included in the files for Title~11 and Title~12, 
	\item inconsistencies regarding the tags \texttt{<statute>} and the comments \texttt{<!-- \/section-head -->}, and
	\item nesting errors in Title 42 in $1999$, $2005$, $2006$ and $2007$ that caused some sections from Chapter~85 ($1999$) or 149 ($2005$--$2007$) to lie outside of their chapter.
\end{itemize}	
	
This data source is nearly identical to the source used in \cite{katz2020}, 
but the inspected time frame is updated.

\vspace*{6pt}
\subsection{United States Regulations}
\label{subsec:uscfr}

For regulations in the United States,
we use the Code of Federal Regulations (CFR) as our data source. 
The CFR is a codification of the general and permanent rules published in the Federal Register by the departments and agencies of the Federal Government, excluding regulations below the federal level and other (e.g., less permanent) regulations.
Like the USC, the CFR is updated continuously, and the codification process can lag behind the state of the law by several years.

Our work is based on the annual bulk data provided by the U.S. Government Publishing Office (GPO) in partnership with the National Archives’ Office of the Federal Register (OFR), which is available from their website.\footnote{\url{https://www.govinfo.gov/bulkdata/CFR/}.}
The data is provided in a documented XML format,\footnote{\url{https://www.govinfo.gov/bulkdata/CFR/resources/CFR-XML_User-Guide_v1.pdf};\newline \url{https://github.com/usgpo/bulk-data}.} 
whose structure is influenced by the printed version of the CFR. 
It contains one file per year and Volume of the (printed) CFR.
The data dates back until $1996$, and it is published with delay. 
Since the first two published years of the data are incomplete, we start our analysis in $1998$ (for all data sources).

\begin{figure}
	\centering
	\includegraphics[height=0.9\textheight]{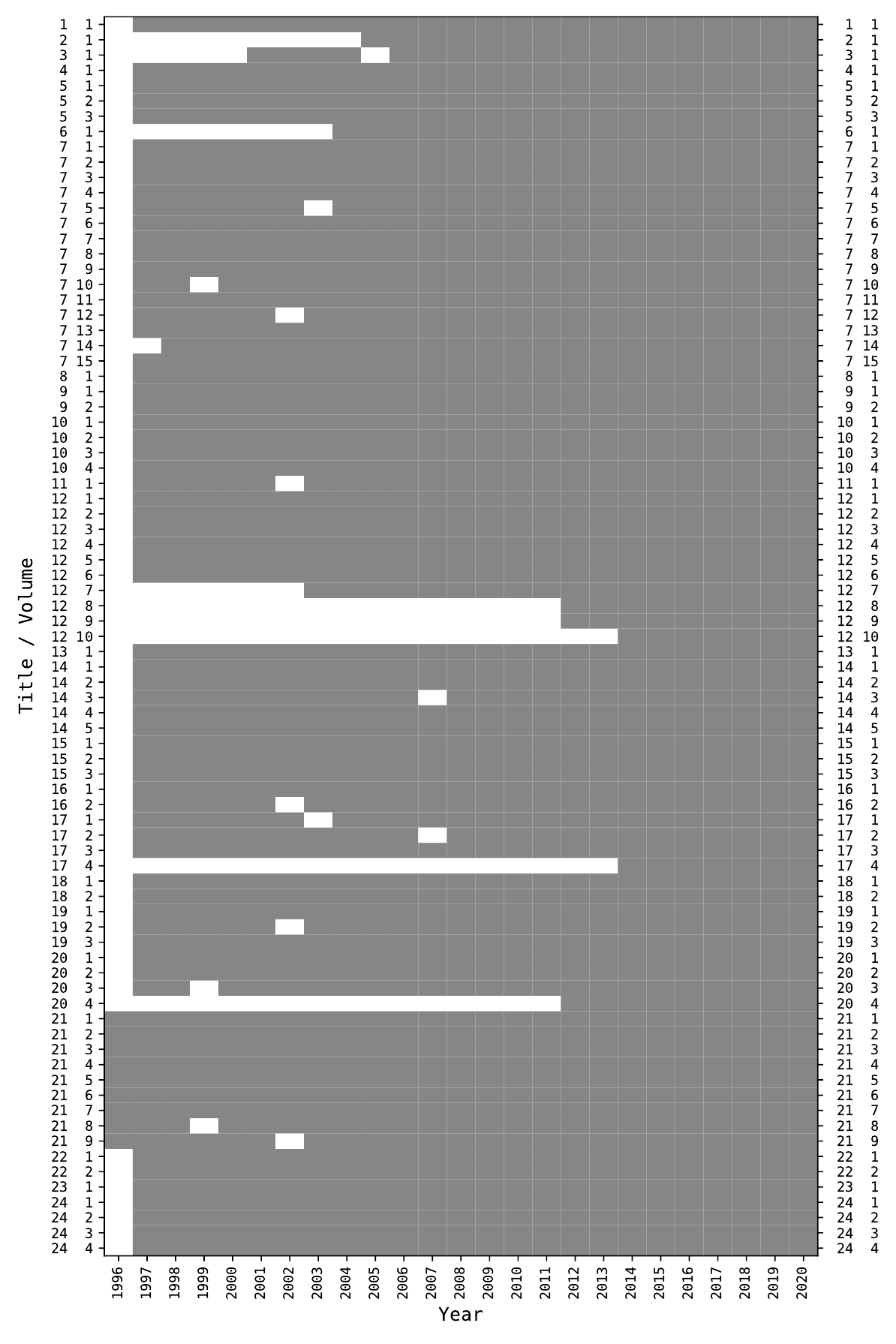}
	\caption{Volumes of the CFR that are included in the data source. The y-axis is ordered by title and volume. (1 of 3)}\label{fig:missing-cfr-volumes-1}
\end{figure}

\begin{figure}
	\centering
	\includegraphics[height=0.9\textheight]{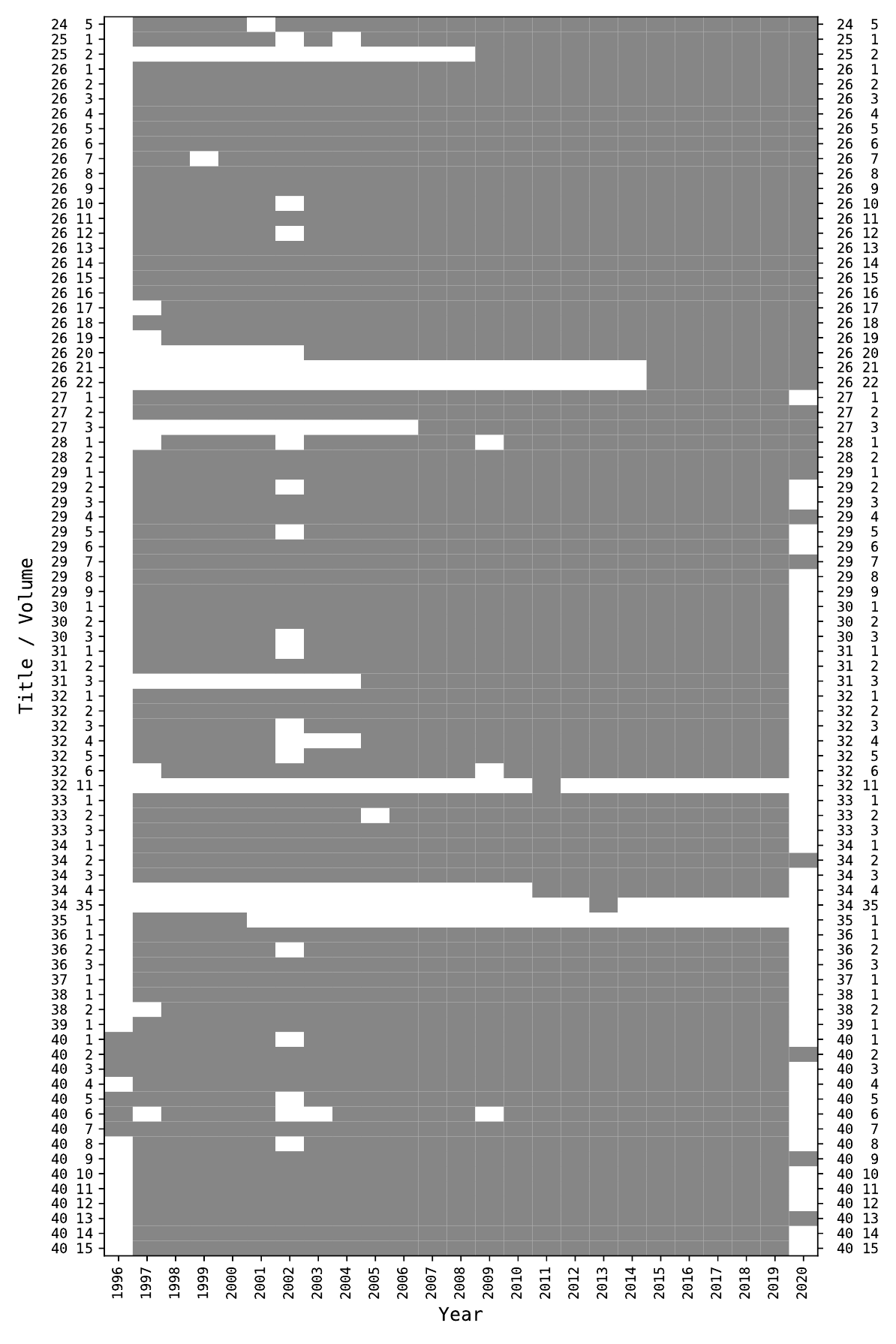}
	\caption{Volumes of the CFR that are included in the data source. The y-axis is ordered by title and volume. (2 of 3)}\label{fig:missing-cfr-volumes-2}
\end{figure}

\begin{figure}
	\centering
	\includegraphics[height=0.9\textheight]{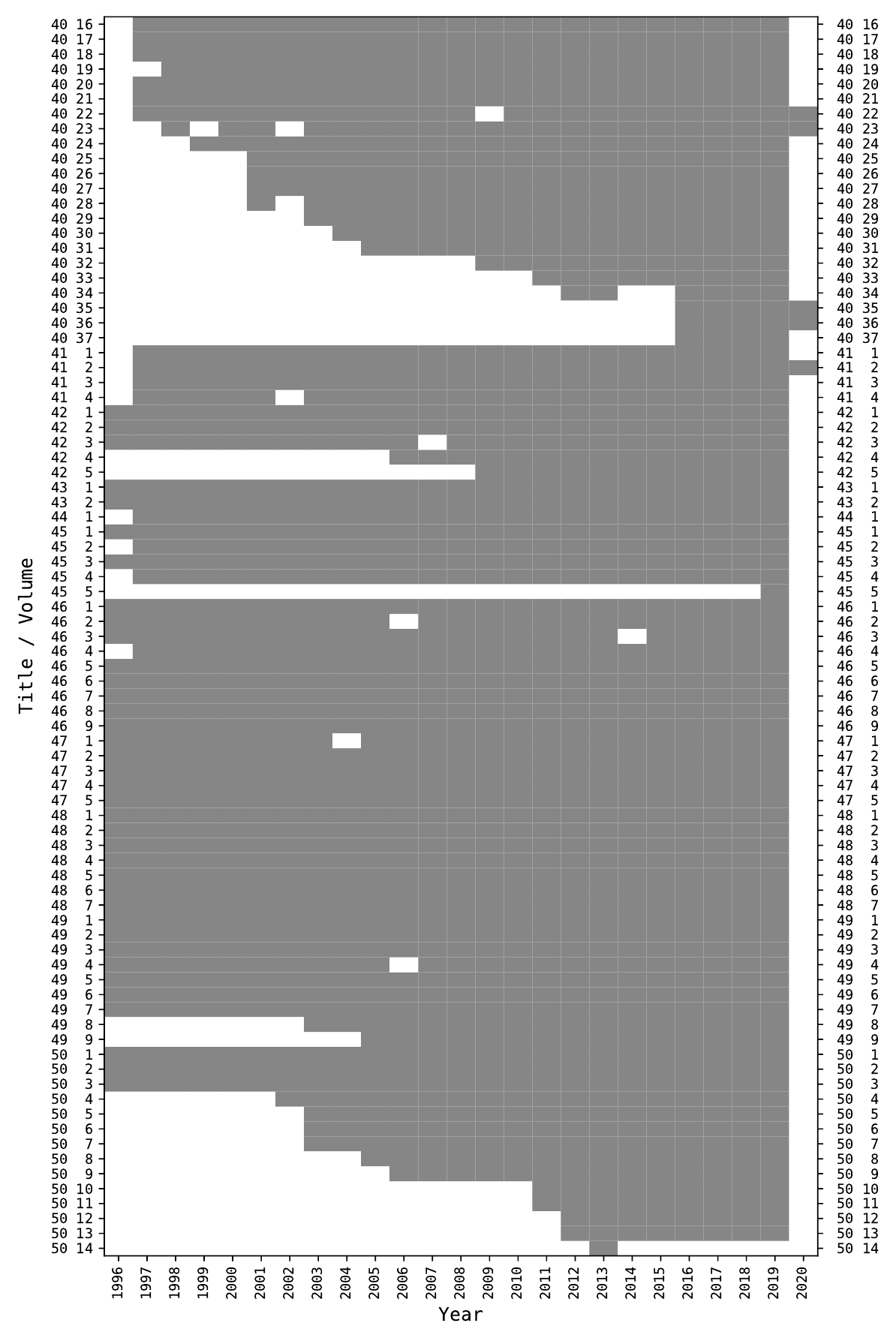}
	\caption{Volumes of the CFR that are included in the data source. The y-axis is ordered by title and volume. (3 of 3)}\label{fig:missing-cfr-volumes-3}
\end{figure}

Unfortunately, in the bulk downloads for some years, especially for the year $2002$, some CFR Volumes are missing although they reappear in later years, i.e., there are gaps in the data. 
We fill these gaps using the data of the closest previous year in which the Volume is available. 
This corresponds to the \emph{forward fill} strategy to complete missing data, with the modification that we only fill \emph{gaps}, 
i.e., we require that a Volume exists in an earlier snapshot and in a later snapshot (e.g., since Title~50 Volume~14 only exists in $2013$, we do not fill any gaps regarding this volume).
In other words, we inspect collections of the most recent versions of CFR Volumes at a given point in time.  
In Figures~\ref{fig:missing-cfr-volumes-1}--\ref{fig:missing-cfr-volumes-3}, we show the Volumes included in the bulk download \emph{before} filling the gaps as described above. 
The figures also illustrate the incompleteness of the data in $1996$, $1997$, and $2020$, 
and the increase in the number of Volumes for Title~40 and Title~50 is striking. 
When interpreting the figures, it is important to note that all Volumes are visualized as rectangles with the same, fixed area, although their text lengths vary widely.

\vspace*{6pt}
\subsection{German Statutes and Regulations}

In Germany,
all federal statutes and regulations are published in the Federal Law Gazette as amending laws,
which often combine introductions of new statutes or regulations with amendments to and repeals of existing laws.
Individual statutes and regulations are officially classified into one of nine substantive categories with currently $73$ sub-categories that contain over $400$ subject areas in the ``Fundstellennachweis~A'' (\emph{Finding Aids A}).
The Finding Aids A are published annually by the Federal Law Gazette upon instruction by the Ministry of Justice.
However, unlike the USC, there is no official data source that provides all compiled general and permanent laws at the federal level along with their historical versions.
Here, we profit from a generous cooperation with the leading German legal data provider, \emph{juris GmbH}, 
to obtain a dataset similar to the annual versions of the USC and CFR. 
Although \emph{juris} is a private company, 
the Federal Republic of Germany is its majority shareholder, 
and all branches of government rely heavily on \emph{juris} to process legislative data. 
According to \emph{juris}, the database contains every German federal statute and regulation since spring $1990$.
The data is not as structured as the USC data or the CFR data.
Instead of maintaining annual consolidated versions,
\emph{juris} stores a new version of a statute and regulation for all changes that take effect on the same day.
The data we use comes in separate files for each law and version.
Thus, for German statutes, our data source is identical to that used in \cite{katz2020},
but the inspected time frame is updated.

We may not share the text content of the German data along with this paper.
However, a website maintained by the German Federal Ministry of Justice and Consumer Protection in collaboration with \emph{juris}
provides almost the entire federal legislation as XML files in the most recent version (i.e., without historical versions).\footnote{\url{https://www.gesetze-im-internet.de}.}
A daily archive of the XML files provided on this website (starting in June $2019$) is maintained by the QuantLaw research group.\footnote{\url{https://github.com/QuantLaw/gesetze-im-internet}.}
This dataset allows a partial reproduction of the research with a similar dataset.
We requested the full dataset from \emph{juris GmbH}, 
which required a dedicated contractual framework and non-disclosure agreement to be signed.

\clearpage

\vspace*{6pt}
\section{Samples of Legal Texts}
\label{sec:textsamples}

The following fragments illustrate the inherent structure of legal texts. 
Hierarchical inclusion relationships are indicated by indentation, 
labels of items on the documents' sequence level (\emph{seqitems}) are typeset in \textsc{small capitals}, 
references are \uline{underlined},
and text content is set in \emph{Italics}.\\

\vspace*{6pt}
\subsection{United States Statutes}

\vspace*{6pt}\noindent\hspace*{0.04\textwidth}\fbox{
	\parbox{0.9\textwidth}{%
		\textbf{United States Code (2018 Main Edition and Supplement I)}
		
		\hspace*{1em}Title 12—Banks and Banking
		
		\hspace*{2em}Chapter 53—Wall Street Reform and Consumer Protection
		
		\hspace*{3em}Subchapter I—Financial Stability
		
		\noindent\hangindent4em\hspace{4em}Part C—Additional Board of Governors Authority for Certain Nonbank Financial Companies and Bank Holding Companies
		
		\hspace*{5em}\textsc{§~5363. Acquisitions}
			
		\hspace*{6em}(a) Acquisitions of Bank; Treatment as a Bank Holding Company
		
		\noindent\hangindent7em\hspace{7em}\emph{For purposes of \uline{section 1842 of this title}, a nonbank financial company supervised by the Board of Governors shall be deemed to be, and shall be treated as, a bank holding company.}
		
		\hspace*{6em}(b) Acquisition of Nonbank Companies
		
		\hspace*{7em}(1) Prior Notice for Large Acquisitions
		
		\noindent\hangindent8em\hspace{8em}\emph{Notwithstanding \uline{section 1843(k)(6)(B) of this title}, a bank holding company with total consolidated assets equal to or greater than \$$250,000,000,000$ or a nonbank financial company supervised by the Board of Governors shall not acquire direct or indirect ownership or control of any voting shares of any company (other than an insured depository institution) that is engaged in activities described in \uline{section 1843(k) of this title} having total consolidated assets of \$$10,000,000,000$ or more, without providing written notice to the Board of Governors in advance of the transaction.}
		
		\hspace*{7em}\dots
		
		\hspace*{6em}\dots
		
		\noindent\hangindent5em\hspace*{5em}\textsc{§~5364. Prohibition against management interlocks between certain financial companies}
		
		\noindent\hangindent6em\hspace{6em}\emph{A nonbank financial company supervised by the Board of Governors shall be treated as a bank holding company for purposes of the Depository Institutions\textsuperscript{[1]} Management Interlocks Act (\uline{12 U.S.C. 3201 et seq.}), except that the Board of Governors shall not exercise the authority provided in section 7\textsuperscript{[2]} of that Act (\uline{12 U.S.C. 3207}) to permit service by a management official of a nonbank financial company supervised by the Board of Governors as a management official of any bank holding company with total consolidated assets equal to or greater than \$$250,000,000,000$, or other nonaffiliated nonbank financial company supervised by the Board of Governors (other than to provide a temporary exemption for interlocks resulting from a merger, acquisition, or consolidation).}
	}
}

\clearpage

\vspace*{6pt}
\subsection{United States Regulations}

\vspace*{6pt}\noindent\hspace*{0.04\textwidth}\fbox{
	\parbox{0.9\textwidth}{%
		\textbf{Code of Federal Regulation (2019 Edition)}
		
		\hspace*{1em}Title 17—Commodity and Securities Exchanges
		
		\hspace*{2em}Chapter 2—Securities and Exchange Commission
		
		\hspace*{3em}Part 200—Organization; Conduct and Ethics, and Information and Requests
		
		\hspace*{4em}Subpart A—Organization and Program Management
		
		\hspace*{5em}General Organization (§§ 200.10 - 200.30-18)
		
		\hspace*{6em}\textsc{§ 200.30-6 Delegation of authority to Regional Directors}
			
		\noindent\hangindent7em\hspace{7em}\emph{Pursuant to the provisions of Pub. L. 87-592, 76 Stat. 394, the Securities and Exchange Commission hereby delegates, until the Commission orders otherwise, the following functions to each Regional Director, to be performed by him or under his direction by such person or persons as may be designated from time to time by the Chairman of the Commission:}
		
		\noindent\hangindent8em\hspace{8em}\emph{(a) With respect to the Securities Exchange Act of 1934, \uline{15 U.S.C. 78} et seq.:}
		
		\noindent\hangindent9em\hspace{9em}\emph{(1) Pursuant to \uline{section 15(b)(2)(C) of the Act (15 U.S.C. 78o(b)(2)(C)}:}
		
		\noindent\hangindent10em\hspace{10em}\emph{(i) To delay until the second six month period from registration with the Commission, the inspection of newly registered broker-dealers that have not commenced actual operations within six months of their registration with the Commission; and}
		
		\hspace*{10em}\dots
		
		\noindent\hangindent9em\hspace{9em}\emph{(2) Pursuant to Rule 0-4 ()\uline{§ 240.0-4 of this chapter}), to disclose to the Comptroller of the Currency, the Board of Governors of the Federal Reserve System and the Federal Deposit Insurance Corporation and to the state banking authorities, information and documents deemed confidential regarding registered clearing agencies and registered transfer agents; Provided That, in matters in which the Commission has entered a formal order of investigation, such disclosure shall be made only with the concurrence of the Director of the Division of Enforcement or his or her delegate, and the General Counsel or his or her delegate.}
		
		\hspace*{8em}\dots
	}
}

\clearpage

\vspace*{6pt}
\subsection{German Statutes}

\vspace*{6pt}\noindent\hspace*{0.04\textwidth}\fbox{
	\parbox{0.9\textwidth}{%
		\textbf{Securities Trading Act (translated version provided by the German Federal Financial Supervisory Authority updated May 16, 2019)\footnotemark}
		
		\hspace{1em}Part 2---Federal Financial Supervisory Authority (BaFin)
		
		\hspace{2em}\textsc{Section 7---Release of communiations data}
		
		\noindent\hangindent3em\hspace{3em}\emph{(1) BaFin can require a telecommunications operator to release existing data traffic records within the meaning of \uline{section 96 (1) of the Telecommunications Act (Telekommunikationsgesetz)} in its possession where certain facts establish a suspicion that a party has infringed Articles 14 or 15 of Regulation (EU) No. 596/2014 or one of the requirements specified in \uline{section 6 (6) sentence 1 numbers 3 and 4}, to the extent this is necessary to investigate the matter. \uline{Section 100a (3) and section 100b (1) to (4) sentence 1 of the Code of Criminal Procedure} apply, with the necessary modifications, provided that BaFin is entitled to make an application. The privacy of correspondence, posts and telecommunications under \uline{Article 10 of the Basic Law} is restricted in this respect.}
		
		\hspace{3em}\dots
		
		\hspace{2em}\dots
		
		\hspace{2em}\textsc{Section 9---Reducing and restricting positions or exposures}	
		
		\noindent\hangindent3em\hspace{3em}\emph{(1) BaFin can require any party to reduce the size of positions or exposures in financial instruments to the extent that this is required for enforcing the prohibitions and requirements of the provisions set out in \uline{section 6 (6) sentence 1 numbers 3 and 4}.}	
		
		\noindent\hangindent3em\hspace{3em}\emph{(2) BaFin can restrict the ability of any party to enter into a position in commodity derivatives to the extent that this is necessary for enforcing the prohibitions and requirements of the provisions set out in \uline{section 6 (6) sentence 1 numbers 3 and 4}.}
	}
}
\footnotetext{\url{https://www.bafin.de/SharedDocs/Veroeffentlichungen/EN/Aufsichtsrecht/Gesetz/WpHG_en.html}.}

\clearpage

\vspace*{6pt}
\subsection{German Regulations}

\vspace*{6pt}\noindent\hspace*{0.04\textwidth}\fbox{%
	\parbox{0.9\textwidth}{%
		\textbf{Regulation Concerning Reports and the Submission of Documentation under the Banking Act (translated version provided by the German Federal Financial Supervisory Authority updated November 2, 2010)\footnotemark}
		
		\hspace{1em}\textsc{Preamble}
		
		\noindent\hangindent2em\hspace{2em}\emph{
			By virtue of 
			\uline{section 24 (4) sentences 1 and 3, also in conjunction with section 2c (1) sentences 2 and 3, of the Banking Act (Kreditwesengesetz)} 
			in the wording of the announcement of 9 September 1998 (Federal Law Gazette I page 2776), of which 
			\uline{section 2c} 
			was last amended by article 1 number 6 and 
			\uline{section 24 (4)} 
			by article 1 number 30 letter (d) of the Act of 17 November 2006 (Federal Law Gazette I page 2606), in conjunction with 
			\uline{section 1 number 5 of the Regulation Transferring the Authority to Issue Statutory Orders to the Federal Financial Supervisory Authority (Verordnung zur Übertragung von Befugnissen zum Erlass von Rechtsverordnungen auf die Bundesanstalt für Finanzdienstleistungsaufsicht)} 
			of 13 December 2002 (Federal Law Gazette 2003 I page 3), as last amended by the Regulation of 24 May 2007 (Federal Law Gazette I page 995), the following Regulation is hereby issued by the Federal Financial Supervisory Authority (Bundesanstalt für Finanzdienstleistungsaufsicht) – hereinafter referred to as BaFin – after consulting the central associations of the institutions and in agreement with the Deutsche Bundesbank.}
	
		\hspace{1em}\textsc{Section 1---Submission procedure}
		
		\noindent\hangindent2em\hspace{2em}\emph{(1) Except as otherwise provided for in this Regulation, one copy of the reports and documentation which are to be filed or submitted under the Banking Act and which are specified in this Regulation shall be submitted both to BaFin and to the Regional Office of the Deutsche Bundesbank responsible for the institution. Reports and documentation from financial holding companies or mixed financial holding companies pursuant to \uline{section 12a (1) sentence 3, (3)} and \uline{section 24 (3a) of the Banking Act} shall be submitted to the Regional Office in whose area the superordinated enterprise pursuant to \uline{section 10a (3) sentence 4 of the Banking Act} in the wording applicable as of 1 January 2007 or the conglomerate enterprise from the banking and investment services sector with the highest balance sheet total is domiciled.}
		
		\hspace{2em}\dots		
	}
}
\footnotetext{\url{https://www.bafin.de/SharedDocs/Veroeffentlichungen/EN/Aufsichtsrecht/Verordnung/AnzV_ba_en.html}.}

\clearpage

\section{Data preprocessing}

We convert the source data into a structured format that facilitates access for our purposes, 
removes unnecessary details (especially most of the style information for the text),
and unifies the data format across all our data sources.
The preprocessing is based on the methods used in \cite{katz2020}. 
We generalize the methods to include regulations and refine them, e.g., reducing their sensitivity to inconsistencies in the source data.

For the purposes of our analysis,
we focus on three properties that characterize the structure of legal texts (illustrated in Section~\ref{sec:textsamples} above):

\begin{enumerate}
	\item They are hierarchically structured (\emph{hierarchy}), 
	e.g., into Titles, Sections, Subsections, Paragraphs, and Subparagraphs.
	\item Their text is placed in containers that are sequentially ordered and can be sequentially labeled (\emph{sequence}),
	e.g., in statutes or regulations, these containers are typically Sections.
	\item Their text may contain explicit citations or cross-references (henceforth: references) to the text in other legal documents or in other parts of the same document (\emph{reference}),
	e.g., one section can reference (the label of) another section in its text.
\end{enumerate}

To make these properties easily accessible in our data, 
we perform the following preprocessing steps on each of our data sources:

\vspace*{6pt}
\subsection{Clean the text}

First, we remove all formatting, annotations, notes, and metadata from the text, 
with the exception of formatting and metadata that we need to extract the hierarchy in the next step. 
Where present, we \emph{include} a document's preamble but \emph{exclude} its appendices.
For the USC, this results in a more conservative definition of legal text than in \cite{bommarito2010}, 
which explains the difference between the reported token counts.
As the German data is not consistently formatted on a level more detailed than text paragraphs (meaning one level below the so-called \emph{paragraphs} or \emph{articles}), we do not preserve formatting below the paragraph level for this dataset.

\vspace*{6pt}
\subsection{Extract the hierarchy}

Using the remaining text, formatting, and metadata, 
we extract the hierarchy, i.e., 
we identify boundaries of elements (Chapters, Parts, Sections, etc.) that structure the code, 
determine their parents (Titles, Chapters, Parts, etc.) and, 
if present, their headings (including their alphanumeric ordering identifiers),
and retrieve their children or textual content. 
With this information, 
we generate XML representations of the legal documents
in which texts and structural elements are nested inside their respective parents.

Our data contains explicit information regarding the boundaries and nesting of structural elements above the section level in its metadata.
We rely on this metadata down to the section level in both Germany
and the United States.
Below the section level, in the USC and the CFR, we exploit the formatting to derive a more detailed nesting of the text (for the CFR, we  also infer the nesting using typical enumeration strategies).

For the CFR, we additionally merge Volumes that contain different parts of the same Title to ensure that the nesting of the data follows its hierarchy (i.e., how a Title is split into multiple Volumes has no influence on the nesting).
We identify a structural element that is continued from a prior Volume by comparing the levels and headings of the last elements and the first elements of adjacent Volumes. 
This strategy captures most continuations correctly, but in some instances (especially in earlier years), it is over-inclusive or under-inclusive.

\subsection{Extract the references}

Next, we extract all explicit references in our preprocessed legal texts that match a common reference format.
To simplify the process, we perform the extraction in three steps:

\begin{enumerate}
	
	\item \emph{Find}: We identify parts of the text that contain a potential reference to another text in the same country.
	To find the referencing parts, we define-country specific regular expressions (\emph{regex}) patterns. 

	Most references in the USC follow a rather simple pattern,
	as references are mostly formatted consistently and include no headings or names but only numbers (and potentially letters of alphanumeric enumerations).
	References within the CFR and from the CFR to the USC are less consistently formatted. 
	Therefore, we had to select a set of patterns for our analysis, 
	and we expect that our extraction method results in a better reference coverage for the USC than for the CFR.
	We do \emph{not} currently recognize, e.g., references by the popular name of the referenced act.
	Due to the variety of citation patterns, some citations are misinterpreted. However, the graphs we base our analysis on only model references whose source \emph{and} target are known.
	Hence, a large part of the misinterpreted references does not affect the model since they are excluded from the graph because the reference target is missing.

	To find the explicit references in the USC and CFR we consider, we distinguish between two reference types.
	The first reference type starts with the number of the Title followed by ``U.S.C.'', ``C.F.R.'', optionally a unit (e.g. ``Section''), and the number of a Section or Part. 
	In some instances, a specific element is referenced in the Section or Part, which is indicated by digits or letters in brackets.
	The second reference type starts with ``Section'', ``§'', or ``Part'', followed by the number of the Section, and optionally digits or letters that reference a specific element of that Section.
	The reference may close by providing the Title or the collection (USC or CFR) of the reference target, too. 
	Either the Title is explicitly mentioned by ``of title ...'' or it must be inferred by the location of the reference (e.g., ``of this title'', ``of this chapter'', ``of this section'', or no reference location is indicated).
	The target collection of the reference may be referenced by ``of the Code of Federal Regulations'' or by ``of the Code of the United States''.

	The precise patterns we use in our reference extraction are provided in the preprocessing code repository in \texttt{statutes\_pipeline\_steps/us\_reference\_areas.py}.
	Examples of extracted citations are
	``part 135 of this chapter'',
	``part 375 of this title'',
	``§ 3.5(b) and § 3.5(c)'',
	``Section 3.5(b)'',
	``49 CFR 1.47(k)'',
	``27 CFR 46.155, 178.152 and 179.182'', and
	``8 U.S.C. 1182(d)(5)''.

	The pattern to extract references in Germany is very similar for statutes and regulations.
	The start of a reference to documents of both types is easy to identify, 
	as references normally begin with ``§'' or ``Art.''.
	The part of the reference that follows may contain numbers (and letters of alphanumeric enumerations) as well as units (e.g., Satz [engl. sentence], Nummer [engl. number]). 
	In the case of a reference to a text in a different law, the reference is followed by the name of the law. 
	A list of the law names is generated from the source data, 
	but it includes only laws valid at the time the analyzed law takes effect.
	Furthermore, 
	references to other national regulations, EU law, etc., are filtered out,
	so that only references within and between federal statutes and regulations remain. 
	Detailed documentation regarding the reference format used in German laws can be found in the ``Handbuch der Rechtsförmlichkeit'' [engl. Handbook of Legal Formalism] of the Federal Ministry of Justice and Consumer Protection.\footnote{\url{https://www.bmjv.de/DE/Themen/RechtssetzungBuerokratieabbau/HDR/HDR_node.html}.}
	Since this guide is not strictly followed by the legislator, 
	we supplement it with the actual data to develop our reference extraction method.
	
	\item \emph{Parse}: We parse the referencing texts and derive citation keys (\emph{cite keys}) that, for the USC, consist of a Title and a Section of the referenced text.
	In the German case, the keys are composed of the abbreviation of the referenced law and the number of the cited § or Article.
	One reference identified in the first step may contain several such citation keys.
	
	\item \emph{Resolve}: We identify the target structural elements of the parsed references.
	To accomplish this, we generate citation keys for each section (i.e., each paragraph or article in Germany) that can be referenced by a specific version of a law.
	In the United States, a section can be referenced if its structural element is part of the same annual version of the USC or the CFR.
	In Germany, the structural element must be part of a valid statute or regulation when the analyzed statute or regulation takes effect.
\end{enumerate}

As sketched in Section~2.2 of the main paper, 
the reference extraction procedure just described is limited to \emph{atomic} and \emph{explicit} references following the specified set of common patterns---%
a problem primarily for the United States data.
To gauge the effect of this restriction, we estimate the number of \emph{explicit} references excluded by our procedure for the USC and the CFR (there is currently no commonly accepted definition of \emph{implicit} references, and hence, no way to estimate their frequency reliably).
We focus on references to items on or below the section level, including \emph{container} references to ranges of sections, \emph{pinpoint} references to one or more sections, and \emph{explicit} references following \emph{uncommon} patterns, but excluding container references and pinpoint references to items above the section level (e.g., references to an entire chapter of the USC).
To approximate the number of excluded references, we use a simple pattern that is contained in almost every reference (matched case-insensitively):  
section(s), sec., sect., part(s), or~§, followed by one or more spaces and (at least) one digit.
We count how often this pattern occurs \emph{outside} the references we extract and compute what fraction of \emph{all} references (unextracted plus extracted) it accounts for. 
This analysis suggests that our model incorporates between $85~\%$ and $90~\%$ of all explicit references to one or more items on or below the section level, i.e., we miss between $10~\%$ and $15~\%$ of those references. 
Consequently, our design decisions regarding reference extraction hardly limit our analysis.
\vspace*{6pt}
\subsection{Generate XML files}

We store each version of each preprocessed Title of the USC or the CFR in the United States and each statute or regulation in Germany in a separate XML file.
The XML files comply with the following XSD specification:

\begin{multicols}{2}
\lstset{basicstyle=\tiny}
\begin{lstlisting}
<xs:schema
  attributeFormDefault="unqualified"
  elementFormDefault="qualified"
  xmlns:xs="http://www.w3.org/2001/XMLSchema"
>
  <xs:element
    name="document"
    type="documentType"
  />
  <xs:complexType name="documentType">
    <xs:choice
      maxOccurs="unbounded"
      minOccurs="0"
    >
      <xs:element
        type="itemType"
        name="item"
      />
      <xs:element
        type="seqitemType"
        name="seqitem"
      />
    </xs:choice>
    <xs:attribute
      type="xs:string"
      name="heading"
    />
    <xs:attribute
      type="xs:string"
      name="key"
      use="required"
    />
    <xs:attribute
      type="xs:unsignedInt"
      name="level"
      use="required"
    />
    <xs:attribute
      type="xs:string"
      name="abbr_1"
    />
    <xs:attribute
      type="xs:string"
      name="abbr_2"
    />
    <xs:attribute
      type="xs:string"
      name="document_type"
    />
    <xs:attribute
      type="xs:string"
      name="title"
    />
    <xs:attribute
      type="xs:string"
      name="year"
    />
  </xs:complexType>
  <xs:complexType name="itemType">
    <xs:choice
      maxOccurs="unbounded"
      minOccurs="0"
    >
      <xs:element
        type="itemType"
        name="item"
      />
      <xs:element
        type="seqitemType"
        name="seqitem"
      />
    </xs:choice>
    <xs:attribute
      type="xs:string"
      name="heading"
    />
    <xs:attribute
      type="xs:string"
      name="key"
      use="required"
    />
    <xs:attribute
      type="xs:unsignedInt"
      name="level"
      use="required"
    />
    <xs:attribute
      type="xs:string"
      name="auth_text"
    />
    <xs:attribute
      type="xs:string"
      name="auth_text_areas"
    />
    <xs:attribute
      type="xs:string"
      name="auth_text_parsed"
    />
  </xs:complexType>
  <xs:complexType name="seqitemType">
    <xs:choice
      maxOccurs="unbounded"
      minOccurs="0"
    >
      <xs:element
        type="textType"
        name="text"
      />
      <xs:element
        type="subseqitemType"
        name="subseqitem"
      />
    </xs:choice>
    <xs:attribute
      type="xs:string"
      name="citekey"
    />
    <xs:attribute
      type="xs:string"
      name="heading"
    />
    <xs:attribute
      type="xs:string"
      name="key"
      use="required"
    />
    <xs:attribute
      type="xs:unsignedInt"
      name="level"
      use="required"
    />
  </xs:complexType>
  <xs:complexType name="subseqitemType">
    <xs:choice
      maxOccurs="unbounded"
      minOccurs="0"
    >
      <xs:element
        type="textType"
        name="text"
      />
      <xs:element
        type="subseqitemType"
        name="subseqitem"
      />
    </xs:choice>
    <xs:attribute
      type="xs:string"
      name="key"
      use="required"
    />
    <xs:attribute
      type="xs:byte"
      name="level"
      use="required"
    />
    <xs:attribute
      type="xs:string"
      name="heading"
    />
  </xs:complexType>
  <xs:complexType
    name="textType"
    mixed="true"
  >
    <xs:sequence>
      <xs:element
        type="referenceType"
        name="reference"
        maxOccurs="unbounded"
        minOccurs="0"
      />
    </xs:sequence>
  </xs:complexType>
  <xs:complexType
    name="referenceType"
    mixed="true"
  >
    <xs:sequence minOccurs="0">
      <xs:element
        type="xs:string"
        name="main"
      />
      <xs:element
        type="xs:string"
        name="suffix"
      />
      <xs:element name="lawname">
        <xs:complexType>
          <xs:simpleContent>
            <xs:extension base="xs:string">
              <xs:attribute
                type="xs:string"
                name="type"
                use="optional"
              />
            </xs:extension>
          </xs:simpleContent>
        </xs:complexType>
      </xs:element>
    </xs:sequence>
    <xs:attribute
      type="xs:string"
      name="parsed"
    />
    <xs:attribute
      type="xs:string"
      name="parsed_verbose"
    />
    <xs:attribute
      type="xs:string"
      name="pattern"
    />
  </xs:complexType>
  <xs:complexType name="lawnameType">
    <xs:simpleContent>
      <xs:extension base="xs:string">
        <xs:attribute
          type="xs:string"
          name="type"
          use="required"
        />
      </xs:extension>
    </xs:simpleContent>
  </xs:complexType>
</xs:schema>
\end{lstlisting}
\end{multicols}

\vspace*{-6pt}
\subsection{Generate graphs}

We use the XML files along with information regarding the annual version the file belongs to (in the United States) 
or the validity period of a specific version of a statute or regulation (in Germany)
to generate our final graphs. 
We produce these graphs for each annual version of the USC and the CFR, 
and for snapshots of the German data taken at the last day of each year, 
from $1998$ to $2019$.
For the German data, we can generally produce graphs for each day in the period under study; 
the chosen dates are designed to match the United States data 
(this is preferable to the strategy used in \cite{katz2020}, where snapshots are created on the \emph{first} day of each year).
All our graphs are generated using \texttt{NetworkX}\footnote{\url{https://networkx.github.io}.} and can be exported, e.g.,
as GraphML\footnote{\url{http://graphml.graphdrawing.org/specification.html}.} files.  
We store larger graphs in two CSV files: one for their nodes and one for their edges. 
The first column in the node CSV represents the key of a node,
which is also used in the first two columns of the edge CSV to indicate the source of an edge (first column) and its target (second column). 
Other columns contain additional attributes of the nodes or the edges represented by each row. 
To optimize storage, the CSV files are compressed, 
and we recommend loading them with \texttt{pandas}\footnote{\url{https://pandas.pydata.org}.}.

\vspace*{6pt}
\section{Supplementary Tables and Figures}

In the following, we provide tables and figures that complement the results presented in the main paper in the order prescribed by the structure of its \emph{Results} section.

\begin{figure}
	\centering
	\includegraphics[width=\textwidth]{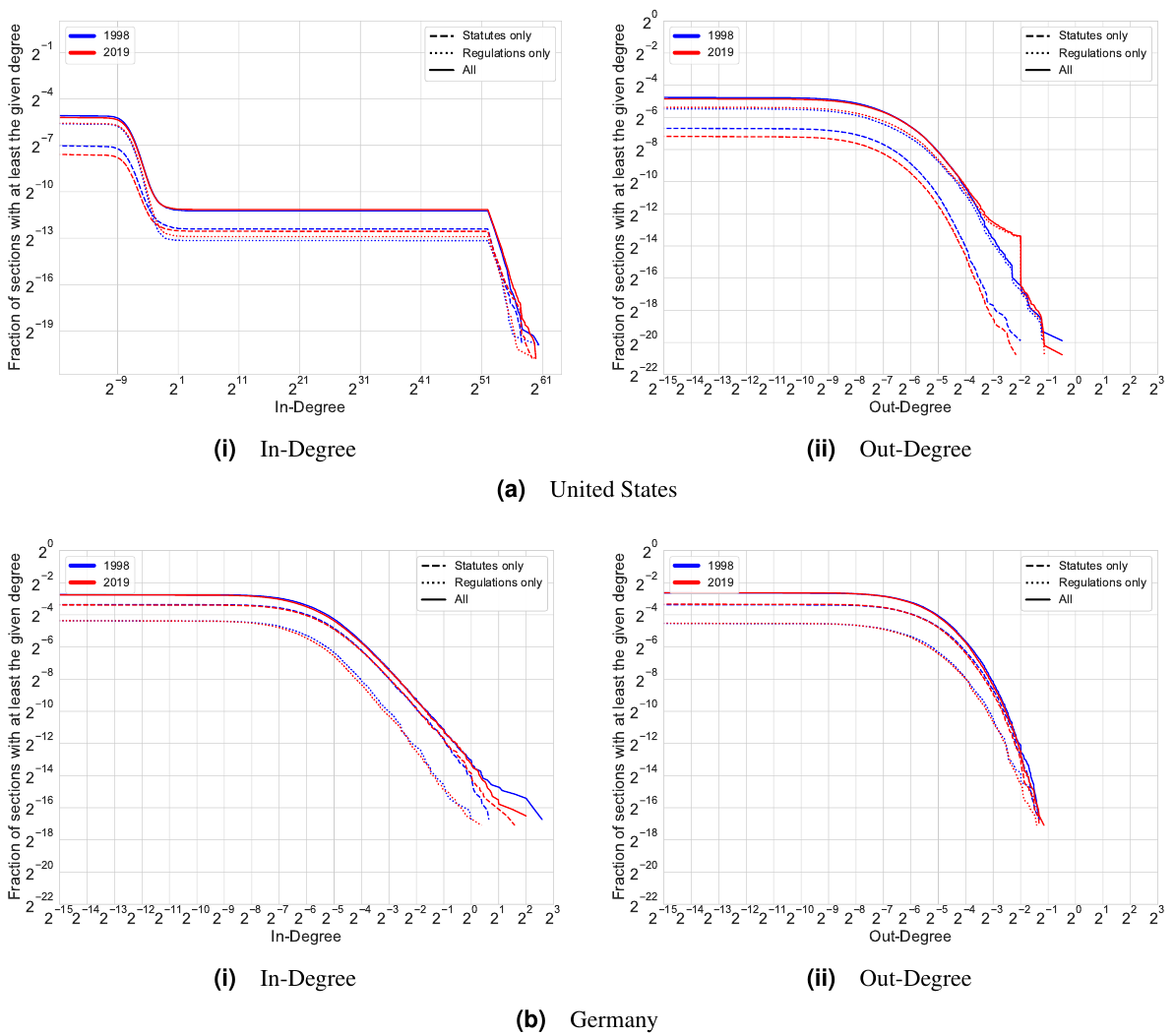}
	\caption{In-degree (left) and out-degree (right) distributions for the United States (top) and Germany (bottom) in $1998$ (blue) and $2019$ (red) when considering statutes only (dashed line), regulations only (dotted line) or statutes and regulations (solid line), normalized by section length (i.e., the plotted distributions are in-degree per token and out-degree per token).}\label{fig:growth-normalized}
\end{figure}

\vspace*{12pt}
\subsection{Growth}

The in-degree and out-degree distributions shown in the main paper do not normalize for section length.
Figure~\ref{fig:growth-normalized} provides a normalized perspective, displaying the in-degree and out-degree distributions after dividing by the sections' lengths in tokens.
The x-axis and the y-axis are shared across all panels except for the top left panel, which shows the normalized in-degree distributions for the United States.
These in-degree distributions are similar to the German in-degree distributions up to an in-degree of two references per token,
but they have a very long right tail due to a tiny number of sections with a very high number of tokens per reference.
Upon closer inspection, these sections turn out to be empty, i.e., they do not have any text at all but are still referenced by other sections in the USC or the CFR.
This is likely an artifact of the incremental codification processes in the United States, in which some Titles can be updated before others that depend on them.

\begin{figure}
	\centering
	\includegraphics[width=\textwidth]{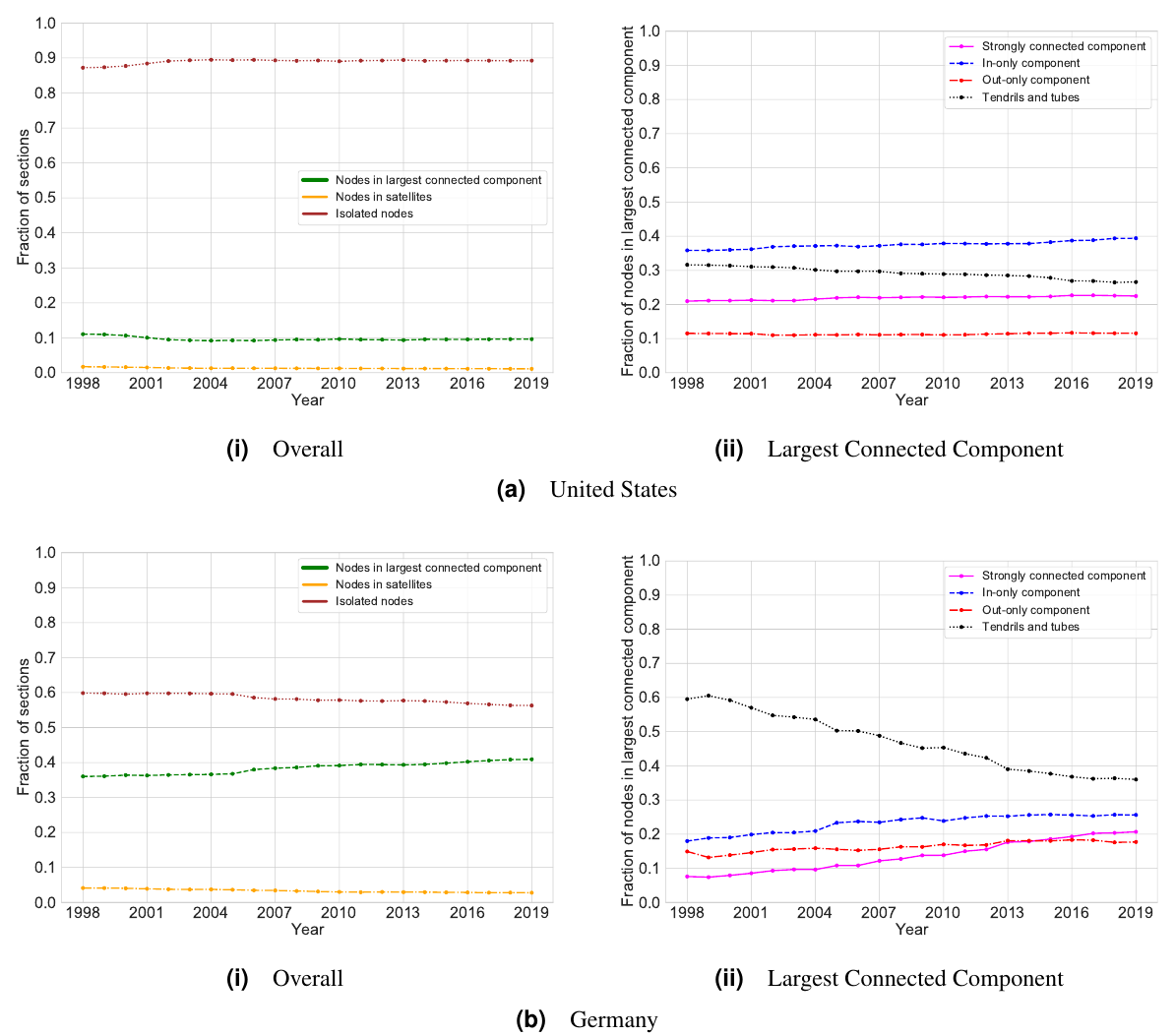}
	\caption{Development of reference connectivity in the United States (top) and Germany (bottom) as measured by the fraction of sections contained in the largest connected component (left) and the internal structure of that component (right) when considering \emph{statute sections only}.}\label{fig:connectivity-statutes}
\end{figure}

\vspace*{12pt}
\subsection{Macro-level connectivity}

\begin{figure}
	\centering
	\includegraphics[width=\textwidth]{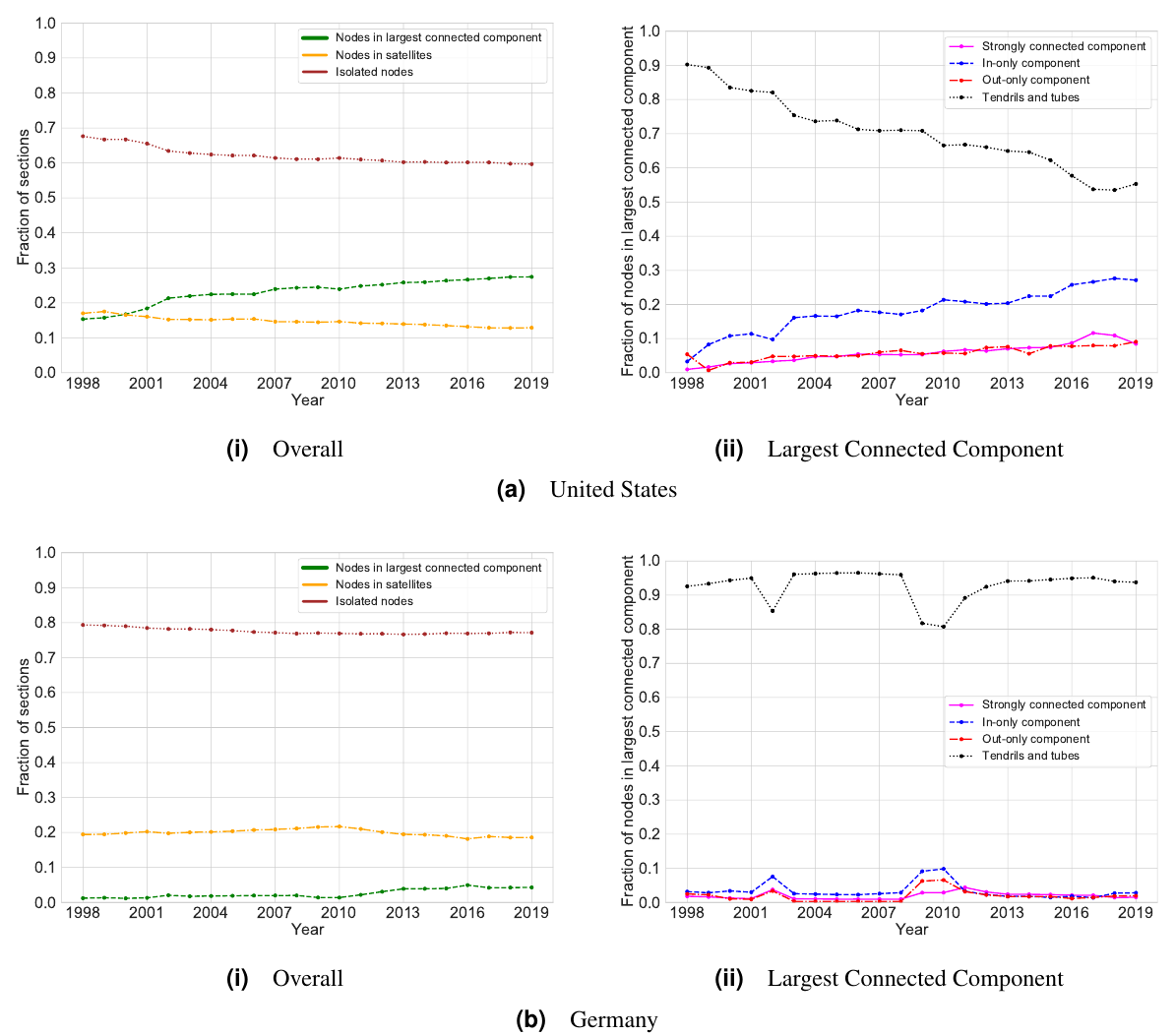}
	\caption{Development of reference connectivity in the United States (top) and Germany (bottom) as measured by the fraction of sections contained in the largest connected component (left) and the internal structure of that component (right) when considering \emph{regulation sections only}.}\label{fig:connectivity-regulations}
\end{figure}

The connectivity figures shown in the main paper depict connectivity in the graphs consisting of statutes and regulations \emph{together}.
Figure~\ref{fig:connectivity-statutes} and Figure~\ref{fig:connectivity-regulations} show connectivity for the graphs consisting of statutes only and regulations only, respectively.
The figures show that the rocket structure is nearly ubiquitous (the exception being German regulations), although the precise rocket shapes differ between the graphs.

\subsection{Meso-level connectivity}

\vspace*{6pt}
\subsubsection{Constructing the cluster family graph}

As input data for the clustering, we use quotient graphs on the chapter level in the United States and on the statute or regulation level (or the book level, if available) in Germany.
Especially in the CFR, there are leaf nodes containing text that do \emph{not} have a chapter as an ancestor.
We merge these nodes into the highest ancestor that does \emph{not} contain a chapter.

Computing the optimal node alignment between two graphs is generally a hard problem,
and methods based on tree edit distance do not scale to legislative trees.
However, we can use the sequence of the most granular structural elements of the statutes and regulations that we extract,
along with the text associated with individual nodes,
and exploit the fact that most rules do not change most of the time
to construct a practical heuristic that greedily computes a partial node alignment $\phi^i_t$ across two snapshots $S^i_t$ and $S^i_{t+1}$ from corpus $i$.
Building on and refining the method proposed in \cite{katz2020}, 
our heuristic operates in at most four sequential passes through these snapshots:
\begin{enumerate}
	\item \emph{First pass:}
	If $v$ is a node in $S^i_t$ and we find exactly one node $w$ in $S^i_{t+1}$ with identical text \emph{and} the text is at least 50 characters long, set $\phi^i_t(v) = w$.
	\item \emph{Second pass:}
	If $v$ is an unmatched node in $S^i_t$ and we find exactly one unmatched node $w$ in $S^i_{t+1}$ with identical key \emph{and} identical text,
	set $\phi^i_t(v) = w$.
	If no key is assigned to the node, the key of the closest ancestor will be used instead.
	\item \emph{Third pass:}
	If $v$ is an unmatched node in $S^i_t$ and we find exactly one unmatched node $w$ in $S^i_{t+1}$ such that (i) the text of $v$ contains the text of $w$ (or the text of $w$ contains the text of $v$) and (ii) the text remaining unmatched in $v$ ($w$) is shorter than the matched part, set $\phi^i_t(v) = w$.
	\item \emph{Fourth pass:}
	If $v$ is an unmatched node in $S^i_t$ and we find a matched node $v'$ in $S^i_t$ in the five-hop neighbourhood of $v$,
	search the five-hop neighborhood of $\phi^i_t(v')$ for the unmatched node $w$ (if any) with the largest Jaro-Winkler string similarity \cite{winkler1990} to $v$;
	if that similarity is above $0.9$, set $\phi^i_t(v) = w$.
	Repeat with all newly matched nodes until no further matches are found.
\end{enumerate}
To compute an alignment between individual clusters, 
we use the node alignment described above, 
leveraging that every node in this alignment belongs to a cluster. 
Therefore, we can approximate the similarity between a cluster $A$ at time $t$ and a cluster $B$ at time $t+1$ by summing up the tokens of the most granular elements that are contained in $A$ at time $t$ and aligned with $B$ at time $t+1$.

\vspace*{12pt}
\subsubsection{Cluster family labeling}
We derive the labels for the ten largest cluster families for each country from a manual inspection of their content, 
leveraging our subject matter expertise and familiarity with the United States and German legal systems.
For each country, we generate a family composition summary. 
This summary lists all clusters contained in a family, details what percentage of the cluster is made up of which particular chapter, book, statute or regulation (measured in tokens), and details the path from the root to each element contained in the family's clusters, 
including the headings of all structural elements in which it is contained.
We show the path down to the level of granularity in which we are interested, i.e., the chapter level of the USC and CFR for the United States and the book (if available), statute, or regulation level for the German data, relying on the headings to provide a description of the content.
Table~\ref{tab:family-summary} shows an excerpt from the summary of the largest cluster family in the United States as used in our manual labeling process.

\begin{table}
	\caption{Excerpt from the summary of the top cluster family, Environmental and Health Protection, as used in the manual labeling process (adapted from the HTML original).
	}\label{tab:family-summary}
	\renewcommand{\arraystretch}{1.15}
\footnotesize
\centering
\begin{tabular}{rlp{0.75\linewidth}}
    \toprule
    \multicolumn{3}{l}{\bfseries Family 0 – 2018\_10}\\
    \midrule
    \multicolumn{3}{l}{\bfseries 1998\_25}\\
    \midrule
89.14 \%&regulation	&Title 40 / Chapter I—Environmental Protection Agency\\
3.92 \%	&statute	&TITLE 42-THE PUBLIC HEALTH AND WELFARE / CHAPTER 85-AIR POLLUTION PREVENTION AND CONTROL\\
2.23 \%	&statute	&TITLE 33-NAVIGATION AND NAVIGABLE WATERS / CHAPTER 26-WATER POLLUTION PREVENTION AND CONTROL\\
1.71 \%	&regulation	&Title 40 / CHAPTER\\
1.50 \%	&statute	&TITLE 42-THE PUBLIC HEALTH AND WELFARE / CHAPTER 82-SOLID WASTE DISPOSAL\\
0.91 \%	&statute	&TITLE 7-AGRICULTURE / CHAPTER 6-INSECTICIDES AND ENVIRONMENTAL PESTICIDE CONTROL\\
0.30 \%	&statute	&TITLE 33-NAVIGATION AND NAVIGABLE WATERS / CHAPTER 27-OCEAN DUMPING\\
0.07 \%	&statute	&TITLE 42-THE PUBLIC HEALTH AND WELFARE / CHAPTER 137-MANAGEMENT OF RECHARGEABLE BATTERIES AND BATTERIES CONTAINING MERCURY\\
0.06 \%	&statute	&TITLE 16-CONSERVATION / CHAPTER 32A-REGIONAL MARINE RESEARCH PROGRAMS\\
0.05 \%	&statute	&TITLE 33-NAVIGATION AND NAVIGABLE WATERS / CHAPTER 41-NATIONAL COASTAL MONITORING\\
    \midrule
    \dots\\
    \midrule
    \multicolumn{3}{l}{\bfseries 2019\_14}\\
    \midrule
    80.70 \%    &regulation	&Title 40 / CHAPTER I—ENVIRONMENTAL PROTECTION AGENCY\\
    10.11 \%	&regulation	&Title 29 / Subtitle B—Regulations Relating to Labor / CHAPTER XVII—OCCUPATIONAL SAFETY AND HEALTH ADMINISTRATION, DEPARTMENT OF LABOR\\
    4.67 \%	    &regulation	&Title 49 / Subtitle B—Other Regulations Relating to Transportation / CHAPTER V—NATIONAL HIGHWAY TRAFFIC SAFETY ADMINISTRATION, DEPARTMENT OF TRANSPORTATION\\
    1.23 \%	    &statute	&TITLE 42-THE PUBLIC HEALTH AND WELFARE / CHAPTER 85-AIR POLLUTION PREVENTION AND CONTROL\\
    0.86 \%	    &statute	&TITLE 33-NAVIGATION AND NAVIGABLE WATERS / CHAPTER 26-WATER POLLUTION PREVENTION AND CONTROL\\
    0.51 \%	    &statute	&TITLE 42-THE PUBLIC HEALTH AND WELFARE / CHAPTER 82-SOLID WASTE DISPOSAL\\
    0.41 \%	    &statute	&TITLE 15-COMMERCE AND TRADE / CHAPTER 53-TOXIC SUBSTANCES CONTROL\\
    0.38 \%	    &statute	&TITLE 7-AGRICULTURE / CHAPTER 6-INSECTICIDES AND ENVIRONMENTAL PESTICIDE CONTROL\\
    0.36 \%	    &regulation	&Title 48 / CHAPTER 15—ENVIRONMENTAL PROTECTION AGENCY\\
    0.15 \%	    &statute	&TITLE 49-TRANSPORTATION / SUBTITLE VI-MOTOR VEHICLE AND DRIVER PROGRAMS / PART A-GENERAL / CHAPTER 301-MOTOR VEHICLE SAFETY\\
    \bottomrule
\end{tabular}

\end{table}

The family composition summaries provide enough dimensionality reduction to enable humans to assign a label, and they are part of the data provided with the paper. 
Supplementing this expert-based approach, 
we provide basic TF-IDF statistics (term frequency-inverse document frequency; for an introduction, see \cite{schutze2008}) for the ten largest cluster families in both countries as CSV files in the data repository accompanying this paper.
Table~\ref{tab:tfidf:part1} and Table~\ref{tab:tfidf:part2} contain the top ten nouns---according to TF-IDF (and excluding nouns describing hierarchical elements in legal texts, e.g., \emph{chapter} or \emph{section})---for the United States cluster families depicted in the main paper. 

\clearpage
 
\begin{table}
	\caption{TF-IDF statistics for the top ten cluster families with translation of abbreviations. (1 of 2)}\label{tab:tfidf:part1}
	\renewcommand{\arraystretch}{1.1}
\small
\centering
\begin{tabular}{p{0.2\linewidth}p{0.1\linewidth}p{0.1\linewidth}p{0.2\linewidth}p{0.2\linewidth}}
    \toprule
    \bfseries Environmental and Health Protection&\bfseries Healthcare and Tax&\bfseries Agriculture and Food&\bfseries Federal Grants and Commercial Activity&\bfseries Maritime Affairs and Transport\\
    \midrule
    hap             &cms       &rus         &fdic           &ocmi\\
    nox             &taxpayer  &fmha        &(o)champus     &tp\\
    cair            &income    &fns         &pbgc           &packagings\\
    emission(s)     &medicare  &aphis       &dod            &longitude\\
    voc             &qio       &premarket   &ots            &cotp\\
    pollutant       &tax       &borrower    &usaid          &commandant\\
    nonattainment   &dpgr      &swine       &occ            &liferaft\\
    cems            &annutiy   &fsa         &depository     &nls\\
    epa             &hmo       &fda         &uss            &ib\\
    kkg             &prs       &flock       &tva            &cargo\\
    \bottomrule
\end{tabular}
\vspace*{18pt}

\footnotesize
\begin{tabular}{ll}
    \toprule
    \multicolumn{2}{c}{\bfseries Environmental and Health Protection}\\
    \midrule
    hap             &Hazardous Air Pollutants\\
    nox             &Nitrogen Oxides\\
    cair            &Clean Air Interstate Rule\\
    voc             &Volatile Organic Compounds\\
    cems            &Continuous Emissions Monitoring System\\
    epa             &Environmental Protection Agency\\
    kkg             &1000 Kilograms\\   
    \midrule
    \multicolumn{2}{c}{\bfseries Healthcare and Tax}\\
    \midrule
    cms             &Centers for Medicare \& Medicaid Services\\
    qio             &Quality Improvement Organization\\
    dpgr            &Domestic Production Gross Receipts\\
    hmo             &Health Maintenance Organization\\
    prs             &Private Retirement Scheme\\
    \midrule
    \multicolumn{2}{c}{\bfseries Agriculture and Food}\\
    \midrule
    rus             &Rural Utilities Service\\
    fmha            &Farmers Home Administration\\
    fns             &Food and Nutrition Service\\
    aphis           &Animal and Plant Health Inspection Service\\
    fsa             &Farm Service Agency\\
    fda             &Food and Drug Administration\\
    \midrule
    \multicolumn{2}{c}{\bfseries Federal Grants and Commercial Activity}\\
    \midrule
    fdic            &Federal Deposit Insurance Corporation\\
    (o)champus      &(Office of) Civilian Health and Medical Programs of the Uniformed Services\\
    pbgc            &Pension Benefit Guaranty Corporation\\
    dod             &Department of Defense\\
    ots             &Officials Tracking System\\
    usaid           &United States Agency for International Development\\
    occ             &Office of the Comptroller of the Currency\\
    uss             &United States Ship\\
    tva             &Tennessee Valley Authority\\
    \midrule
    \multicolumn{2}{c}{\bfseries Maritime Affairs and Transport}\\
    \midrule
    ocmi            &Officer in Charge, Marine Inspection\\
    tp              &Time in Port\\
    cotp            &Captain of the Port\\
    nls             &Noxious Liquid Substances\\
    ib              &Invoice Book\\
    \bottomrule
\end{tabular}
\end{table}

\begin{table}
	\caption{TF-IDF statistics for the top ten cluster families with translation of abbreviations. (2 of 2)}\label{tab:tfidf:part2}
	\renewcommand{\arraystretch}{1.1}
\small
\centering

\begin{tabular}{p{0.125\linewidth}p{0.2\linewidth}p{0.15\linewidth}p{0.175\linewidth}p{0.175\linewidth}}
    \toprule
    \bfseries Financial Regulation&\bfseries Public Procurement and Funding&\bfseries Energy&\bfseries Telecommunications&\bfseries Housing\\
    \midrule
    swap(s)         &sba        &nrc                &(m$|$k$|$g)hz  &pha\\
    futures         &nasa       &doe                &antenna        &hud\\
    issuer          &hubzone    &reactor            &fcc            &mortgag(or$|$ee$|$e)\\
    registrant      &wosb       &ofe                &lec            &cdbg\\
    broker          &hsar       &uranium            &(d)tv          &homeownership\\
    dealer          &contractor &licensee           &e(i)rp         &hfa\\
    securities      &nmvc       &decommissioning    &bandwidth      &phas\\
    counterparty    &usaid      &isfsi              &transmitter    &gse\\
    sipc            &offeror    &ffd                &subscriber     &homebuyer\\
    trading         &pgi        &hrp                &px             &pae\\
    \bottomrule
\end{tabular}
\vspace*{18pt}

\footnotesize
\begin{tabular}{ll}
    \toprule
    \multicolumn{2}{c}{\bfseries Financial Regulation}\\
    \midrule
    sipc            &Securities Investor Protection Corporation\\
    \midrule
    \multicolumn{2}{c}{\bfseries Public Procurement and Funding}\\
    \midrule
    sba             &Small Business Act\\
    nasa            &National Aeronautics and Space Administration\\
    wosb            &Women-Owned Small Business\\
    hsar            &Homeland Security Acquisition Regulation\\
    nmvc            &New Model Venture Capital\\
    usaid           &United States Agency for International Development\\
    pgi             &Procedures, Guidance, and Information\\
    \midrule
    \multicolumn{2}{c}{\bfseries Energy}\\
    \midrule
    nrc             &Nuclear Regulatory Commission\\
    doe             &Department of Energy\\
    ofe             &Office of Fossil Energy\\
    isfsi           &Independent Spent Fuel Storage Installation\\
    ffd             &Fitness for Duty\\
    hrp             &Hazardous Air Pollutant\\
    \midrule
    \multicolumn{2}{c}{\bfseries Telecommunications}\\
    \midrule
    (m$|$k$|$g)hz   &(Mega$|$Kilo$|$Giga)Hertz\\
    fcc             &Federal Communications Commission\\
    lec             &Local Exchange Carrier\\
    (d)tv           &(Digital) Television\\
    e(i)rp          &Effective Isotropic Radiated Power\\
    px              &Peak Power\\
    \midrule
    \multicolumn{2}{c}{\bfseries Housing}\\
    \midrule
    pha             &Public Housing Agency\\
    hud             &U.S. Department of Housing and Urban Development\\
    cdbg            &Community Development Block Grant\\
    hfa             &Housing Finance Agency\\
    phas            &Public Housing Assessment System\\
    gse             &Government-Sponsored Enterprise\\
    pae             &Participating Administrative Entity\\
    \bottomrule
\end{tabular}
\end{table}

\clearpage

\vspace*{6pt}
\subsubsection{Cluster family categorization}

\begin{figure}
	\includegraphics[width=\textwidth]{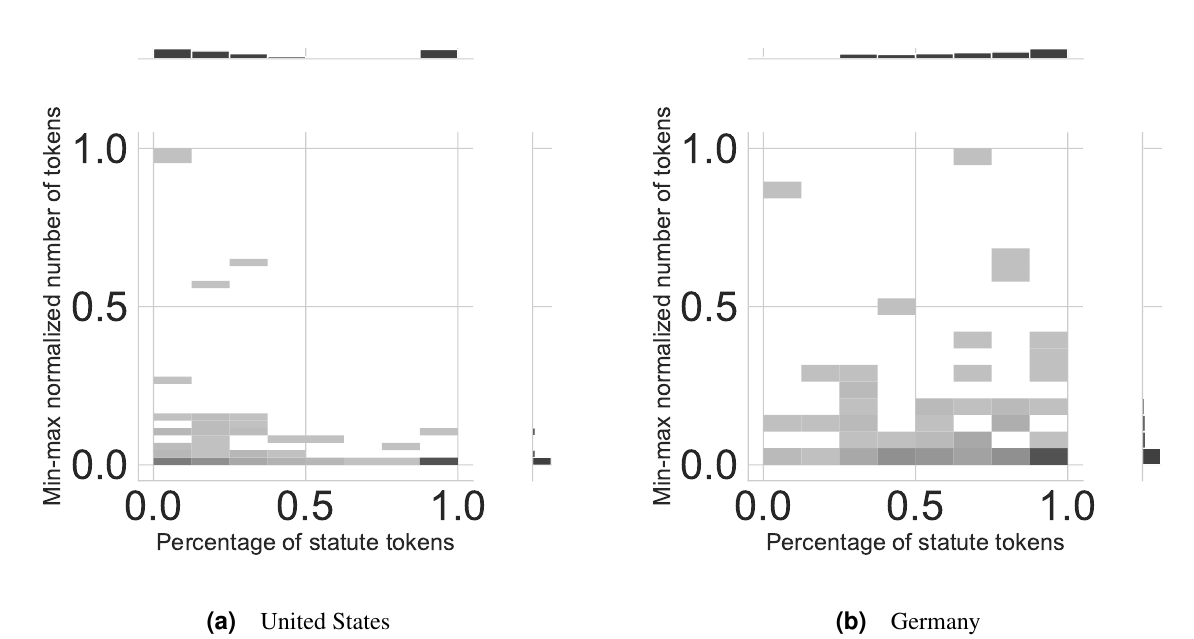}
	\caption{Percentage of statute tokens versus min-max normalized cluster family size in tokens for the $100$ largest families (measured in tokens) in $2019$ for the United States (left) and Germany (right).}\label{fig:families-top-100}
\end{figure}

We show the exact numbers behind the classification of the top ten cluster families by their average composition or their growth,
using either the $4:1$ threshold or a simple majority rule, 
in Table~\ref{fig:family-composition} and Table~\ref{fig:family-growth}. 
As can be seen from Figure~\ref{fig:families-top-100}, 
the trends found in the inspection of the top ten families hold for the top hundred families as well. 
In the United States, the vast majority of cluster families contain less than $40~\%$ statute tokens, with a mode below $1/8$. 
Note that a substantial number of families contains statutes only, 
but while these small families make up the majority of families by number, they only account for a small fraction of the overall tokens.
In Germany, most families are also small and contain more statute tokens than regulation tokens.
Here, the larger the family, the higher the chance that it is dominated by statute tokens,
with the majority of the largest clusters containing more than $50~\%$ statutes and a mode above $7/8$. 

\begin{table}
	\caption{Family classification by average composition}\label{fig:family-composition}
	\renewcommand{\arraystretch}{1.075}
	\small\centering
	\begin{tabular}{lrrrrrrrll}
\toprule
 Family                                 &   $\mu$ &   $\sigma$ &   $\min$ &   $25~\%$ &   $50~\%$ &   $75~\%$ &   $\max$ & Cat.   & Maj.   \\
\midrule
 Environmental and Health Protection    &    0.05 &       0.02 &     0.03 &      0.04 &      0.05 &      0.06 &     0.1  & R      & R      \\
 Healthcare and Tax                     &    0    &       0    &     0    &      0    &      0    &      0    &     0.01 & R      & R      \\
 Agriculture and Food                   &    0.31 &       0.03 &     0.27 &      0.29 &      0.29 &      0.31 &     0.39 & M      & R      \\
 Federal Grants and Commercial Activity &    0.12 &       0.02 &     0.08 &      0.11 &      0.12 &      0.13 &     0.15 & R      & R      \\
 Maritime Affairs and Transport         &    0.06 &       0.01 &     0.05 &      0.06 &      0.06 &      0.07 &     0.09 & R      & R      \\
 Financial Regulation                   &    0.22 &       0.06 &     0.14 &      0.19 &      0.2  &      0.22 &     0.4  & M      & R      \\
 Public Procurement and Funding         &    0.14 &       0.02 &     0.1  &      0.13 &      0.14 &      0.15 &     0.18 & R      & R      \\
 Energy                                 &    0.29 &       0.07 &     0.21 &      0.26 &      0.27 &      0.28 &     0.58 & M      & R      \\
 Telecommunications                     &    0.15 &       0.08 &     0.11 &      0.11 &      0.12 &      0.12 &     0.36 & R      & R      \\
 Housing                                &    0.34 &       0.13 &     0.26 &      0.28 &      0.29 &      0.34 &     0.79 & M      & R      \\
\bottomrule
\end{tabular}
	
	{\vspace*{6pt}\small \textbf{\textsf{(a)}}\quad United States}\vspace*{12pt}

	\begin{tabular}{lrrrrrrrll}
\toprule
 Family                             &   $\mu$ &   $\sigma$ &   $\min$ &   $25~\%$ &   $50~\%$ &   $75~\%$ &   $\max$ & Cat.   & Maj.   \\
\midrule
 Corporations and Banking           &    0.63 &       0.04 &     0.56 &      0.59 &      0.63 &      0.67 &     0.68 & M      & S      \\
 Vocational Training                &    0.05 &       0.01 &     0.04 &      0.04 &      0.05 &      0.05 &     0.07 & R      & R      \\
 Social Security                    &    0.78 &       0.01 &     0.74 &      0.77 &      0.78 &      0.78 &     0.8  & M      & S      \\
 Courts and Data Protection         &    0.88 &       0.03 &     0.82 &      0.86 &      0.88 &      0.9  &     0.93 & S      & S      \\
 Environmental and Workplace Safety &    0.36 &       0.07 &     0.26 &      0.31 &      0.36 &      0.42 &     0.47 & M      & R      \\
 Criminal Law and Justice           &    0.9  &       0.02 &     0.84 &      0.9  &      0.91 &      0.91 &     0.93 & S      & S      \\
 Personal and Consumption Taxes     &    0.58 &       0.05 &     0.53 &      0.55 &      0.56 &      0.59 &     0.69 & M      & S      \\
 Corporate Taxes                    &    0.9  &       0.02 &     0.86 &      0.88 &      0.89 &      0.91 &     0.94 & S      & S      \\
 Environmental Protection           &    0.29 &       0.06 &     0.2  &      0.23 &      0.3  &      0.33 &     0.39 & M      & R      \\
 Property                           &    0.89 &       0.02 &     0.84 &      0.89 &      0.9  &      0.91 &     0.93 & S      & S      \\
\bottomrule
\end{tabular}
	
	{\vspace*{6pt}\small \textbf{\textsf{(b)}}\quad Germany} 
\end{table}

\begin{table}
	\caption{Family classification by growth}\label{fig:family-growth}
	\renewcommand{\arraystretch}{1.075}
	\small\centering
	\begin{tabular}{lrrrrrll}
\toprule
 Family                                 &   $\Delta$ &   $\Delta_S$ &   $\Delta_S/\Delta$ &   $\Delta_R$ &   $\Delta_R/\Delta$ & Cat.   & Maj.   \\
\midrule
 Environmental and Health Protection    &   10805178 &       217268 &                0.02 &     10587910 &                0.98 & R      & R      \\
 Healthcare and Tax                     &    7642791 &       -15410 &               -0    &      7658201 &                1    & R      & R      \\
 Agriculture and Food                   &    3746072 &       411719 &                0.11 &      3334353 &                0.89 & R      & R      \\
 Federal Grants and Commercial Activity &    3025038 &       447388 &                0.15 &      2577650 &                0.85 & R      & R      \\
 Maritime Affairs and Transport         &    2490268 &       156868 &                0.06 &      2333400 &                0.94 & R      & R      \\
 Financial Regulation                   &    1391745 &       119549 &                0.09 &      1272196 &                0.91 & R      & R      \\
 Public Procurement and Funding         &    1010815 &       161332 &                0.16 &       849483 &                0.84 & R      & R      \\
 Energy                                 &    1709216 &       287289 &                0.17 &      1421927 &                0.83 & R      & R      \\
 Telecommunications                     &    1663230 &        59303 &                0.04 &      1603927 &                0.96 & R      & R      \\
 Housing                                &     398553 &        31541 &                0.08 &       367012 &                0.92 & R      & R      \\
\bottomrule
\end{tabular}
	
	{\vspace*{6pt}\small \textbf{\textsf{(a)}}\quad United States}\vspace*{12pt}

	\begin{tabular}{lrrrrrll}
\toprule
 Family                             &   $\Delta$ &   $\Delta_S$ &   $\Delta_S/\Delta$ &   $\Delta_R$ &   $\Delta_R/\Delta$ & Cat.   & Maj.   \\
\midrule
 Corporations and Banking           &     449339 &       289601 &                0.64 &       159738 &                0.36 & M      & S      \\
 Vocational Training                &     234346 &         3267 &                0.01 &       231079 &                0.99 & R      & R      \\
 Social Security                    &     305047 &       257164 &                0.84 &        47883 &                0.16 & S      & S      \\
 Courts and Data Protection         &     252257 &       187826 &                0.74 &        64431 &                0.26 & M      & S      \\
 Environmental and Workplace Safety &     340812 &       199907 &                0.59 &       140905 &                0.41 & M      & S      \\
 Criminal Law and Justice           &     104295 &       101712 &                0.98 &         2583 &                0.02 & S      & S      \\
 Personal and Consumption Taxes     &      32480 &        76191 &                2.35 &       -43711 &               -1.35 & S      & S      \\
 Corporate Taxes                    &      80372 &        71571 &                0.89 &         8801 &                0.11 & S      & S      \\
 Environmental Protection           &     144036 &        43994 &                0.31 &       100042 &                0.69 & M      & R      \\
 Property                           &     100231 &        90412 &                0.9  &         9819 &                0.1  & S      & S      \\
\bottomrule
\end{tabular}
	
	{\vspace*{6pt}\small \textbf{\textsf{(b)}}\quad Germany} 
	\vspace*{-6pt}
\end{table}

\vspace*{12pt}
\subsubsection{Clustering algorithm parametrisation}

In the following, we analyze the performance of our clustering algorithm under different parameter choices to ensure that our results are not artifacts of our parametrization. 
The statistics we report are based on the Normalized Mutual Information (NMI) and Adjusted Rand Index (ARI)%
---two metrics that are commonly used for pairwise comparisons of clustering results.

The NMI is an information-theoretic measure expressing how much information is shared between two clusterings.
It is scaled to range between $0$ (not similar at all) and 1 (identical), and defined as
\begin{align*}
	\text{NMI}(X;Y) = \frac{I(X;Y)}{\sqrt{H(X)H(Y)}}~,
\end{align*}
where $I(X;Y) = H(X;Y)-H(X\mid Y)-H(Y\mid X)$ is the mutual information between $X$ and $Y$, $H(X;Y)$ is the joint entropy of $X$ and $Y$, 
$H(X)$ and $H(Y)$ are the individual entropies, 
and $H(X\mid Y)$ and $H(Y\mid X)$ are the conditional entropies.
For more information on this measure, see \cite{strehl2002}.

The ARI is variant of the Rand Index (RI) adjusted for chance. 
The Rand Index is based on counting how many pairs of nodes are in the same clusters or in different clusters in both clusterings. 
It is defined as
\begin{align*}
	\text{RI}(X;Y) = \frac{a+b}{a+b+c+d}~,
\end{align*}
where $a$ is the number of node pairs that appear in the same cluster in both clusterings, 
$b$ is the number of node pairs that appear in different clusters in both clusterings, 
$c$ is the number of node pairs that appear in the same cluster in $X$ but in different clusters in $Y$, and
$d$ is the number of node pairs that appear in different clusters in $X$ but in the same cluster in $Y$. 
The ARI is defined as
\begin{align*}
	\text{ARI}(X;Y) = \frac{\text{RI}(X;Y) - \mathbb{E}[\text{RI}(X;Y)]}{1 - \mathbb{E}[\text{RI}(X;Y)]}~,
\end{align*}
where $\mathbb{E}[\text{RI}(X;Y)]$ is the expected RI when assuming that the $X$ and $Y$ partitions are constructed randomly, 
subject to having the original number of clusters and the original number of nodes in each cluster.
While the RI ranges between $0$ and $1$, the ARI is bounded from above by $1$ but may take negative values when the agreement between the clusterings is less than expected. 
Unrelated clusterings have an ARI close to $0$ and identical clusterings have an ARI of $1$. 
More information on this measure can be found in \cite{hubert1985}.

\vspace*{12pt}
\paragraph{Sensitivity analysis}

\begin{figure}
	\center
	\includegraphics[width=0.9\textwidth]{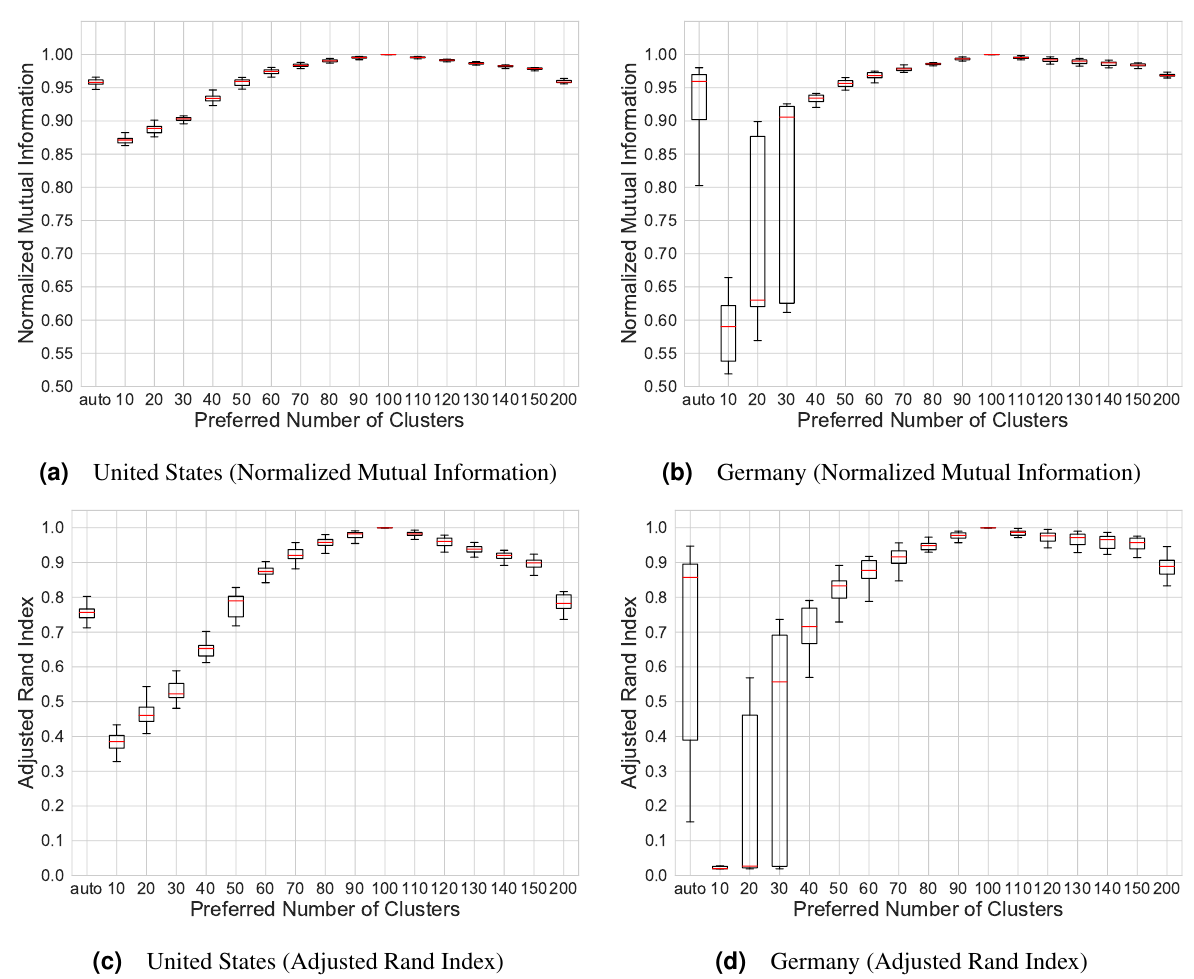}
	\caption{Distribution of pairwise similarities between clusterings with different preferred cluster sizes in the same year over the $22$ years from $1998$ to $2019$. 
		\emph{Auto} indicates that the \emph{Infomap} algorithm chooses the preferred number of clusters.
		Note that only the box plots labeled $10$ through $150$ are equidistant to each other on the real line. 
		The $y$-coordinates of the box boundaries indicate the second and fourth quartile, while the red line indicates the median. 
		Upper whiskers extend to the last data point less than $1.5$ times the box height above the fourth quartile, 
		while lower whiskers extend to the first data point less than $1.5$ times the box height below the first quartile.
	}\label{fig:sensitivity-preferred-clusters}
\end{figure}

Figure~\ref{fig:sensitivity-preferred-clusters} shows how the clustering results change when we alter the preferred number of clusters, 
with our chosen number of clusters as the baseline. 
As preferred numbers of clusters, we test all numbers divisible by $10$ from $10$ to $150$ as well as the number $200$. 
In one experiment, labeled \emph{auto}, we let the \emph{Infomap} algorithm choose the preferred number of clusters.
Unsurprisingly, the box plots show that clusterings become more similar to our baseline clustering with $100$ preferred clusters as we approach this number. 
At the same time, clusterings with $50$ or $200$ preferred clusters are already relatively similar to the baseline, with NMI values over $0.9$ and ARI values over $0.7$.

Note that the spread in clustering similarities is large for comparisons of the baseline with \emph{auto}, i.e., the clusterings in which \emph{Infomap} chooses the preferred number of clusters.
This is likely due to the jumps in clustering granularity that sometimes occur in \emph{Infomap} due to small differences in the minimum description length of competing models with different resolutions. 
These jumps also likely cause the spread for clusterings of the German data with a preferred number of clusters of around $20$ to $30$.
Avoiding these jumps is our primary motivation for specifying a preferred number of clusters.


\begin{figure}
	\center
	\includegraphics[width=0.9\textwidth]{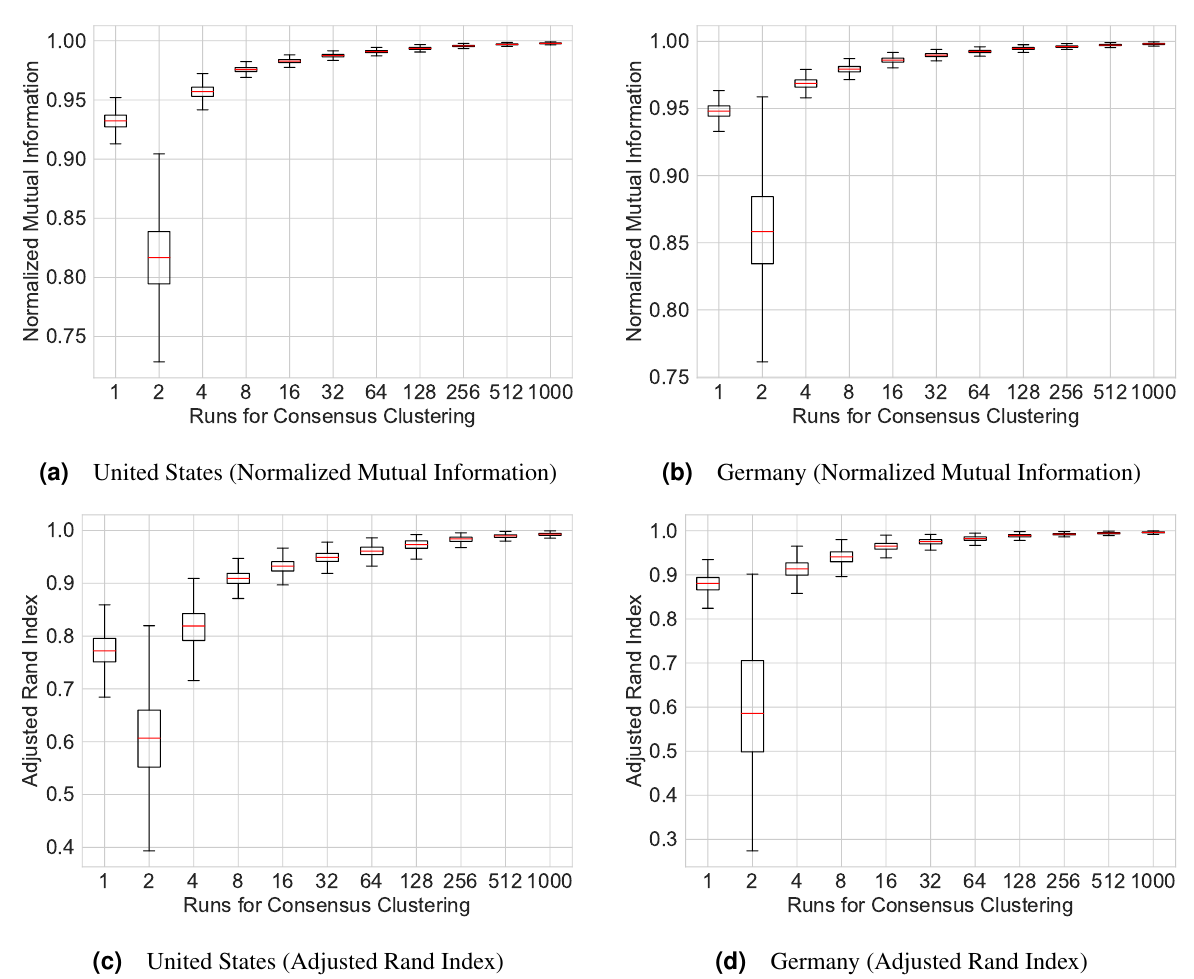}
	\caption{Pairwise similarity between $100$ consensus clustering results by number of clusterings used for finding the consensus (box plot interpretation as described in the caption of Figure~\ref{fig:sensitivity-preferred-clusters}).}
	\label{fig:consensus-effect}
\end{figure}

\vspace*{6pt}
\paragraph{Robustness checks}

Figure~\ref{fig:consensus-effect} shows the distribution of pairwise similarities between $100$ consensus clustering results for different numbers of clusterings used in the consensus. 
The plots show that using a higher number of clusterings to form the consensus increases the overall similarity level and reduces the spread between the observed similarities. 
When choosing $1000$ clusterings to form the consensus (as we do in the main paper), 
the consensus clusterings we obtain in different runs are almost identical.

\begin{figure}
	\center
	\includegraphics[width=0.9\textwidth]{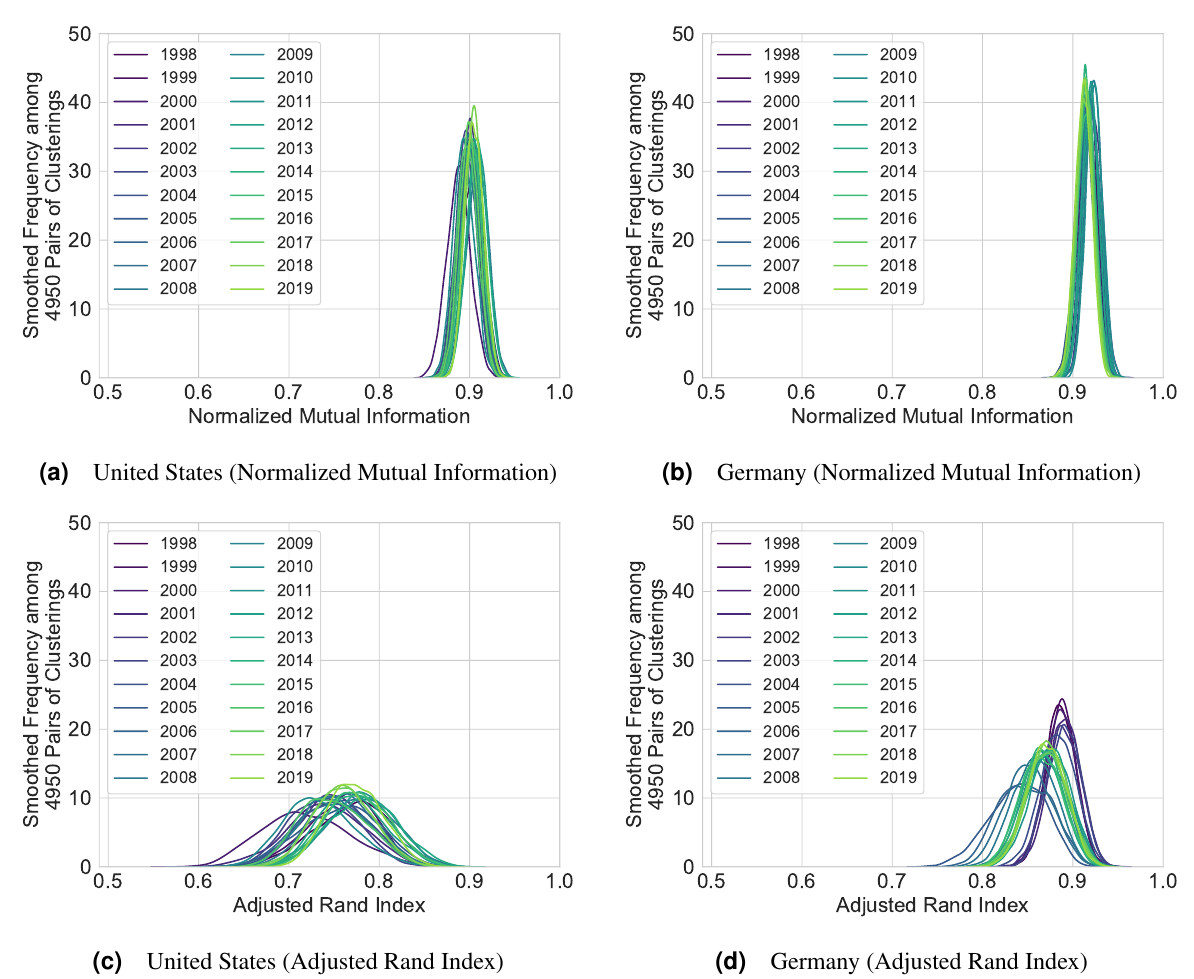}
			\caption{Pairwise similarity between $100$ clusterings with $100$ as the preferred number of clusters, depicted as kernel-density estimates rather than frequency histograms to reduce visual clutter.}
	\label{fig:consensus-within}
\end{figure}

Figure~\ref{fig:consensus-within} shows the distribution of pairwise similarities between $100$ pairs of clusterings (i.e., a total of $4950$ similarities) with $100$ as the preferred number of clusters. 
The NMI values for the United States clusterings mostly range between $0.85$ and $0.95$, while the NMI values for Germany mostly range between $0.88$ and $0.95$. 
The ARI values for the United States clusterings mostly range between $0.55$ and $0.85$ (with the majority lying between $0.60$ and $0.90$), 
while the ARI values for Germany mostly range between $0.75$ and $0.95$.
All similarity distributions seem to shift toward the left over time, 
i.e., clusterings in earlier years tend to be more similar to each other than clusterings in later years. 
This is likely due to the growth in complexity reported in the main paper.

\vspace*{12pt}
\subsubsection{Cluster family evolution}

\begin{figure}
	\centering
	\vspace*{-12pt}\includegraphics[width=0.9\textwidth]{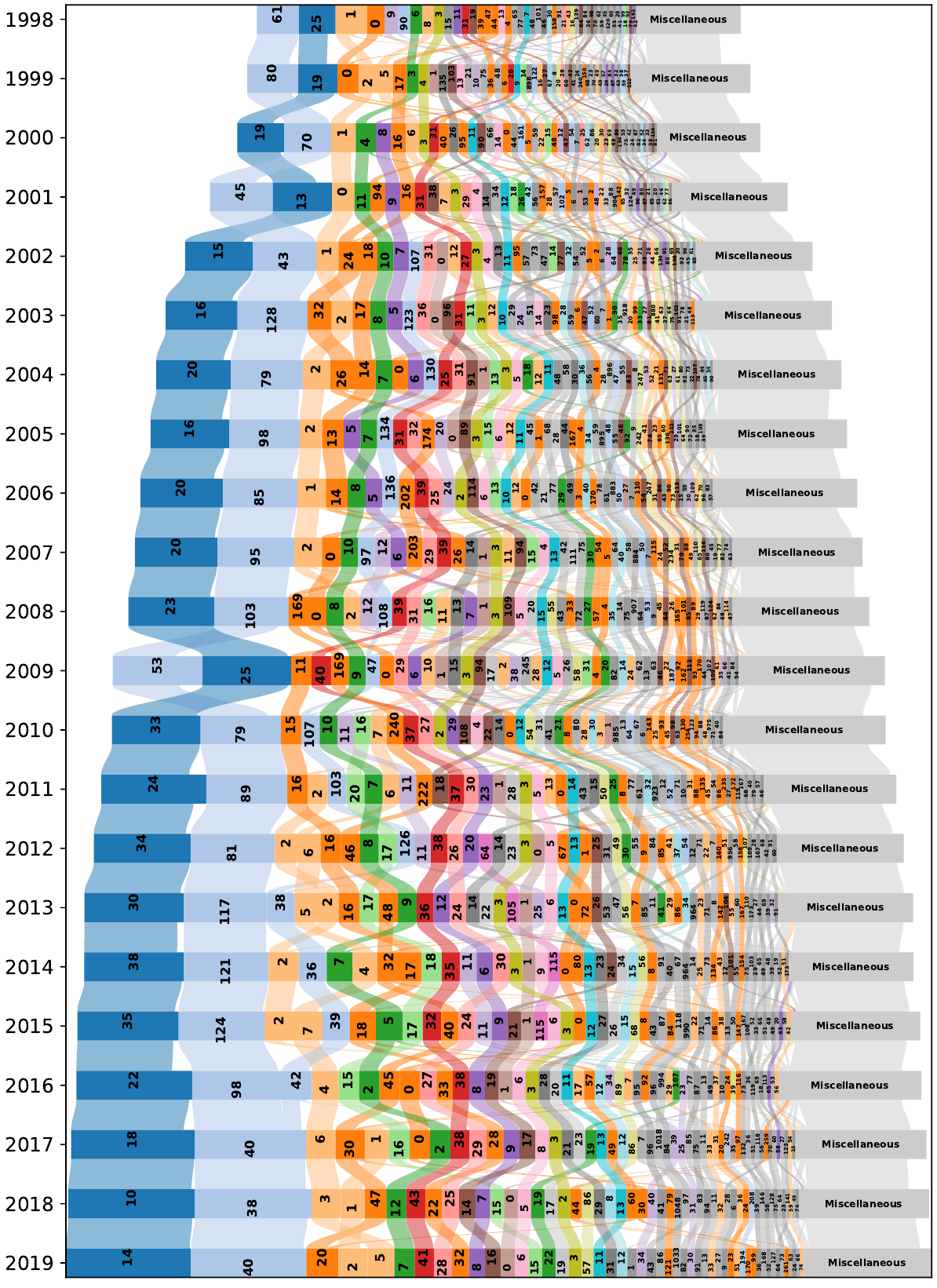}
	\caption{%
		Federal statutes and regulations in the United States by cluster (1998--2019), 
		with cluster numbers to enable content inspection. 
		Each block in each year represents a cluster. 
		Clusters are ordered from left to right by decreasing size (measured in tokens) and colored by the cluster family to which they belong, 
		where clusters not in one of the $20$ largest cluster families are colored in alternating greys. 
		Small clusters are summarized in one miscellaneous cluster, which is always the rightmost cluster and colored in light grey.
	}
	\label{fig:sankey-us-labels}
\end{figure}

\begin{figure}
	\centering
	\vspace*{-12pt}\includegraphics[width=0.9\textwidth]{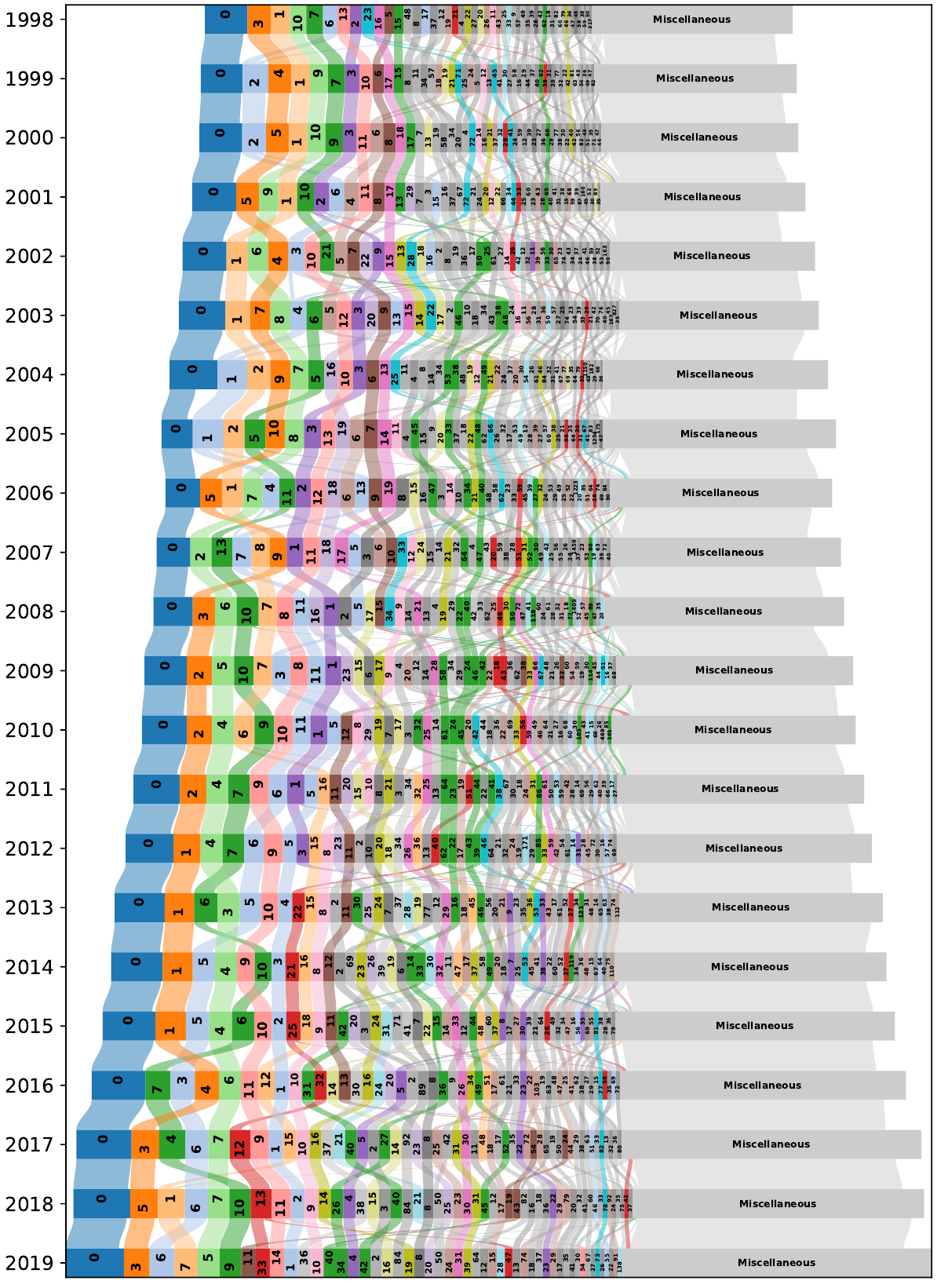}
	\caption{%
		Federal statutes and regulations in Germany by cluster (1998--2019), 
		with cluster numbers to enable content inspection. 
		Each block in each year represents a cluster. 
		Clusters are ordered from left to right by decreasing size (measured in tokens) and colored by the cluster family to which they belong, 
		where clusters not in one of the $20$ largest cluster families are colored in alternating greys. 
		Small clusters are summarized in one miscellaneous cluster, which is always the rightmost cluster and colored in light grey.
	}
	\label{fig:sankey-de-labels}
\end{figure}

In the main paper, we show how the composition of the top ten cluster families in the United States and Germany changes over time. 
Complementing this picture, Figure~\ref{fig:sankey-us-labels} and Figure~\ref{fig:sankey-de-labels} show the evolution of the top hundred cluster families in both countries.
The figures are generated using a procedure developed in \cite{katz2020}, where it is applied to data containing only statutes.

\vspace*{12pt}
\subsection{Profiles}
\begin{figure}
	\centering
	\includegraphics[width=\textwidth]{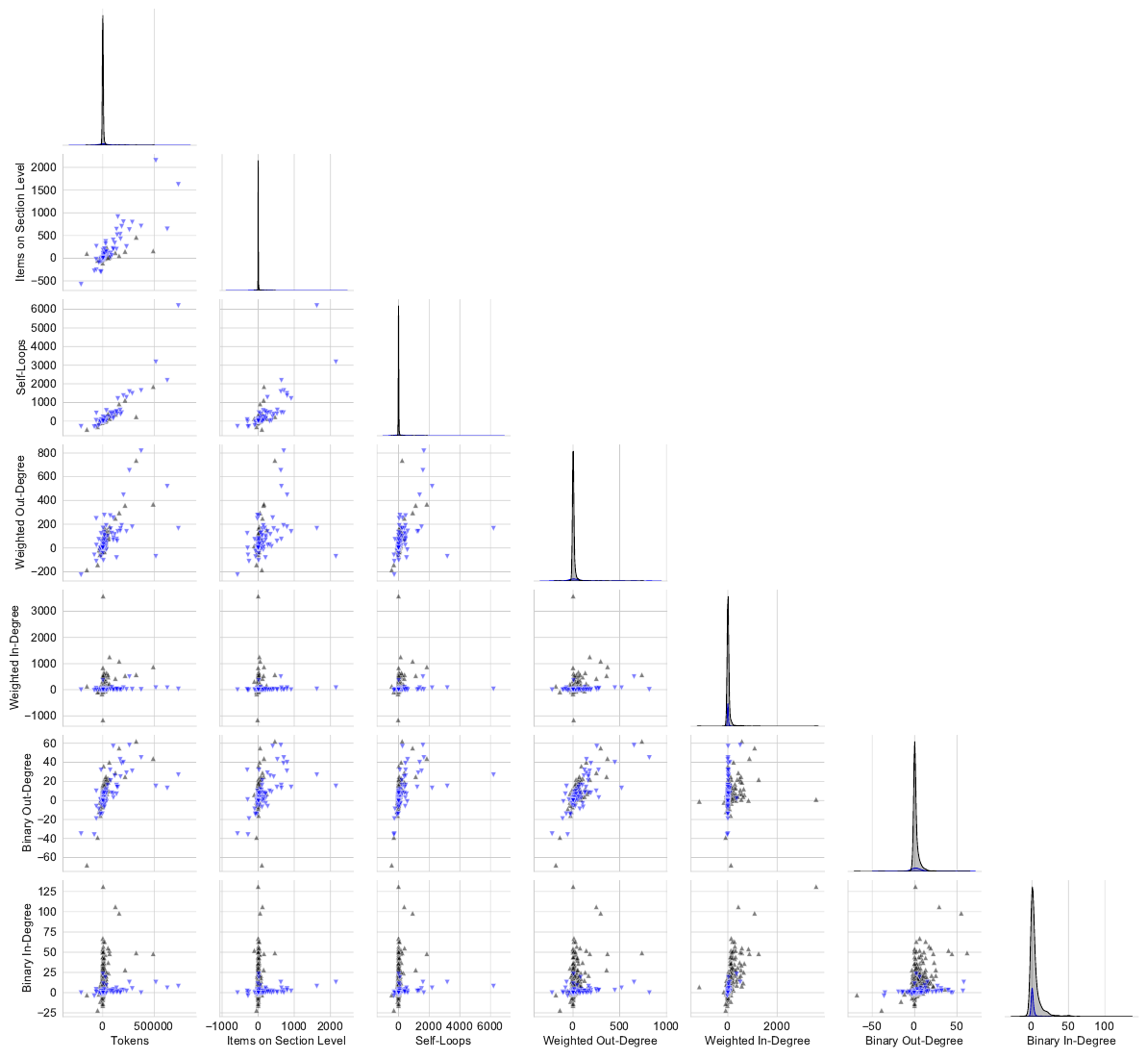}
	\caption{%
	Individual distribution (diagonal) and pairwise joint distribution (lower triangle) of changes in profile variables for USC (black) and CFR (blue) chapters present in $1998$ and $2019$, for seven of the ten indicators introduced in the main paper.
	}\label{fig:si-profile-trends}
\end{figure}

In the main paper, we show the profiles of selected statutes and regulations in a case study on financial regulation. 
Complementing this analysis, Figure~\ref{fig:si-profile-trends} depicts the macro-level trends we observe in the United States data as indicated by the individual and pairwise joint distributions of the changes (value in $2019$ minus value in $1998$) in seven of our ten indicators, for all USC and CFR chapters that exist in both $1998$ and $2019$.
By excluding chapters added after $1998$, we ensure that we measure growth over the same interval for all chapters.
Figure~\ref{fig:si-profile-trends} reveals, inter alia, that the change distributions are generally more concentrated for statutes than they are for regulations, and that the joint change distribution of \emph{binary in-degree} and \emph{items on section level} best (visually) separates statutes and regulations.
Furthermore, outliers in the (growth of) size and self-referentiality are mostly regulations, while outliers in in-degree indicators are mostly statutes, and outliers in out-degree indicators are of mixed document types.
For illustration, Table~\ref{tab:si-profile-trends} lists the USC or CFR chapters that experienced the largest absolute increase from $1998$ to $2019$ in one of our ten indicators.

\begin{table}
	\centering
	\small
	\caption{%
	Chapters of the USC and the CFR that experienced the largest absolute increase from $1998$ to $2019$ in one of the ten indicators introduced in the main paper, with CFR chapters marked grey.}\label{tab:si-profile-trends}
	\begin{tabular}{lp{0.625\textwidth}r}
		\toprule
		\textbf{Indicator}&\textbf{Chapter}&\textbf{$\Delta$}\\
		\midrule
		\rowcolor{lightgray!30}Tokens & 10 CFR Ch. I—NUCLEAR REGULATORY COMMISSION & 734462\\
		\rowcolor{lightgray!30}Unique tokens & 49 CFR Ch. V—NATIONAL HIGHWAY TRAFFIC SAFETY ADMINISTRATION, DEPARTMENT OF TRANSPORTATION & 27090\\\midrule
		\rowcolor{lightgray!30}Items above Section Level & 14 CFR Ch. I—FEDERAL AVIATION ADMINISTRATION, DEPARTMENT OF TRANSPORTATION & 301\\
		\rowcolor{lightgray!30}Items on Section Level & 14 CFR CHAPTER I—FEDERAL AVIATION ADMINISTRATION, DEPARTMENT OF TRANSPORTATION & 2152\\
		Items below Section Level & 42 USC Ch. 7—SOCIAL SECURITY & 22188\\\midrule
		\rowcolor{lightgray!30}Self-Loops & 10 CFR Ch. I—NUCLEAR REGULATORY COMMISSION & 6193\\\midrule
		\rowcolor{lightgray!30}Weighted Out-Degree & 12 CFR Ch. III—FEDERAL DEPOSIT INSURANCE CORPORATION & 817\\
		Weighted In-Degree & 5 USC Ch. 5—ADMINISTRATIVE PROCEDURE & 3580\\\midrule
		Binary Out-Degree & 42 USC Ch. 6A—PUBLIC HEALTH SERVICE & 62\\
		Binary In-Degree & 5 USC Ch. 5—ADMINISTRATIVE PROCEDURE & 131\\
		\bottomrule
	\end{tabular}
\end{table}

\begin{figure}
	\centering
	\vspace*{0pt}\includegraphics[width=\textwidth]{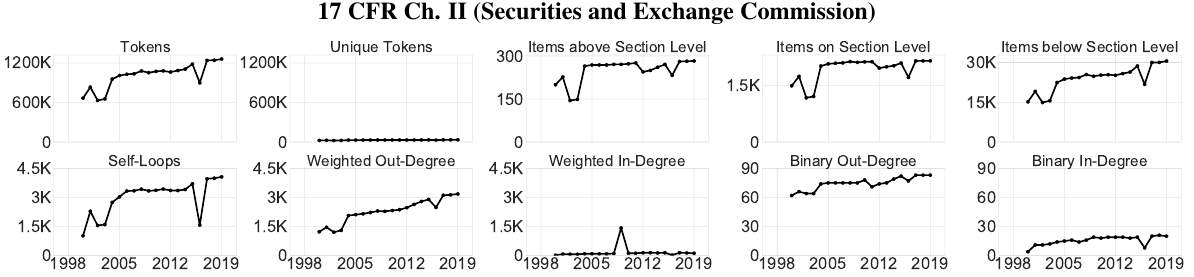}
	\caption{%
		Profile tracking the evolution of 17 CFR Ch. II (Securities and Exchange Commission),
		including several indicators highlighting data quality problems.
	}
	\label{fig:si-profile-17-cfr-ii}
\end{figure}

The profiles of individual statutes and regulations primarily serve to track their development over time.
Over the course of the analysis, however, we found that they are also helpful in highlighting potential problems with our dataset.
As visualized in Figure~\ref{fig:si-profile-17-cfr-ii},
17~CFR~Ch.~II, which deals with the Securities and Exchange Commission, displays some astonishing dynamics:
All metrics drop to varying degrees in $2002$ and $2003$, then rebound in $2004$.
Similarly, in $2016$, all these statistics fall sharply before rising back in $2017$.

Investigating these shifts, we find that they are indeed contained in the data.
In $2003$, large parts of 17~CFR~Ch.~II are contained within a tag designated \texttt{<SUPERSED>}, which---based on a comment in the document type definition---is used for ``superseded material''.
Since the relevant parts reappear in $2004$, we suspect that the closing tag is misplaced or missing.
As our extraction correctly disregards text within \texttt{<SUPERSED>} tags, all metrics drop depending on their sensitivity to lost text within this specific chapter.
The drop in $2016$ can be explained by a similar problem with a presumably accidentally extended \texttt{<REVTXT>} tag,
which---based on a comment in the document type definition---is used for ``text that is revised''.
This text, too, is not included in a \texttt{<REVTXT>} tag in the following year.
Both dynamics are therefore artifacts, which---once identified---could be mitigated by applying suitable rules or using the \emph{forward fill} strategy devised in Section~\ref{subsec:uscfr}.
We choose a separate regulation to demonstrate our framework in the main text and detail our observations here to help improve data quality over time.

A separate issue is indicated by the sharp spike in weighted in-degree, accompanied by a modest rise in binary in-degree, in $2009$.
Upon closer inspection, this increase stems from the fact that our data for this year contains a separate node titled ``17~CFR~Ch.~II~(CONTINUED)'' indicating that-–-as happens from time to time-–-the raw data contains two separate files that both contain text for the chapter in question.
We estimate the overall effect of this unexpected phenomenon, which occurs a total of fifteen times in the data,\footnote{%
The affected (sub)chapters are (in chronological, then numerical order):
43~CFR~Ch.~II ($1998$),
26~CFR~Subch.~A ($1998$),
26~CFR~Ch.~1 ($2000$-$2002$),
46~CFR~Ch.~I ($2006$, twice),
43~CFR~Ch~.~II ($2008$),
17~CFR~Ch.~II ($2009$),
42~CFR~Ch.~IV ($2009$),
10~CFR~Ch.~I ($2009$),
21~CFR~Ch.~I ($2009$), and
26~CFR~Ch.~I ($2009$, thrice).
} to be negligible for the overall analysis.
However, this needs to be addressed in future work, and it serves as a reminder that, despite the best efforts of the bodies managing the codification processes, the data quality can still be improved.
In the past, the respective public offices have been open toward and grateful for deficiencies pointed out by quantitative researchers,
and we will work with the U.S. Government Publishing Office to remove the errors we identify.

\newpage

\bibliographystyle{frontiersinHLTH&FPHY} 
\bibliography{bibliography.bib}